\documentclass{article}

\usepackage[left = 20 mm, right = 20 mm, top = 30 mm, bottom = 20 mm]{geometry}
\usepackage{amsmath, amssymb, verbatim, graphicx, subfig, caption, slashed, makecell, empheq, cite, jheppub}
\usepackage[toc,page]{appendix}

\newcommand{\minus}{\scalebox{0.75}[1.0]{$-$}}

\allowdisplaybreaks

\begin{document}

\title{Flavor mixing and renormalization in a perturbation theory}

\author{Chang-Hun Lee}
\affiliation{National Center for Theoretical Sciences, \\
101, Section 2, Kuang-Fu Road, Hsinchu, \\
Republic of China (Taiwan)}
\emailAdd{changhun@terpmail.umd.edu}

\begin{abstract}
{The renormalization of theories with flavor mixing is discussed, and it is shown that the physical unstable particles should be interpreted as quasiparticles which cannot be regarded as external states. Several popular beliefs on renormalization are disproved accordingly, and the limitations of physical renormalization schemes are discussed. In addition, the properties of unstable particles with flavor mixing such as decay widths are studied from scattering mediated by them.}
\end{abstract}

\maketitle

\section{Introduction}
In high-energy physics, many particles come in multiple flavors, and their mixing produces rich phenomenology. The mixing of mesons such as kaons, for example, causes CP violation in their decays. Heavy sterile Majorana neutrinos can mix with massless neutrinos of the Standard Model to generate small non-zero masses of light neutrinos by the seesaw mechanism. Moreover, the CP violation effects caused by the flavor mixing of heavy Majorana neutrinos might be able to explain the origin of matter by the mechanism called leptogenesis. In multi-Higgs models, the mixing of multiple scalar fields is also inevitable in general. Hence, the physics of flavor mixing has been one of the interesting topics in particle physics.

Nonetheless, there still exists subtle issues in the flavor mixing of unstable particles. In reference \cite{FlavMixGaugeInvWFR}, it was claimed that the fields of in- and out-states of a physical process require different field-strength renormalization matrices when they are unstable particles that undergo flavor mixing due to the absorptive part of the self-energy. Accepting such a result, references \cite{RenPropFermionFlavMix, PropMixRenMaj} discussed the renormalization of the propagators of the unstable Dirac and Majorana particles with flavor mixing up to the infinite order in perturbation. However, if two different field-strength renormalization matrices are needed for each physical unstable particle, we must have two different bare fields for each particle in order to have a single renormalized field (or vice versa). In that case, it is unclear how the Lagrangian density can be written in terms of those two bare fields and how we can obtain the bare propagator itself that is going to be renormalized. The standard Feynman rules cannot be used to calculate a dressed bare propagator from two different bare fields, which implies that this approach does not fully resolve all the subtleties therein. Besides, in more recent works \cite{BWApproxMixUnsP, LSZResNonDiagProp} that discussed the flavor mixing of unstable particles, the result of reference \cite{FlavMixGaugeInvWFR} was not appropriately considered. In reference \cite{BWApproxMixUnsP}, finite field-strength renormalization matrices were introduced to the purpose of fully eliminating mixing among external unstable particles. Reference \cite{LSZResNonDiagProp} tried to generalize and simplify the procedure to calculate the field-strength renormalization matrices. In those works, however, a single field-strength renormalization matrix was used to diagonalize a bare propagator matrix of unstable particles. Furthermore, in many works which studied the decays of unstable particles, a popular way to calculate the decay width is defining effective couplings to incorporate all the outcomes of mixing, and the loop-corrected decay widths are obtained by replacing the couplings in the tree-level formula of the decay width with those effective couplings. For example, the decay widths of heavy Majorana neutrinos are usually calculated in such a way by using effective Yukawa couplings \cite{CPVMaj, ResLepto, CPMajDec, CPResLepto}. By the essentially identical reason that required two different field-strength renormalization matrices in reference \cite{FlavMixGaugeInvWFR}, the decay and inverse decay gets determined by two different effective Yukawa coupling matrices that are not simply related to each other by Hermitian conjugation, which results in CP violation.

The common approach in all those works is to regard the unstable particles as external states of a physical process. However, in reference \cite{UnitUnsP} which studied a single flavor of an unstable particle, it was shown that the one-particle state of an unstable particle as an external state does not contribute to the unitarity cut of a diagram such as figure \ref{fig:UnitCut} mediated by the field of the unstable particle, and the unitarity cut is determined by the multiparticle states consisting of the stable particles of the decay product. This implies that the understanding of a decay as a phenomenon in which a one-particle state evolves in time and decays into stable particles is not correct. Instead, it should be understood from the scattering mediated by the unstable particle. This is a natural result since a decaying particle cannot be an external state which is an asymptotically  free state that exists at $t = \pm \infty$ in a perturbation theory. Furthermore, in the Lehmann-Symanzik-Zimmermann (LSZ) reduction formula which relates the correlation function of a physical process to a finite $S$-matrix element, there exist divergences at the poles of external fields in the correlation function, and they can be canceled out to give a non-vanishing finite $S$-matrix element, only when the divergences exist at real poles, \textit{i.e.}, only when the external states are stable particles. Hence, even though we can obtain, for example, a correct decay width in the case of a single flavor by taking the unstable particle as an external state and using equation \ref{eq:DWFormula} while neglecting the loop corrections to the field strength as conventionally done, such a method cannot be rigorously justified by itself and it is unclear whether that approach is applicable to the case of multiple flavors as well.

The purpose of this paper is providing the answers to all those problems. In order to have the canonical expressions of the Lagrangian density in terms of the canonically quantized bare as well as renormalized fields to which standard Feynman rules can be applied, a single field-strength renormalization matrix should be used for the fields of unstable particles. Instead, two different finite mixing matrices will be needed to diagonalize a renormalized non-diagonal propagator of unstable particles. In other words, two different mixing matrices are involved in each of physical particles associated with the complex poles, and thus it cannot be related to a single renormalized field. Such a particle will be interpreted as a \textit{quasiparticle}, \textit{i.e.}, an emergent particle generated by the interactions of various fields in the theory. Such particles cannot be regarded as external states, and their properties can be correctly studied only from a physical process in which they appear as intermediate particles. Specifically, we will discuss how to calculate the decay widths of such physical particles from scattering mediated by those unstable particles, using the unitarity cut that generate those scattering processes, and will explicitly see that the popular method of calculating decay widths by using effective couplings as mentioned above is wrong. Furthermore, it will turn out that each of physical particles is in a sense unphysical, because it is impossible to separately observe each of them and any observable quantity related to them is a result of their interferences. Since the only consistent way of obtaining the physical particles is treating them as intermediate states, the on-shell or complex-mass renormalization schemes, \textit{i.e.}, the physical renormalization schemes, cannot be applied to the fields of unstable particles with flavor mixing. Even though each of physical particles propagates like a free particle until it decays as the one associated with each component of the diagonalized propagator, it cannot be related to a single one-particle state in the Hilbert space, as explained. This means that the unstable particles with flavor mixing cannot be described in the frameworks of quantum mechanics, in which a particle is regarded as a single state vector in the Hilbert space that satisfies the Schr\"odigner equation. This has an important implication, since quantum mechanics has been the standard way to study the mixing of neutral mesons in the literature and several anomalies have been reported in the decays of mesons. The limitations of quantum mechanics in the physics of neutral meson mixing were discussed in detail in reference \cite{MixingQFT}.

In addition, the effect of mass differences between flavors has not been appropriately examined in the works cited above. In particular, we will see that, if the mass differences are small, there must exists a large difference between the fields of physical unstable particles and the canonically quantized renormalized fields in order not to ruin perturbativity, which is another reason why the physical renormalization schemes are inapplicable to multiple flavors. Furthermore, it will be shown that, depending on the mass differences, the loop corrections to the self-energy of the unstable particles can generate subleading or leading corrections to the $S$-matrix element or mixing matrices. Hence, it is important not to introduce perturbative assumptions in diagonalizing propagators to correctly obtain the mixing matrices and to discern the effect of mass differences on observable quantities such as decay widths and CP violation.

This paper in a sense aims at disproving some popular beliefs on renormalization. To be specific, we will prove that the following statements are false when there exists mixing among multiple flavors of unstable particles:

\begin{enumerate}
	\item It is always possible to find a single renormalized field associated with each physical particle.

	\item A theory can always be renormalized in a physical scheme such that the Lagrangian density is written in terms of the fields of physical particles.

	\item A bare propagator matrix of bare fields can always be renormalized into a diagonal form by a single field-strength renormalization matrix. In the case of complex scalars, a bare propagator $i \Delta_{\Phi_0} (p^2)$ of bare complex scalar fields $\Phi_{0 \alpha}$ can always be renormalized by a field-strength renormalization matrix $Z_\Phi$ such that
	\begin{align}
		i (\Delta_{\Phi_0})_{\beta \alpha} (p^2) = \sum_\gamma (Z_\Phi^{\frac{1}{2}})_{\beta \gamma}
			\frac{i}{p^2 - p_{\Phi_\gamma}^2}
			(Z_\Phi^{\frac{1}{2} \dag})_{\gamma \alpha} + \cdots,
		\label{eq:RenPropWrong}
	\end{align}
	where $p_{\Phi_\gamma}^2$ is a complex-valued pole of the propagator and the ellipsis denotes the subleading terms of the Laurent series. Note that this statement has already been disproved in reference \cite{FlavMixGaugeInvWFR} although it does not appear to be widely accepted yet.
	
	\item The fields of unstable particles can always be renormalized such that those particles can be regarded as external states, and the matrix elements of their decays can be calculated by defining effective vertices $\widehat{V}$ as follows:
	\begin{align*}
		\parbox{16 mm}{\includegraphics[width = 16 mm]{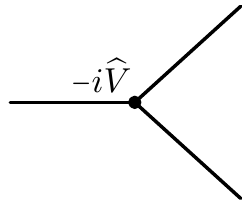}} \quad = \quad
		\parbox{16 mm}{\includegraphics[width = 16 mm]{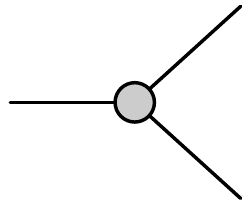}} + \
		\parbox{20 mm}{\includegraphics[width = 20 mm]{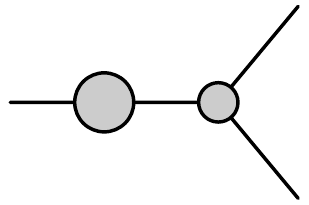}} + \
		\parbox{28 mm}{\includegraphics[width = 28 mm]{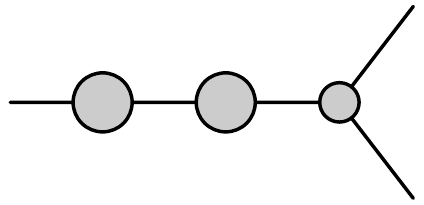}} + \ \cdots.
	\end{align*}
	
\end{enumerate}

In this paper, we will only consider a simple toy model of complex scalar fields that mix with each other by the loop corrections induced by Yukawa interactions. The Dirac or Majorana fields introduce complication especially in chiral theories, which obscures the subtleties in flavor mixing. To clearly understand the physics of unstable particles, we will carefully discuss the mixing of stable particles as well for comparison. Many common errors in the literature originate from mistakenly applying a theoretical treatment only appropriate for stable particles to unstable ones. Moreover, we will work in the plane-wave limit in which an external particle of a physical process is described by a momentum eigenstate, \textit{i.e.}, a one-particle state in the Hilbert space. A wave-packet description of a particle and its effect are beyond the scope of this study.

This paper is organized as follows: in section \ref{sec:QuasiFlavMixing}, a toy model of complex scalar fields is introduced, and it is renormalized in a non-physical scheme. The dressed propagator of the scalar fields is diagonalized to obtain the fields of physical particles. In section \ref{sec:Quant&Phys}, the properties of physical particles are discussed, and it is explained why unstable particles should be studied in a very different way from stable ones. In section \ref{sec:RenScheme}, the limitations of physical renormalization schemes are discussed. In section \ref{sec:Example}, the case of two flavors are studied analytically and numerically, and several examples are provided to confirm the facts derived in the paper.

\section{Physical particles in the presence of flavor mixing}		\label{sec:QuasiFlavMixing}
In this section, we discuss the flavor mixing of complex scalar fields using a simple toy model in which the scalar fields mix with each other by the loop corrections induced by Yukawa interactions. The theory will be renormalized step-by-step, and the fields of physical particles will be identified. It will be shown that, in the case of unstable particles, each of physical particles cannot be related to a single renormalized field, and it should be interpreted as a quasiparticle. Accordingly, some popular beliefs about renormalization will turn out to be false. Moreover, when the mass differences between flavors are small, \textit{i.e.}, $|m_{\Phi_\beta} - m_{\Phi_\alpha}| \lesssim m_{\Phi_\alpha} \mathcal{O} (\alpha)$ where $\alpha$ of $\mathcal{O} (\alpha)$ is a small parameter of perturbation, we will see that the flavor mixing becomes a non-perturbative effect, and such a case should be carefully studied not to ruin the perturbativity.

\subsection{Toy model and its renormalization}		\label{sec:ToyRen}
Here, we discuss the renormalization of the toy model. The calculation will be performed only up to the one-loop order in the self-energy, not to unnecessarily complicate the discussion while showing the subtleties that exist in the case of multiple flavors. However, it should be emphasized that the diagonalization of the propagator in section \ref{sec:Quasi} is correct up to the infinite order in perturbation.

The toy model consists of complex scalars $\Phi_\alpha$ and fermions $\chi_i$, $\xi$. Note that, in this paper, the Greek indices will denote the flavors of complex scalars while the Latin indices will be used for the flavors of fermions. In the toy model, they interact with each other by Yukawa interactions
\begin{align}
	\mathcal{L}_\text{int} = \minus \sum_{i, \alpha} f_{i \alpha} \overline{\chi_i} \xi \Phi_\alpha + \text{H.c.},
\end{align}
which induces mixing among flavors of $\Phi$ and $\chi$. Here, $\Phi_\alpha (x) \coloneqq e^{i H t} \Phi_\alpha (\mathbf{x}) e^{-i H t}$ denotes the field in the Heisenberg picture, where $H$ is the Hamiltonian and $\Phi_\alpha (\mathbf{x}) = \Phi_\alpha (0, \mathbf{x})$ is the field in the Schr\"odinger picture that creates $| \Phi (m_{\Phi_\alpha}; \mathbf{p}) \rangle$ from the vacuum $| 0 \rangle$ of the free theory. Note that $| \Phi (m_{\Phi_\alpha}; \mathbf{p}) \rangle$ is the one-particle state of $\Phi$ with mass $m_{\Phi_\alpha}$ and three-momentum $\mathbf{p}$. The field $\Phi_\alpha (x)$ at $t = x^0 \neq 0$ does not necessarily create $| \Phi (m_{\Phi_\alpha}; \mathbf{p}) \rangle$ due to the flavor mixing.

\begin{figure}[t]
	\centering
	\subfloat[$(\Sigma_{\Phi_0})_{\beta \alpha} (p^2)$]{
		\includegraphics[width = 45 mm]{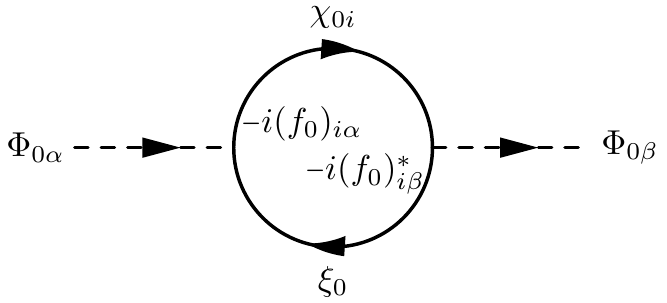}
		\label{fig:SEPBare}
	} \qquad
	\subfloat[$(\Sigma_{\Phi_0^*})_{\beta \alpha} (p^2)$]{
		\includegraphics[width = 45 mm]{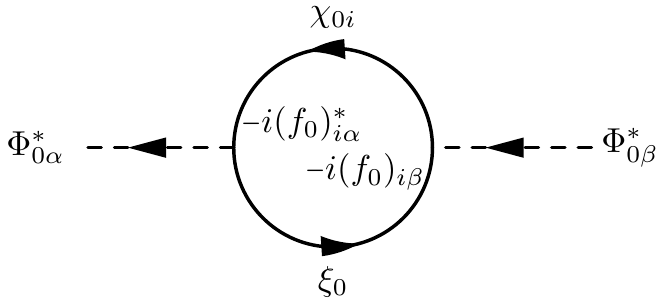}
		\label{fig:SEPcBare}
	}
	\caption{One-loop contributions to the self-energies of $\Phi_0$ and $\Phi_0^*$.}
	\label{fig:SEBare1Loop}
\end{figure}

\begin{figure}[t]
	\centering
	\subfloat[$(\Sigma_{0 \Phi})_{\beta \alpha} (p^2)$]{
		\includegraphics[width = 45 mm]{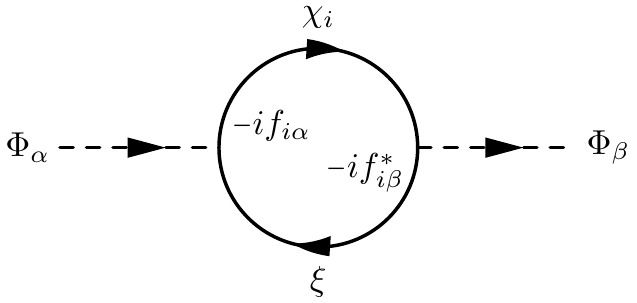}
		\label{fig:SEP}
	} \qquad
	\subfloat[$(\Sigma_{0 \Phi^*})_{\beta \alpha} (p^2)$]{
		\includegraphics[width = 45 mm]{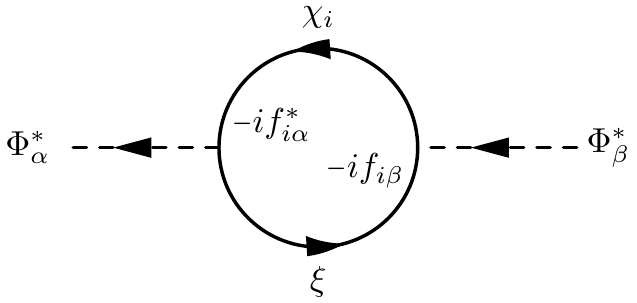}
		\label{fig:SEPc}
	}
	\caption{One-loop contributions to the self-energies of $\Phi$ and $\Phi^*$.}
	\label{fig:SE1Loop}
\end{figure}

To discuss renormalization, let us introduce bare scalar fields $\Phi_{0 \alpha}$ whose mass is $m_{\Phi_{0 \alpha}}$, and write the Lagrangian density which involves the complex scalar fields as
\begin{align}
	\mathcal{L} = \sum_\alpha \partial^\mu \Phi_{0 \alpha}^\dag \partial_\mu \Phi_{0 \alpha} - \sum_\alpha m_{\Phi_{0 \alpha}}^2 \Phi_{0 \alpha}^\dag \Phi_{0 \alpha}
		- \sum_{i, \alpha} (f_0)_{i \alpha} \overline{\chi_{0 i}} \xi_0 \Phi_{0 \alpha} - \sum_{i, \alpha} (f_0)_{i \alpha}^* \overline{\xi_0} \chi_{0 i} \Phi_{0 \alpha}^\dag,
\end{align}
where $f_0$ is the bare Yukawa coupling matrix. For some given renormalized fields $\Phi_\alpha$, $\chi_i$, and $\xi$, the field-strength renormalization matrices $Z_\Phi$, $Z_\chi$, and $Z_\xi$ are defined by
\begin{align}
	\Phi_{0 \beta} \eqqcolon \sum_\alpha (Z_\Phi^\frac{1}{2})_{\beta \alpha} \Phi_\alpha, \qquad
	\chi_{0 j} \eqqcolon \sum_i (Z_\chi^\frac{1}{2})_{ji} \chi_i, \qquad
	\xi_0 \eqqcolon Z_\xi^\frac{1}{2} \xi,
\end{align}
and the counterterm matrices $\delta_\Phi$ and $\delta_\xi$ are
\begin{align}
	Z_\Phi^\frac{1}{2} \eqqcolon U_\Phi \bigg( 1 + \frac{1}{2} \delta_\Phi \bigg), \qquad
	Z_\chi^\frac{1}{2} \eqqcolon U_\chi \bigg( 1 + \frac{1}{2} \delta_\xi \bigg).
\end{align}
Here, $U_\Phi$ and $U_\chi$ are unitary matrices, which are required because, in the basis where the mass matrices of $\Phi_0$ and $\chi_0$ are diagonal, $Z_\Phi^\frac{1}{2}$ and $Z_\chi^\frac{1}{2}$ cannot be written as small perturbative corrections to the identity matrix in general. Moreover, for the given diagonal bare mass matrix $(M_{\Phi_0})_{\beta \alpha} = m_{\Phi_{0 \alpha}} \delta_{\beta \alpha}$ and a renormalized mass matrix $(M_\Phi)_{\beta \alpha} = m_{\Phi_\alpha} \delta_{\beta \alpha}$, let us define the mass renormalization matrix $Z_M$ and mass counterterm matrix $\delta M_\Phi^2$ as
\begin{align}
	M_{\Phi_0}^2 \eqqcolon Z_M^\dag M_\Phi^2 Z_M, \qquad
	\delta M_\Phi^2 \coloneqq Z_\Phi^{\frac{1}{2} \dag} Z_M^\dag M_\Phi^2 Z_M Z_\Phi^\frac{1}{2} - M_\Phi^2,
	\label{eq:MassRen}
\end{align}
where $\delta M_\Phi^2$ is a Hermitian matrix that is non-diagonal in general. In addition, for a given renormalized Yukawa coupling $f$, we also define the vertex counterterm $\delta f$ as
\begin{align}
	Z_\xi^\frac{1}{2} (Z_\chi^{\frac{1}{2} \dag} f_0 Z_\Phi^\frac{1}{2})_{i \alpha} \eqqcolon f_{i \alpha} + \delta f_{i \alpha}.
\end{align}
The Lagrangian density can then be rewritten as
\begin{align}
	\mathcal{L} &= \sum_\alpha \partial^\mu \Phi_{0 \alpha}^\dag \partial_\mu \Phi_{0 \alpha} - \sum_{\alpha} m_{\Phi_{0 \alpha}}^2 \Phi_{0 \alpha}^\dag \Phi_{0 \alpha}
		- \sum_{i, \alpha} (f_0)_{i \alpha} \overline{\chi_{0 i}} \xi_0 \Phi_{0 \alpha} - \sum_{i, \alpha} (f_0)_{i \alpha}^* \overline{\xi_0} \chi_{0 i} \Phi_{0 \alpha}^\dag \nonumber \\
	&= \sum_{\alpha, \beta} (Z_\Phi^{\frac{1}{2} \dag} Z_\Phi^\frac{1}{2})_{\beta \alpha} \partial^\mu \Phi_\beta^\dag \partial_\mu \Phi_\alpha - \sum_{\alpha, \beta} (Z_\Phi^{\frac{1}{2} \dag} Z_M^\dag M_\Phi^2 Z_M Z_\Phi^\frac{1}{2})_{\beta \alpha} \Phi_\beta^\dag \Phi_\alpha \nonumber \\
		&\qquad - \sum_{i, \alpha} Z_\xi^\frac{1}{2} (Z_\chi^{\frac{1}{2} \dag} f_0 Z_\Phi^\frac{1}{2})_{i \alpha} \overline{\chi_i} \xi \Phi_\alpha - \sum_{i, \alpha} Z_\xi^{\frac{1}{2} *} (Z_\chi^{\frac{1}{2} \dag} f_0 Z_\Phi^\frac{1}{2})_{i \alpha}^* \overline{\xi} \chi_i \Phi_\alpha^\dag \nonumber \\
	&= \sum_\alpha \partial^\mu \Phi_\alpha^\dag \partial_\mu \Phi_\alpha - \sum_\alpha m_{\Phi_\alpha}^2 \Phi_\alpha^\dag \Phi_\alpha - \sum_{i, \alpha} f_{i \alpha} \overline{\chi_i} \xi \Phi_\alpha - \sum_{i, \alpha} f_{i \alpha}^* \overline{\xi} \chi_i \Phi_\alpha^\dag \nonumber \\
		&\qquad + \sum_{\alpha, \beta} \frac{1}{2} (\delta_\Phi^\dag + \delta_\Phi + \cdots)_{\beta \alpha} \partial^\mu \Phi_\beta^\dag \partial_\mu \Phi_\alpha
		- \sum_{\alpha, \beta} (\delta M_\Phi^2)_{\beta \alpha} \Phi_\beta^\dag \Phi_\alpha \nonumber \\                             
		&\qquad - \sum_{i, \alpha} \delta f_{i \alpha} \overline{\chi_i} \xi \Phi_\alpha - \sum_{i, \alpha} \delta f_{i \alpha}^* \overline{\xi} \chi_i \Phi_\alpha^\dag.
\end{align}
In this paper, we will assume that $m_{\chi_i}$ are so small that we may regard $\chi_i$ as massless to a given precision of perturbative calculations. In dimensional regularization, the self-energy of $\Phi_0$ up to the one-loop order is given by
\begin{align}
	\Sigma_{\Phi_0} (p^2) = p^2 \Sigma_{\Phi_0}' (p^2) + \delta \Sigma_{\Phi_0},
\end{align}
where
\begin{align}
	(\Sigma_{\Phi_0}')_{\beta \alpha} &= \sum_i \frac{(f U_\Phi^\dag)_{i \beta}^* (f U_\Phi^\dag)_{i \alpha}}{16 \pi^2} \bigg[ \frac{2}{\epsilon} + \frac{3}{2} - \log{\frac{m_\xi^2}{\widetilde{\mu}^2}} - \bigg( 1 - \frac{5 m_\xi^2}{p^2} + \frac{6 m_\xi^4}{p^4} - \frac{2 m_\xi^6}{p^6} \bigg) \log{\frac{m_\xi^2 - p^2}{m_\xi^2}} \bigg] + \mathcal{O} (\epsilon),
		\label{eq:SEBare} \\
	(\delta \Sigma_{\Phi_0})_{\beta \alpha} &= \minus \sum_i \frac{(f U_\Phi^\dag)_{i \beta}^* (f U_\Phi^\dag)_{i \alpha}}{8 \pi^2} m_\xi^2 \bigg( \frac{2}{\epsilon} + 1 - \log{\frac{m_\xi^2}{\widetilde{\mu}^2}} \bigg) + \mathcal{O} (\epsilon).
		\label{eq:SEBare1}
\end{align}
Here, we have used $f_0 = U_\chi f U_\Phi^\dag$ which is valid up to the leading order and also used the identity $(f U_\Phi^\dag)^\dag (f U_\Phi^\dag) = (U_\chi f U_\Phi^\dag)^\dag (U_\chi f U_\Phi^\dag)$. We have also introduced $\widetilde{\mu}^2 \coloneqq 4 \pi e^{-\gamma_E} \mu^2$, where $\gamma_E$ is the Euler–Mascheroni constant and $\mu^{2 - d/2} f$ is the Yukawa coupling in the $d = 4 - \epsilon$ dimension. This expression of the self-energy is explicitly derived in appendix \ref{sec:SECal}. We can also deduce the self-energy of $\Phi_0^*$, where $\Phi_{0 \alpha}^* \coloneqq \Phi_{0 \alpha}^\dag$ is the charge conjugate of $\Phi_{0 \alpha}$, and it can be written as
\begin{align}
	\Sigma_{\Phi_0^*} (p^2) = p^2 \Sigma_{\Phi_0^*}' (p^2) + \delta \Sigma_{\Phi_0^*}
	= \Sigma_{\Phi_0}^{\mathsf{T}} (p^2).
\end{align}
Here, the superscript $\mathsf{T}$ denotes the transpose of a matrix. Note that $(\Sigma_{\Phi_0})_{\beta \alpha}$ is related to the transition $\Phi_{0 \alpha} \to \Phi_{0 \beta}$, while $(\Sigma_{\Phi_0^*})_{\beta \alpha}$ is to the transition $\Phi_{0 \alpha}^* \to \Phi_{0 \beta}^*$. The associated diagrams are shown in figure \ref{fig:SEBare1Loop}.
The renormalized self-energy is defined by
\begin{align}
	\boxed{\Sigma_\Phi (p^2) \coloneqq p^2 \Sigma_\Phi' (p^2) + \delta \Sigma_\Phi - \delta M_\Phi^2,}
	\label{eq:SERenwCT}
\end{align}
where
\begin{align}
	\boxed{1 + \Sigma_\Phi' (p^2) \coloneqq Z_\Phi^{\frac{1}{2} \dag} [1 + \Sigma_{\Phi_0}' (p^2)] Z_\Phi^\frac{1}{2}, \qquad
	\delta \Sigma_\Phi \coloneqq Z_\Phi^{\frac{1}{2} \dag} \delta \Sigma_{\Phi_0} Z_\Phi^\frac{1}{2},}
	\label{eq:SERen}
\end{align}
Up to the one-loop order, we can write
\begin{align}
	\boxed{\Sigma_\Phi' (p^2) = (U_\Phi^\dag \Sigma_{\Phi_0}' U_\Phi) (p^2) + \delta_\Phi^H, \qquad
	\delta \Sigma_\Phi = U_\Phi^\dag \delta \Sigma_{\Phi_0} U_\Phi.}
	\label{eq:SERen1Loop}
\end{align}
Here, the superscript $H$ denotes the Hermitian part of the corresponding matrix. Note that a matrix can always be decomposed into Hermitian and skew-Hermitian parts: for an arbitary matrix $X$, we can write $X = X^H + X^S$ where $X^H \coloneqq (X + X^\dag) / 2$ and $X^S \coloneqq (X - X^\dag) / 2$. Hence, the absorptive (skew-Hermitian) parts of $\Sigma_\Phi' (p^2)$ and $(U_\Phi^\dag \Sigma_{\Phi_0}' U_\Phi) (p^2)$ must be identical because they are not affected by renormalization since $\delta_\Phi^H$, $\delta \Sigma_\Phi$, and $\delta M_\Phi^2$ are all Hermitian. The skew-Hermitian part of $\delta_\Phi$ has no role in renormalization, and it may therefore be chosen to be a Hermitian matrix, \textit{i.e.}, $\delta_\Phi = \delta_\Phi^H$. Moreover, we may also choose $\delta M_\Phi^2 = \delta \Sigma_\Phi$ such that
\begin{align}
	\boxed{\Sigma_\Phi (p^2) = p^2 \Sigma_\Phi' (p^2).}
\end{align}
Alternatively, we can calculate a non-renormalized self-energy $\Sigma_{0 \Phi} (p^2)$ of $\Phi$ rather than the self-energy $\Sigma_{\Phi_0} (p^2)$ of $\Phi_0$ by absorbing the field-strength renormalization matrices in $f_0$ into the bare fields in figure \ref{fig:SEBare1Loop} and amputating the external fields, as is usually done. The one-loop diagrams of $\Sigma_{0 \Phi} (p^2)$ and $\Sigma_{0 \Phi^*} (p^2)$ are shown in figure \ref{fig:SE1Loop}.
The coupling is $f$ rather than $f U_\Phi^\dag$ in that approach, and the non-renormalized self-energy is then given by
\begin{align}
	\boxed{\Sigma_{0 \Phi} (p^2) = p^2 \Sigma_{0 \Phi}' (p^2) + \delta \Sigma_\Phi, \qquad
	\Sigma_{0 \Phi}' (p^2) \coloneqq (U_\Phi^\dag \Sigma_{\Phi_0}' U_\Phi) (p^2).}
	\label{eq:SENonRen}
\end{align}
Those two alternative self-energies will be used interchangeably.

In this paper, we consider two different mass hierarchies: (i) $m_{\Phi_\alpha} \gg m_{\chi_i} \sim m_\xi \sim 0$; (ii) $m_\xi > m_{\Phi_\alpha} \gg m_{\chi_i} \sim 0$. In the first case, all the flavors of $\Phi_\alpha$ are unstable because of the decay $\Phi_\alpha \to \chi_i \xi^c$, and the self-energy has an absorptive part. In contrast, $\chi_i$ and $\xi$ can be considered to be stable up to the one-loop order in the self-energy, since the decay such as $\chi_i \to (\Phi_\alpha \xi \to) \chi_j \xi^c \xi$ comes from the cutting of a two-loop self-energy diagram of $\chi_i$. For simplicity, we further assume that the masses of $\chi_i$ and $\xi$ are negligible to working precision, which can always be satisfied if their tree-level masses are small enough. In the second case, $\Phi_\alpha$ and $\chi_i$ can be regarded as stable up to the one-loop order in the self-energy by the same reason. The mass of $\chi_i$ will be assumed to be negligible in this case. When $\boxed{m_{\Phi_\alpha} \gg m_{\chi_i} \sim m_\xi \sim 0,}$ the choice of the counterterms
\begin{align}
	(\delta_\Phi)_{\beta \alpha} = \sum_i \frac{f_{i \beta}^* f_{i \alpha}}{16 \pi^2} \bigg( \minus \frac{2}{\epsilon} - \frac{3}{2} + \log{\frac{m_{\Phi_\beta} m_{\Phi_\alpha}}{\widetilde{\mu}^2}} \bigg), \qquad
	\delta M_\Phi^2 = 0
\end{align}
gives the renormalized self-energy
\begin{align}
	\boxed{(\Sigma_\Phi')_{\beta \alpha} (p^2) = \sum_i \frac{f_{i \beta}^* f_{i \alpha}}{16 \pi^2} \bigg[ \minus \log{\bigg( \frac{p^2}{m_{\Phi_\beta} m_{\Phi_\alpha}} \bigg)} + i \pi \bigg], \qquad
	\Sigma_{\Phi^*}' (p^2) = \Sigma_\Phi'^{\mathsf{T}} (p^2).}
	\label{eq:SE1LoopUnstable}
\end{align}
When $\boxed{m_\xi > m_{\Phi_\alpha} \gg m_{\chi_i} \sim 0,}$ we can choose the counterterms
\begin{align}
	(\delta_\Phi)_{\beta \alpha} = \sum_i \frac{f_{i \beta}^* f_{i \alpha}}{16 \pi^2} \bigg( \minus \frac{2}{\epsilon} - \frac{3}{2} + \log{\frac{m_\xi^2}{\widetilde{\mu}^2}} \bigg), \quad
	(\delta M_\Phi^2)_{\beta \alpha} = \minus \sum_i \frac{f_{i \beta}^* f_{i \alpha}}{8 \pi^2} m_\xi^2 \bigg( \frac{2}{\epsilon} + 1 - \log{\frac{m_\xi^2}{\widetilde{\mu}^2}} \bigg),
	\label{eq:CTStable}
\end{align}
such that the renormalized self-energy is given by
\begin{align}
	\boxed{(\Sigma_\Phi')_{\beta \alpha} (p^2)
	= \minus \sum_i \frac{f_{i \beta}^* f_{i \alpha}}{16 \pi^2} \bigg( 1 - \frac{5 m_\xi^2}{p^2} + \frac{6 m_\xi^4}{p^4} - \frac{2 m_\xi^6}{p^6} \bigg) \log{\frac{m_\xi^2 - p^2}{m_\xi^2}}, \qquad
	\Sigma_{\Phi^*}' (p^2) = \Sigma_\Phi'^{\mathsf{T}} (p^2).}
	\label{eq:SE1LoopStable}
\end{align}
The choices of counterterms as given above are not the most general ones, and there exist cases which are inconsistent with these choices of renormalization conditions.

For convenience, we define the order of perturbative corrections as
\begin{align}
	\boxed{\mathcal{O} (\alpha) \coloneqq \mathcal{O} (\Sigma_\Phi').}
\end{align}
For unstable $\Phi_\alpha$, we have $\mathcal{O} (\alpha) = \mathcal{O} (f^2 / 16 \pi)$ where $f$ denotes the typical magnitude of $f_{i \alpha}$. In contrast, for stable $\Phi_\alpha$, the parameter $\alpha$ as defined above can have any value in the range $(0, \infty)$. For example, if $m_\xi = 10 \, m_{\Phi_\alpha}$, we have $\mathcal{O} (\alpha) \sim \mathcal{O} (10^4 f^2 / 16 \pi)$. To have a reasonably simple perturbation theory in this paper, we assume $m_\xi \sim m_{\Phi_\alpha}$ so that $\mathcal{O} (\alpha) \sim \mathcal{O} (f^2 / 16 \pi)$. It is needless to say that a perturbative analysis with a small parameter such as $\alpha$ has a limitation when there exist hierarchies in the model parameters themselves such as Yukawa couplings and masses.

\subsection{Quasiparticles as physical particles}		\label{sec:Quasi}
Now let us find the physical particles by diagonalizing the dressed propagator and identifying the complex scalar fields associated with its components. Note that the diagonalization procedure in this section is correct up to the infinite order in perturbation. Even though we have imposed $\delta M_\Phi^2 = \delta \Sigma_\Phi$ to obtain the renormalized self-energy in the form of $\Sigma_\Phi (p^2) = p^2 \Sigma_\Phi' (p^2)$, it is always allowed for appropriately chosen $M_\Phi$.

The dressed propagator $i \Delta_\Phi (p^2)$ of $\Phi$ is calculated from a geometric series as follows:
\begin{align}
	i \Delta_\Phi (p^2) &= i (p^2 - M_\Phi^2)^{-1} + i (p^2 - M_\Phi^2)^{-1} [i \Sigma_\Phi (p^2)] i (p^2 - M_\Phi^2)^{-1} \nonumber \\
		&\qquad + i (p^2 - M_\Phi^2)^{-1} [i \Sigma_\Phi (p^2)] i (p^2 - M_\Phi^2)^{-1} [i \Sigma_\Phi (p^2)] i (p^2 - M_\Phi^2)^{-1} + \cdots	\label{eq:PropResum} \\
	&= \sum_{n = 0}^\infty i (p^2 - M_\Phi^2)^{-1} \big\{ [i \Sigma_\Phi (p^2)] i (p^2 - M_\Phi^2)^{-1} \big\}^n \nonumber \\
	&= i \big\{ [1 + \Sigma_\Phi' (p^2)] p^2 - M_\Phi^2 \big\}^{-1}.
\end{align}
The component $(\Delta_\Phi)_{\beta \alpha} (p^2)$ is associated with the transition $\Phi_\alpha \to \Phi_\beta$, and thus each component assumes a specific direction of $p$. We first find a momentum-dependent mixing matrix $C (p^2)$ which diagonalizes $M_\Phi [1 + \Sigma_\Phi' (p^2)]^{-1} M_\Phi$ as follows:
\begin{align}
	\boxed{P^2 (p^2) \coloneqq C (p^2) M_\Phi [1 + \Sigma_\Phi' (p^2)]^{-1} M_\Phi C^{-1} (p^2),}
	\label{eq:PDiag}
\end{align}
where $P^2 (p^2)$ is a diagonal matrix. Since
\begin{align}
	\Delta_\Phi (p^2) &= \big\{ [1 + \Sigma_\Phi' (p^2)] p^2 - M_\Phi^2 \big\}^{-1}
	= (M_\Phi^{-1} C^{-1} M_\Phi) M_\Phi^{-2} P^2 (p^2 - P^2)^{-1} (M_\Phi C M_\Phi^{-1}),
\end{align}
we can write
\begin{align}
	\Delta_\Phi (p^2) = C_f M_\Phi^{-2} P^2 (p^2 - P^2)^{-1} C_i^\mathsf{T},
\end{align}
where
\begin{align}
	\boxed{C_f (p^2) \coloneqq M_\Phi^{-1} C^{-1} (p^2) M_\Phi, \qquad
	C_i (p^2) \coloneqq M_\Phi^{-1} C^\mathsf{T} (p^2) M_\Phi.}
	\label{eq:CfCi}
\end{align}
In addition, we also have
\begin{align}
	\Delta_{\Phi^*} (p^2) = \big\{ [1 + \Sigma_\Phi'^{\mathsf{T}} (p^2)] p^2 - M_\Phi^2 \big\}^{-1}
	= \Delta_\Phi^\mathsf{T} (p^2),
\end{align}
which implies
\begin{align}
	\Delta_{\Phi^*} (p^2) = C_i M_\Phi^{-2} P^2 (p^2 - P^2)^{-1} C_f^\mathsf{T}.
\end{align}
Hence, the diagonalized propagator can be written as
\begin{empheq}[box=\fbox]{align}
	\Delta_{\widehat{\Phi}} (p^2) &\coloneqq M_\Phi^{-2} P^2 (p^2) [p^2 - P^2 (p^2)]^{-1} \nonumber \\[5pt]
	&= C_f^{-1} (p^2) \Delta_\Phi (p^2) (C_i^\mathsf{T})^{-1} (p^2)
	= C_i^{-1} (p^2) \Delta_{\Phi^*} (p^2) (C_f^\mathsf{T})^{-1} (p^2).
	\label{eq:PropDiag}
\end{empheq}
Note that the CPT-conjugate of $(\Delta_\Phi)_{\beta \alpha} (p^2)$ is $(\Delta_{\Phi^*})_{\alpha \beta} (p^2)$, \textit{i.e.}, $\Delta_{\Phi^*} (p^2) = \Delta_\Phi^\mathsf{T} (p^2)$ as long as the CPT symmetry is conserved. This in turn implies that $\Sigma_{\Phi^*} (p^2) = \Sigma_\Phi^\mathsf{T} (p^2)$ is true up to the infinite order in perturbation. In other words, the diagonalization given by equation \ref{eq:PropDiag} which depends on $\Sigma_{\Phi^*} (p^2) = \Sigma_\Phi^\mathsf{T} (p^2)$ is exact up to the infinite order.

The mixing matrices $C_i$ and $C_f$ can alternatively be defined such that the factor $M_\Phi^{-2} P^2$ in $\Delta_{\widehat{\Phi}}$ are absorbed into them, which would make $\Delta_{\widehat{\Phi}}$ have a simpler form. However, the definitions given by equation \ref{eq:CfCi} is more useful, not only because the diagonalized self-energy $p^2 \Sigma_{\widehat{\Phi}}' (p^2)$ can be defined from them, but also because they allow a transparent distinction between the cases where $C$ is unitary and non-unitary as we will see later. The diagonalization as given above implies that $\Delta_{\widehat{\Phi}}$ is related to the field $\widehat{\Phi}$ transformed from $\Phi$ by $C_f^{-1}$ and $C_i^{-1}$, and let us write the flavor of $\widehat{\Phi}$ as $\widehat{\Phi}_{\widehat{\alpha}}$ for clarity.

A diagonal matrix $\Sigma_{\widehat{\Phi}}' (p^2)$ can be defined by
\begin{align}
	\boxed{1 + \Sigma_{\widehat{\Phi}}' (p^2) \coloneqq C_i^\mathsf{T} (p^2) [1 + \Sigma_\Phi' (p^2)] C_f (p^2),}
	\label{eq:SEDiag}
\end{align}
in terms of which we can write
\begin{align}
	\boxed{P^2 (p^2) = M_\Phi^2 [1 + \Sigma_{\widehat{\Phi}}' (p^2)]^{-1}.}
	\label{eq:P}
\end{align}
Note that we have $(\Sigma_{\widehat{\Phi}}')_{\widehat{\beta} \widehat{\alpha}} (p^2) \sim \mathcal{O} (\alpha)$ because
\begin{align}
	(C_i^\mathsf{T} C_f)_{\widehat{\beta} \widehat{\alpha}} (p^2) - \delta_{\widehat{\beta} \widehat{\alpha}} \sim \mathcal{O} (\alpha), \qquad
	(C_i C_f^\mathsf{T})_{\beta \alpha} (p^2) - \delta_{\beta \alpha} \sim \mathcal{O} (\alpha).
	\label{eq:CiTCf}
\end{align}
These relations can be roughly proved as follows: when the mass differences are large, \textit{i.e.}, $|m_{\Phi_\beta} - m_{\Phi_\alpha}| \gg m_{\Phi_\alpha} \mathcal{O} (\alpha)$, equation \ref{eq:PDiag} implies $P^2 = C M_\Phi^2 [1 + \mathcal{O} (\alpha)] C^{-1}$, and thus $C = 1 + \mathcal{O} (\alpha)$ to obtain a diagonal matrix $P^2$.  This in turn means that $C_f = M_\Phi^{-1} C^{-1} M_\Phi = 1 + \mathcal{O} (\alpha)$ and $C_i = M_\Phi^{-1} C^\mathsf{T} M_\Phi = 1 + \mathcal{O} (\alpha)$. Hence, $C_i^\mathsf{T} C_f = 1 + \mathcal{O} (\alpha)$ and $C_i C_f^\mathsf{T} = 1 + \mathcal{O} (\alpha)$. On the other hand, when the mass differences are small, \textit{i.e.}, $|m_{\Phi_\beta} - m_{\Phi_\alpha}| \lesssim m_{\Phi_\alpha} \mathcal{O} (\alpha)$, we can write $M_\Phi = m_{\Phi_\alpha} [1 + \mathcal{O} (\alpha)]$, and thus $C_f = M_\Phi^{-1} C^{-1} M_\Phi = C^{-1} + \mathcal{O} (\alpha)$ and $C_i = M_\Phi^{-1} C^\mathsf{T} M_\Phi = C^\mathsf{T} + \mathcal{O} (\alpha)$. This implies $C_i^\mathsf{T} C_f = 1 + \mathcal{O} (\alpha)$ and $C_i C_f^\mathsf{T} = 1 + \mathcal{O} (\alpha)$.

The poles of the propagator are the solutions of
\begin{align}
	p^2 = P_{\widehat{\alpha}}^2 (p^2)
	= m_{\Phi_{\widehat{\alpha}}}^2 [1 + (\Sigma_{\widehat{\Phi}}')_{\widehat{\alpha}} (p^2)]^{-1},
\end{align}
where we have introduced a shorthand notation for the index of a diagonal matrix, \textit{e.g.}, $P_{\widehat{\alpha}}^2 \coloneqq P_{\widehat{\alpha} \widehat{\alpha}}^2$. Note that $m_{\Phi_{\widehat{\alpha}}}$ is the tree-level mass $m_{\Phi_\alpha}$ with its index replaced by $\widehat{\alpha}$. Let us call the states corresponding to the poles \textit{physical particles}, and they are associated with the fields $\widehat{\Phi}_{\widehat{\alpha}}$ which will be identified later. Each pole of the propagator is in the form of
\begin{align}
	p_{\widehat{\Phi}_{\widehat{\alpha}}}^2 = m_{\widehat{\Phi}_{\widehat{\alpha}}}^2 - i m_{\widehat{\Phi}_{\widehat{\alpha}}} \Gamma_{\widehat{\Phi}_{\widehat{\alpha}}},
\end{align}
where $m_{\widehat{\Phi}_{\widehat{\alpha}}}$ and $\Gamma_{\widehat{\Phi}_{\widehat{\alpha}}}$ are the pole mass and total decay width of $\widehat{\Phi}_{\widehat{\alpha}}$. The diagonal matrix $p^2 (\Sigma_{\widehat{\Phi}}')_{\widehat{\alpha}} (p^2)$ can be interpreted as the self-energy of $\widehat{\Phi}_{\widehat{\alpha}}$, and $\Sigma_{\widehat{\Phi}}' (p^2)$ up to $\mathcal{O} (\alpha)$ satisfies
\begin{align}
	\text{Re} [(\Sigma_{\widehat{\Phi}}')_{\widehat{\alpha}} (p_{\widehat{\Phi}_{\widehat{\alpha}}}^2)]
		= 1 - \frac{m_{\widehat{\Phi}_{\widehat{\alpha}}}^2}{m_{\Phi_{\widehat{\alpha}}}^2}, \qquad
	\text{Im} [(\Sigma_{\widehat{\Phi}}')_{\widehat{\alpha}} (p_{\widehat{\Phi}_{\widehat{\alpha}}}^2)]
		= \frac{\Gamma_{\widehat{\Phi}_{\widehat{\alpha}}}}{m_{\Phi_{\widehat{\alpha}}}}.
	\label{eq:SEDiagReIm}
\end{align}
Now we define mixing matrices
\begin{align}
	\boxed{(C_{\widehat{\Phi}^f})_{\beta {\widehat{\alpha}}} \coloneqq |R_{\widehat{\Phi}_{\widehat{\alpha}}}|^\frac{1}{2} (C_f)_{\beta {\widehat{\alpha}}} (p_{\widehat{\Phi}_{\widehat{\alpha}}}^2), \qquad
	(C_{\widehat{\Phi}^i})_{\beta {\widehat{\alpha}}} \coloneqq |R_{\widehat{\Phi}_{\widehat{\alpha}}}|^\frac{1}{2} (C_i)_{\beta {\widehat{\alpha}}} (p_{\widehat{\Phi}_{\widehat{\alpha}}}^2),}
	\label{eq:MixingMat}
\end{align}
where $R_{\widehat{\Phi}_{\widehat{\alpha}}}$ is the residue of the pole $p_{\widehat{\Phi}_{\widehat{\alpha}}}^2$ given by
\begin{align}
	R_{\widehat{\Phi}_{\widehat{\alpha}}} \coloneqq \lim_{p^2 \to p_{\widehat{\Phi}_{\widehat{\alpha}}}^2} (p^2 - p_{\widehat{\Phi}_{\widehat{\alpha}}}^2) (\Delta_{\widehat{\Phi}})_{\widehat{\alpha}} (p^2)
	= \frac{p_{\widehat{\Phi}_{\widehat{\alpha}}}^2}{m_{\Phi_{\widehat{\alpha}}}^2} \bigg[ 1 - \frac{dP_{\widehat{\alpha}}^2}{dp^2} (p_{\widehat{\Phi}_{\widehat{\alpha}}}^2) \bigg]^{-1}
	\eqqcolon |R_{\widehat{\Phi}_{\widehat{\alpha}}}| e^{i \theta_{\widehat{\Phi}_{\widehat{\alpha}}}},
	\label{eq:Res}
\end{align}
which is in the form of $R_{\widehat{\Phi}_{\widehat{\alpha}}} = 1 + \mathcal{O} (\alpha)$. For unstable particles, we have $\theta_{\widehat{\Phi}_{\widehat{\alpha}}} \sim \mathcal{O} (\alpha) \neq 0$ since $p_{\widehat{\Phi}_{\widehat{\alpha}}}^2$ and $P_{\widehat{\alpha}}^2$ are complex-valued. In contrast, for stable particles, we have $\theta_{\widehat{\Phi}_{\widehat{\alpha}}} = 0$ since $p_{\widehat{\Phi}_{\widehat{\alpha}}}^2 = m_{\widehat{\Phi}_{\widehat{\alpha}}}^2$ and $P_{\widehat{\alpha}}^2$ is a real-valued function obtained by diagonalizing a Hermitian matrix. In equation \ref{eq:MixingMat}, the complex phase of $R_{\widehat{\Phi}_{\widehat{\alpha}}}$ has not been absorbed into the mixing matrices, the reason of which will be explained later. In other words, the fields have not been renormalized such that their propagator has a unit residue at each pole. Note that the motivation of a unit residue is to regard it as an external on-shell state. Note also that a different pole is used to define each different column of $C_{\widehat{\Phi}^f}$ and $C_{\widehat{\Phi}^i}$ in equation \ref{eq:MixingMat}. In addition, the effective Yukawa couplings are defined as
\begin{align}
	\boxed{\widehat{f}_{i \widehat{\alpha}} \coloneqq (f C_{\widehat{\Phi}^f})_{i \widehat{\alpha}}, \qquad
	\widehat{f}_{i \widehat{\alpha}}^c \coloneqq (f^* C_{\widehat{\Phi}^i})_{i \widehat{\alpha}}.}
\end{align}
The effective vertices incorporating vertex-loop corrections as well can also be defined as follows:
\begin{align}
	\boxed{(\widehat{V}_{\widehat{\Phi}^f})_{i \widehat{\alpha}} (p^2) \coloneqq (V C_{\widehat{\Phi}^f})_{i \widehat{\alpha}} (p^2), \qquad
	(\widehat{V}_{\widehat{\Phi}^i})_{i \widehat{\alpha}} (p^2) \coloneqq (V^* C_{\widehat{\Phi}^i})_{i \widehat{\alpha}} (p^2),}
	\label{eq:Veff}
\end{align}
where $V (p^2)$ is the vertex function which is given in appendix \ref{sec:VertexCal}. The component of the diagonalized propagator is now written as
\begin{align}
	i (\Delta_{\widehat{\Phi}})_{\widehat{\alpha}} (p^2) = \frac{i R_{\widehat{\Phi}_{\widehat{\alpha}}}}{p^2 - p_{\widehat{\Phi}_{\widehat{\alpha}}}^2} + \cdots,
	\label{eq:PropMD}
\end{align}
and the non-diagonal dressed propagators can therefore be expressed as the leading terms of the Laurent expansions around its poles as follows:
\begin{empheq}[box=\fbox]{align}
	i (\Delta_\Phi)_{\beta \alpha} (p^2) = \sum_{\widehat{\gamma}} (C_{\widehat{\Phi}^f})_{\beta \widehat{\gamma}} \frac{i e^{i \theta_{\widehat{\Phi}}}}{p^2 - p_{\widehat{\Phi}_{\widehat{\gamma}}}^2} (C_{\widehat{\Phi}^i})_{\alpha \widehat{\gamma}} + \cdots,
		\label{eq:PropPMDExp} \\
	i (\Delta_{\Phi^*})_{\beta \alpha} (p^2) = \sum_{\widehat{\gamma}} (C_{\widehat{\Phi}^i})_{\beta \widehat{\gamma}} \frac{i e^{i \theta_{\widehat{\Phi}}}}{p^2 - p_{\widehat{\Phi}_{\widehat{\gamma}}}^2} (C_{\widehat{\Phi}^f})_{\alpha \widehat{\gamma}} + \cdots.
		\label{eq:PropAMDExp}
\end{empheq} \\

Let us identify the physical particles associated with the components of the diagonal propagator. The two-point correlation functions corresponding to the non-diagonal propagators are
\begin{align}
	\int \frac{d^4 p}{(2 \pi)^4} e^{-i p \cdot (x - y)} i (\Delta_\Phi)_{\beta \alpha} (p^2)
		&= \langle \Omega | \Phi_\beta (x) \Phi_\alpha^\dag (y) | \Omega \rangle \quad (x^0 > y^0), \\
	\int \frac{d^4 p}{(2 \pi)^4} e^{-i p \cdot (x - y)} i (\Delta_{\Phi^*})_{\beta \alpha} (p^2)
		&= \langle \Omega | \Phi_\beta^\dag (y) \Phi_\alpha (x) | \Omega \rangle \quad (y^0 > x^0),
\end{align}
where the correlation functions should be already time-ordered since $\Delta_\Phi$ and $\Delta_{\Phi^*}$ have been defined with specific directions of energy transfer: $(\Delta_\Phi)_{\beta \alpha}$ for $\Phi_\alpha \to \Phi_\beta$ and $(\Delta_{\Phi^*})_{\beta \alpha}$ for $\Phi_\alpha^* \to \Phi_\beta^*$. Hence, using equations \ref{eq:PropPMDExp} and \ref{eq:PropAMDExp}, we can write
\begin{align}
	\boxed{\int \frac{d^4 p}{(2 \pi)^4} e^{-i p \cdot (x - y)} \frac{i e^{i \theta_{\widehat{\Phi}_{\widehat{\alpha}}}}}{p^2 - p_{\widehat{\Phi}_{\widehat{\alpha}}}^2} + \cdots
	= \langle \Omega | T \{\widehat{\Phi}_{\widehat{\alpha}}^f (x) \widehat{\Phi}_{\widehat{\alpha}}^{i \dag} (y)\} | \Omega \rangle,}
	\label{eq:CorrDiag}
\end{align}
where
\begin{align}
	\boxed{\widehat{\Phi}_{\widehat{\alpha}}^i (x) \coloneqq \sum_\beta (C_{\widehat{\Phi}^i}^{-1})_{\widehat{\alpha} \beta}^* \Phi_\beta (x), \quad
	\widehat{\Phi}_{\widehat{\alpha}}^f (x) \coloneqq \sum_\beta (C_{\widehat{\Phi}^f}^{-1})_{\widehat{\alpha} \beta} \Phi_\beta (x).}
\end{align}
Note that the definitions of mixing matrices given by equation \ref{eq:CfCi} show that $C_i = C_f^*$ if and only if $C$ is unitary, \text{i.e.}, $\widehat{\Phi}_{\widehat{\alpha}}^f = \widehat{\Phi}_{\widehat{\alpha}}^i$ if and only if $C$ is unitary. Moreover, if we had absorbed the phase $e^{i \theta_{\widehat{\Phi}_{\widehat{\alpha}}}}$ into the mixing matrices in equation \ref{eq:MixingMat}, then we would not have a simple expression of the diagonalized propagator in terms of the time-ordering as in equation \ref{eq:CorrDiag}.

In the case of stable particles, we therefore have $C_i = C_f^*$ since $C$ is unitary, and thus $\widehat{\Phi}_{\widehat{\alpha}} = \widehat{\Phi}_{\widehat{\alpha}}^f = \widehat{\Phi}_{\widehat{\alpha}}^i$. In addition, we always have $\theta_{\widehat{\Phi}_{\widehat{\alpha}}} = 0$ for stable particles so that each physical particle can be identified as a canonically normalized one-particle state with mass $m_{\widehat{\Phi}_{\widehat{\alpha}}}$, which propagates like a free particle in the interacting theory as the state associated with each component of the diagonalized propagator. In the case of unstable particles, we have $\widehat{\Phi}_{\widehat{\alpha}}^f \neq \widehat{\Phi}_{\widehat{\alpha}}^i$ since $C$ is non-unitary in general due to the absorptive part of the self-energy, and the physical particles propagate like free particles with masses $m_{\widehat{\Phi}_{\widehat{\alpha}}}$ until they decay. However, the interpretation of those physical particles is non-trivial, since equation \ref{eq:CorrDiag} implies that each of them is a particle created by $\widehat{\Phi}_{\widehat{\alpha}}^{i \dag}$ and annihilated by $\widehat{\Phi}_{\widehat{\alpha}}^f$ $(\widehat{\Phi}_{\widehat{\alpha}}^f \neq \widehat{\Phi}_{\widehat{\alpha}}^i)$, or its antiparticle created by $\widehat{\Phi}_{\widehat{\alpha}}^f$ and annihilated by $\widehat{\Phi}_{\widehat{\alpha}}^{i \dag}$. In other words, it is not an excitation of a single linear combination of the fields $\Phi_\alpha$. Furthermore, we have $\theta_{\widehat{\Phi}_{\widehat{\alpha}}} \neq 0$ for unstable particles, \textit{i.e.}, each of physical particles is like a state with a complex-valued norm, which is unphysical by itself. Such a particle should be interpreted as a \textit{quasiparticle}, \textit{i.e.}, an emergent particle dynamically generated by interactions of various fields in the theory. The properties of physical particles will be discussed in more details in section \ref{sec:Quant&Phys}.

\begin{figure}[t]
	\centering
	\includegraphics[width = 75 mm]{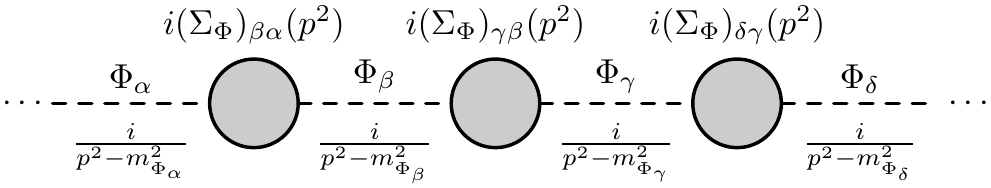}
	\caption{A non-perturbative effect is generated when $\Phi_\alpha$ goes almost on-shell if $|m_{\Phi_\beta} - m_{\Phi_\alpha}| \lesssim m_{\Phi_\alpha} \mathcal{O} (\alpha)$.}
	\label{fig:NonPert}
\end{figure}

When the mass differences between flavors are small, \textit{i.e.}, $|m_{\Phi_\beta} - m_{\Phi_\alpha}| \lesssim m_{\Phi_\alpha} \mathcal{O} (\alpha)$, the difference between $\Phi_\alpha$ and $\widehat{\Phi}_{\widehat{\alpha}}$ can be hugely enhanced whether $\Phi_\alpha$ is stable or unstable. This occurs because the small mass differences make the quantum corrections to $\Phi_\alpha$ non-perturbative when $\Phi_\alpha$ goes almost on-shell. Note that the factor $\Sigma_{\beta \alpha} (p^2) / (p^2 - m_{\Phi_\beta}^2)~(\beta \neq \alpha)$ in figure \ref{fig:NonPert} can be hugely enhanced for $p^2 \sim m_{\Phi_\beta}^2$ so that the collective effects of those factors on flavor mixing in the dressed propagator go beyond the typical perturbative correction $\mathcal{O} (\alpha)$ of the theory.
Especially in the case of unstable particles, the deviation of $C$ from unitarity can be much larger than $\mathcal{O} (\alpha)$, and accordingly the difference between $\widehat{\Phi}_{\widehat{\alpha}}^f$ and $\widehat{\Phi}_{\widehat{\alpha}}^i$ can also be beyond $\mathcal{O} (\alpha)$, \textit{i.e.}, $C_{\widehat{\Phi}^f}^{-1} C_{\widehat{\Phi}^i}^* \neq 1$ even up to $\mathcal{O} (1)$. A numerical example of such a case will be presented in section \ref{sec:Example}.

\subsection{Revisiting renormalization of the propagator and Lagrangian density}		\label{sec:RevRen}
Now we return to renormalization. The non-diagonal bare and renormalized propagators are related to the two-point correlation functions as follows:
\begin{align}
	\int \frac{d^4 p}{(2 \pi)^4} e^{-i p \cdot (x - y)} i (\Delta_{\Phi_0})_{\beta \alpha} (p^2)
		&= \langle \Omega | \Phi_{0 \beta} (x) \Phi_{0 \alpha}^\dag (y) | \Omega \rangle \quad (x^0 > y^0), \\
	\int \frac{d^4 p}{(2 \pi)^4} e^{-i p \cdot (x - y)} i (\Delta_\Phi)_{\beta \alpha} (p^2)
		&= \langle \Omega | \Phi_\beta (x) \Phi_\alpha^\dag (y) | \Omega \rangle \quad (x^0 > y^0).
\end{align}
Hence, the bare propagator can be written in terms of the renormalized propagator as
\begin{align}
	\Delta_{\Phi_0} (p^2) &= \big\{ p^2 - M_{\Phi_0}^2 + \Sigma_{\Phi_0} (p^2) \big\}^{-1}
	= \big\{ [1 + \Sigma_{\Phi_0}' (p^2)] p^2 + \delta \Sigma_{\Phi_0} - M_{\Phi_0}^2 \big\}^{-1} \nonumber \\
	&= Z_\Phi^\frac{1}{2} \Delta_\Phi (p^2) Z_\Phi^{\frac{1}{2} \dag}
	= Z_\Phi^\frac{1}{2} C_f (p^2) \Delta_{\widehat{\Phi}} (p^2) C_i^\mathsf{T} (p^2) Z_\Phi^{\frac{1}{2} \dag},
\end{align}
and thus
\begin{align}
	\boxed{(\Delta_{\Phi_0})_{\beta \alpha} (p^2) = \sum_{\widehat{\gamma}} (Z_\Phi^\frac{1}{2} C_{\widehat{\Phi}^f})_{\beta \widehat{\gamma}}
		\frac{e^{i \theta_{\widehat{\Phi}_{\widehat{\gamma}}}}}{p^2 - p_{\widehat{\Phi}_{\widehat{\gamma}}}^2}
		(Z_\Phi^\frac{1}{2} C_{\widehat{\Phi}^i}^*)^\dag_{\widehat{\gamma} \alpha} + \cdots.}
	\label{eq:RenProp}
\end{align}
This procedure is consistent with the renormalization of the self-energy discussed in section \ref{sec:ToyRen} because of equations \ref{eq:MassRen} and \ref{eq:SERen}. \\

A popular belief about renormalization is that the bare propagator $\Delta_{\Phi_0}$ can always be diagonalized by a single field-strength renormalization matrix $Z^\frac{1}{2}$ as follows:
\begin{align}
	(\Delta_{\Phi_0})_{\beta \alpha} (p^2) = \sum_{\widehat{\gamma}} (Z^\frac{1}{2})_{\beta \widehat{\gamma}}
		\frac{1}{p^2 - p_{\widehat{\Phi}_{\widehat{\gamma}}}^2}
		(Z^{\frac{1}{2} \dag})_{\widehat{\gamma} \alpha} + \cdots.
\end{align}
However, we have shown in equation \ref{eq:RenProp} that it is not generally possible in theories of unstable particles with flavor mixing, since $C_{\widehat{\Phi}^f}^{-1} C_{\widehat{\Phi}^i}^* \neq 1$ sometimes even up to $\mathcal{O} (1)$, although $\theta_{\widehat{\Phi}_{\widehat{\gamma}}} \sim \mathcal{O} (\alpha)$ implies that it can be set to zero at least up to the leading order in perturbation.

Moreover, neither $Z_\Phi^\frac{1}{2} C_{\widehat{\Phi}^f}$ nor $Z_\Phi^\frac{1}{2} C_{\widehat{\Phi}^i}^*$ had better be regarded as a field-strength renormalization matrix when $C_{\widehat{\Phi}^f}$ and $C_{\widehat{\Phi}^i}$ are non-unitary matrices whose deviations from unitarity are large, since it causes a complicated issue in perturbation. For example, let us try to define a field-strength renormalization matrix $Z_{\widehat{\Phi}}^{\frac{1}{2}}$ as
\begin{align}
	(Z_{\widehat{\Phi}}^{\frac{1}{2}})_{\beta \widehat{\alpha}} \coloneqq (Z_\Phi^\frac{1}{2} C_{\widehat{\Phi}^f})_{\beta \widehat{\alpha}}.
\end{align}
Since
\begin{align}
\Phi_{0 \alpha} = \sum_\beta (Z_\Phi^\frac{1}{2})_{\alpha \beta} \Phi_\beta
	= \sum_{\widehat{\gamma}} (Z_{\widehat{\Phi}}^{\frac{1}{2}})_{\alpha \widehat{\gamma}} \widehat{\Phi}_{\widehat{\gamma}}^f,
\end{align}
the Lagrangian density can be rewritten as
\begin{align}
	\mathcal{L} &= \sum_\alpha \partial^\mu \Phi_{0 \alpha}^\dag \partial_\mu \Phi_{0 \alpha} - \sum_\alpha m_{\Phi_{0 \alpha}}^2 \Phi_{0 \alpha}^\dag \Phi_{0 \alpha} + \cdots
		\label{eq:LagBare} \\
	&= \sum_\alpha \partial^\mu \Phi_\alpha^\dag \partial_\mu \Phi_\alpha - \sum_\alpha m_{\Phi_\alpha}^2 \Phi_\alpha^\dag \Phi_\alpha
		+ \sum_{\alpha, \beta} (\delta_\Phi^H)_{\beta \alpha} \partial^\mu \Phi_\beta^\dag \partial_\mu \Phi_\alpha
		- \sum_{\alpha, \beta} (\delta M_\Phi^2)_{\beta \alpha} \Phi_\beta^\dag \Phi_\alpha + \cdots
		\label{eq:LagRen} \\
	&= \sum_{\widehat{\alpha}} \partial^\mu \widehat{\Phi}_{\widehat{\alpha}}^{f \dag} \partial_\mu \widehat{\Phi}_{\widehat{\alpha}}^f
		- \sum_{\widehat{\alpha}} m_{\widehat{\Phi}_{\widehat{\alpha}}}^2 \widehat{\Phi}_{\widehat{\alpha}}^{f \dag} \widehat{\Phi}_{\widehat{\alpha}}^f \nonumber \\
		&\qquad + \sum_{\widehat{\alpha}, \widehat{\beta}} \big[ (C_{\widehat{\Phi}^f}^\dag C_{\widehat{\Phi}^f})_{\widehat{\beta} \widehat{\alpha}} - \delta_{\widehat{\beta} \widehat{\alpha}} \big] \partial^\mu \widehat{\Phi}_{\widehat{\beta}}^{f \dag} \partial_\mu \widehat{\Phi}_{\widehat{\alpha}}^f
		- \sum_{\widehat{\alpha}, \widehat{\beta}} \big[ (C_{\widehat{\Phi}^f}^\dag M_\Phi^2 C_{\widehat{\Phi}^f})_{\widehat{\beta} \widehat{\alpha}} - m_{\widehat{\Phi}_{\widehat{\alpha}}}^2 \delta_{\widehat{\beta} \widehat{\alpha}} \big] \widehat{\Phi}_{\widehat{\beta}}^{f \dag} \widehat{\Phi}_{\widehat{\alpha}}^f \nonumber \\
		&\qquad + \sum_{\widehat{\alpha}, \widehat{\beta}} (C_{\widehat{\Phi}^f}^\dag \delta_\Phi^H C_{\widehat{\Phi}^f})_{\widehat{\beta} \widehat{\alpha}} \partial^\mu \widehat{\Phi}_{\widehat{\beta}}^{f \dag} \partial_\mu \widehat{\Phi}_{\widehat{\alpha}}^f
		- \sum_{\widehat{\alpha}, \widehat{\beta}} (C_{\widehat{\Phi}^f}^\dag \delta M_\Phi^2 C_{\widehat{\Phi}^f})_{\widehat{\beta} \widehat{\alpha}} \widehat{\Phi}_{\widehat{\beta}}^{f \dag} \widehat{\Phi}_{\widehat{\alpha}}^f
		+ \cdots,
		\label{eq:LagPhys1} \\
	&= \sum_{\widehat{\alpha}} \partial^\mu \widehat{\Phi}_{\widehat{\alpha}}^{f \dag} \partial_\mu \widehat{\Phi}_{\widehat{\alpha}}^f
		- \sum_{\widehat{\alpha}} m_{\widehat{\Phi}_{\widehat{\alpha}}}^2 \widehat{\Phi}_{\widehat{\alpha}}^{f \dag} \widehat{\Phi}_{\widehat{\alpha}}^f
		+ \sum_{\widehat{\alpha}, \widehat{\beta}} (\delta_{\widehat{\Phi}}^H)_{\widehat{\beta} \widehat{\alpha}} \partial^\mu \widehat{\Phi}_{\widehat{\beta}}^{f \dag} \partial_\mu \widehat{\Phi}_{\widehat{\alpha}}^f
		- \sum_{\widehat{\alpha}, \widehat{\beta}} (\delta M_{\widehat{\Phi}}^2)_{\widehat{\beta} \widehat{\alpha}} \widehat{\Phi}_{\widehat{\beta}}^{f \dag} \widehat{\Phi}_{\widehat{\alpha}}^f \nonumber \\
		&\qquad - \sum_{i, \widehat{\alpha}} \widehat{f}_{i \widehat{\alpha}} \overline{\chi_i} \xi \widehat{\Phi}_\alpha^f - \sum_{i, \alpha} \widehat{f}_{i \widehat{\alpha}}^* \overline{\xi} \chi_i \widehat{\Phi}_{\widehat{\alpha}}^{f \dag}
		- \sum_{i, \widehat{\alpha}} \delta \widehat{f}_{i \widehat{\alpha}} \overline{\chi_i} \xi \widehat{\Phi}_{\widehat{\alpha}}^f - \sum_{i, \widehat{\alpha}} \delta \widehat{f}_{i \widehat{\alpha}}^* \overline{\xi} \chi_i \widehat{\Phi}_{\widehat{\alpha}}^{f \dag}
		+ \cdots,
		\label{eq:LagPhys}
\end{align}
where
\begin{align}
	(\delta_{\widehat{\Phi}}^H)_{\widehat{\beta} \widehat{\alpha}}
		&\coloneqq (Z_{\widehat{\Phi}}^{\frac{1}{2} \dag} Z_{\widehat{\Phi}}^\frac{1}{2})_{\widehat{\beta} \widehat{\alpha}} - \delta_{\widehat{\beta} \widehat{\alpha}}
		= (C_{\widehat{\Phi}^f}^\dag Z_\Phi^{\frac{1}{2} \dag} Z_\Phi^\frac{1}{2} C_{\widehat{\Phi}^f})_{\widehat{\beta} \widehat{\alpha}} - \delta_{\widehat{\beta} \widehat{\alpha}} \nonumber \\
		&= (C_{\widehat{\Phi}^f}^\dag \delta_\Phi^H C_{\widehat{\Phi}^f})_{\widehat{\beta} \widehat{\alpha}}
			+ (C_{\widehat{\Phi}^f}^\dag C_{\widehat{\Phi}^f})_{\widehat{\beta} \widehat{\alpha}} - \delta_{\widehat{\beta} \widehat{\alpha}} + \cdots,
		\label{eq:CTPhys1} \\
	(\delta M_{\widehat{\Phi}}^2)_{\widehat{\beta} \widehat{\alpha}}
		&\coloneqq \big[ C_{\widehat{\Phi}^f}^\dag (M_\Phi^2 + \delta M_\Phi^2) C_{\widehat{\Phi}^f} \big]_{\widehat{\beta} \widehat{\alpha}}
			- m_{\widehat{\Phi}_{\widehat{\alpha}}}^2 \delta_{\widehat{\beta} \widehat{\alpha}}.
		\label{eq:CTPhys2}
\end{align}
Here, equation \ref{eq:LagBare} consists of the kinetic and mass terms of bare fields $\Phi_{0 \alpha}$, and equation \ref{eq:LagRen} is composed of the kinetic and mass terms of renormalized fields $\Phi_\alpha$ and associated counterterms. Equations \ref{eq:LagPhys1} and \ref{eq:LagPhys} are the expressions of the Lagrangian density written in terms of $\widehat{\Phi}_{\widehat{\alpha}}^f$. The first line of equation \ref{eq:LagPhys1} is the kinetic and mass terms of $\widehat{\Phi}_{\widehat{\alpha}}^f$, and its second line consists of new interaction terms generated from the kinetic and mass terms of $\Phi_\alpha$. In equation \ref{eq:LagPhys}, those new interaction terms are absorbed into the counterterms. As long as the coefficients of new interaction terms are in the perturbative regime, \textit{e.g.}, roughly $\big| \delta_{\widehat{\beta} \widehat{\alpha}} - (C_{\widehat{\Phi}^f}^\dag C_{\widehat{\Phi}^f})_{\widehat{\beta} \widehat{\alpha}} \big| < \sqrt{4 \pi}$, it might be thought that we can still treat the theory in terms of $\widehat{\Phi}_{\widehat{\alpha}}^f$ as a valid perturbation theory. Especially for stable particles, $\widehat{\Phi}_{\widehat{\alpha}} = \widehat{\Phi}_{\widehat{\alpha}}^f = \widehat{\Phi}_{\widehat{\alpha}}^i$ is always satisfied, and thus equation \ref{eq:LagPhys} may be considered to be the Lagrangian density written in terms of the fields of physical particles. In other words, equation \ref{eq:LagPhys} is the expression of the Lagrangian density we want to obtain in the on-shell renormalization scheme.

However, when the new interaction terms have large coefficients, \textit{e.g.},
\begin{align}
	\big| \delta_{\widehat{\beta} \widehat{\alpha}} - (C_{\widehat{\Phi}^f}^\dag C_{\widehat{\Phi}^f})_{\widehat{\beta} \widehat{\alpha}} \big|
		&\sim \mathcal{O} (1) \gg \mathcal{O} (\alpha), \\
	\bigg| \big[ C_{\widehat{\Phi}^f}^\dag (M_\Phi^2 + \delta M_\Phi^2) C_{\widehat{\Phi}^f} \big]_{\widehat{\beta} \widehat{\alpha}} - m_{\widehat{\Phi}_{\widehat{\alpha}}}^2 \delta_{\widehat{\beta} \widehat{\alpha}} \bigg|
		&\sim m_{\widehat{\Phi}_{\widehat{\alpha}}}^2 \mathcal{O} (1)
		\gg m_{\widehat{\Phi}_{\widehat{\alpha}}}^2 \mathcal{O} (\alpha),
\end{align}
the counterterms $\delta_{\widehat{\Phi}}^H$ and $\delta M_{\widehat{\Phi}}^2$ are no longer corrections of $\mathcal{O} (\alpha)$. Hence, in order to keep the precision $\mathcal{O} (\alpha)$ of the original theory, we must consider higher-order contributions of those counterterms. In that case, for example, equations \ref{eq:SERenwCT} and \ref{eq:SERen1Loop} are no longer valid expressions of the self-energy up to $\mathcal{O} (\alpha)$, since they assume $(\delta_\Phi^H)_{\beta \alpha} \sim \mathcal{O} (\alpha)$ and $(\delta M_\Phi^2)_{\beta \alpha} \sim m_{\Phi_\alpha}^2 \mathcal{O} (\alpha)$. This approach complicates order-by-order renormalization, and has no practical advantage at all. In section \ref{sec:Example}, we will explicitly see that the new interaction terms can indeed have large coefficients in the case of unstable particles with small mass differences. On the contrary, in the case of stable particles, $\delta_{\widehat{\Phi}}^H$ and $\delta M_{\widehat{\Phi}}^2$ generated by those new interaction terms are always small, since, for stable particles, we can always write
\begin{align}
	C_{\widehat{\Phi}^f} = U + \mathcal{O} (\alpha)
	\label{eq:CfStable}
\end{align}
where $U$ is a unitary matrix. This is because equation \ref{eq:CiTCf} implies $C_f^\dag C_f = 1 + \mathcal{O} (\alpha)$ as $C_f = C_i^*$ for stable particles, and $C_f^\dag C_f = 1 + \mathcal{O} (\alpha)$ is true if and only if $C_f (p^2) = U' (p^2) + \mathcal{O} (\alpha)$ for a unitary matrix $U' (p^2)$. It follows that $C_{\widehat{\Phi}^f} = U + \mathcal{O} (\alpha)$ where $U = U' (m_{\widehat{\Phi}_{\widehat{\alpha}}}^2)$ is unitary. Since each component of $U$ is at most $\mathcal{O} (1)$, it is clear that $\delta_{\widehat{\Phi}}^H$ given by equation \ref{eq:CTPhys1} is always of $\mathcal{O} (\alpha)$. We can also show $(\delta M_{\widehat{\Phi}}^2)_{\widehat{\beta} \widehat{\alpha}} \sim m_{\Phi_{\widehat{\alpha}}}^2 \mathcal{O} (\alpha)$, using $m_{\widehat{\Phi}_{\widehat{\alpha}}}^2 = m_{\Phi_{\widehat{\alpha}}}^2 + \mathcal{O} (\alpha)$ and the unitarity of $C$ for stable particles.

\section{Quantization of fields and properties of physical particles}		\label{sec:Quant&Phys}
In this section, let us examine how the one-particle states in the Fock space are related to the physical particles we have been discussing. We will begin with the quantization of fields and their time evolution, and see that the physical unstable particles as quasiparticles can neither be simply related to the one-particle states nor be regarded as external states. Their properties will be read from the unitarity cut of the scattering in which the particles of $\Phi$ appear as intermediate states.

\subsection{Stable particles}
We first study the case of stable particles. Let us introduce $\Phi_\alpha (\mathbf{x})$ which are the fields in the Schr\"odinger picture. These fields are canonically quantized, and they are written as
\begin{align}
	\Phi_\alpha (\mathbf{x}) = \int \frac{d^3 \mathbf{p}}{(2 \pi)^3} \frac{1}{\sqrt{2 E_{\alpha \mathbf{p}}}} (a_{\alpha \mathbf{p}} e^{i \mathbf{p} \cdot \mathbf{x}} + b_{\alpha \mathbf{p}}^\dag e^{-i \mathbf{p} \cdot \mathbf{x}}),
\end{align}
where $E_{\alpha \mathbf{p}} \coloneqq \sqrt{m_{\Phi_\alpha}^2 + |\mathbf{p}|^2}$. The only non-zero commutation relations between ladder operators are
\begin{align}
	[a_\mathbf{q}, a_\mathbf{p}^\dag] = [b_\mathbf{q}, b_\mathbf{p}^\dag] = (2 \pi)^3 \delta^3 (\mathbf{q} - \mathbf{p}),
\end{align}
and those operators create or annihilate one-particle states according to
\begin{align}
	| \Phi (m_{\Phi_\alpha}; \mathbf{p}) \rangle = \sqrt{2 E_{\alpha \mathbf{p}}} a_{\alpha \mathbf{p}}^\dag | 0 \rangle, \qquad
	| \Phi^* (m_{\Phi_\alpha}; \mathbf{p}) \rangle = \sqrt{2 E_{\alpha \mathbf{p}}} b_{\alpha \mathbf{p}}^\dag | 0 \rangle,
\end{align}
where $| \Phi (m_{\Phi_\alpha}; \mathbf{p}) \rangle$ and $| \Phi^* (m_{\Phi_\alpha}; \mathbf{p}) \rangle$ are the states of a particle and its anti-particles with mass $m_{\Phi_\alpha}$ and three-momentum $\mathbf{p}$. The Hamiltonian can be decomposed into two parts:
\begin{align}
	H (t) = H_0 + H_I (t),
\end{align}
where $H_0$ is the free Hamiltonian for $\Phi_\alpha$ and $H_I$ is the interaction part. The time evolution of the ladder operators is given by
\begin{align}
	e^{i H_0 t} a_{\alpha \mathbf{p}} e^{-i H_0 t} = a_{\alpha \mathbf{p}} e^{-i E_{\alpha \mathbf{p}} t}, \qquad
	e^{i H_0 t} b_{\alpha \mathbf{p}} e^{-i H_0 t} = b_{\alpha \mathbf{p}} e^{-i E_{\alpha \mathbf{p}} t}.
\end{align}
However, the full Hamiltonian $H$ mixes flavors, and thus the time-dependent ladder operators
\begin{align}
	a_{\alpha \mathbf{p}} (t) \coloneqq e^{i H t} a_{\alpha \mathbf{p}} e^{-i H t}, \qquad
	b_{\alpha \mathbf{p}} (t) \coloneqq e^{i H t} b_{\alpha \mathbf{p}} e^{-i H t},
\end{align}
do not necessarily create or annihilate $| \Phi (m_{\Phi_\alpha}; \mathbf{p}) \rangle$ and $| \Phi^* (m_{\Phi_\alpha}; \mathbf{p}) \rangle$. As mentioned before, the same is true for the fields in the Heisenberg picture defined by
\begin{align}
	\Phi_\alpha (x) \coloneqq e^{i H t} \Phi_\alpha (\mathbf{x}) e^{-i H t}.
\end{align}
The fields of physical particles in the Heisenberg picture are written as
\begin{align}
	\widehat{\Phi}_{\widehat{\alpha}} (x) = e^{i H t} \widehat{\Phi}_{\widehat{\alpha}} (\mathbf{x}) e^{-i H t}
	 = e^{i P \cdot x} \widehat{\Phi}_{\widehat{\alpha}} (0) e^{-i P \cdot x},
\end{align}
where $P = (H, \mathbf{P})$ is the spacetime translation operator and
\begin{align}
	\widehat{\Phi}_{\widehat{\alpha}} (\mathbf{x}) = \sum_\beta (C_{\widehat{\Phi}}^{-1})_{\widehat{\alpha} \beta} \Phi_\beta (\mathbf{x})
\end{align}
are the fields in the Schr\"odinger picture. In contrast to $\Phi_\alpha (x)$, the fields $\widehat{\Phi}_{\widehat{\alpha}} (x)$ create or annihilate the one-particle states of physical particles, $| \Phi (m_{\widehat{\Phi}_{\widehat{\alpha}}}; \mathbf{p}) \rangle$ and $| \Phi^* (m_{\widehat{\Phi}_{\widehat{\alpha}}}; \mathbf{p}) \rangle$, at any time. A perturbation theory depends on the Dyson series which uses the interaction Hamiltonian and the fields in the interaction picture defined by
\begin{align}
	\Phi_{I \alpha} (x) \coloneqq e^{i H_0 t} \Phi_\alpha (\mathbf{x}) e^{-i H_0 t}
	= \int \frac{d^3 \mathbf{p}}{(2 \pi)^3} \frac{1}{\sqrt{2 E_{\alpha \mathbf{p}}}} (a_{\alpha \mathbf{p}} e^{-i p \cdot x} + b_{\alpha \mathbf{p}}^\dag e^{i p \cdot x}).
\end{align}
For convenience, let us introduce simplified notations of various one-particle states of the scalar fields $\Phi$ and $\Phi^*$ with momentum $\mathbf{p}$ as follows:
\begin{alignat}{2}
	&| \Phi_\alpha (\mathbf{p}) \rangle \coloneqq | \Phi (m_{\Phi_\alpha}; \mathbf{p}) \rangle, \qquad
	&&| \Phi_\alpha^* (\mathbf{p}) \rangle \coloneqq | \Phi^* (m_{\Phi_\alpha}; \mathbf{p}) \rangle, \\
	&| \widehat{\Phi}_{\widehat{\alpha}} (\mathbf{p}) \rangle \coloneqq | \Phi (m_{\widehat{\Phi}_{\widehat{\alpha}}}; \mathbf{p}) \rangle, \qquad
	&&| \widehat{\Phi}_{\widehat{\alpha}}^* (\mathbf{p}) \rangle \coloneqq | \Phi^* (m_{\widehat{\Phi}_{\widehat{\alpha}}}; \mathbf{p}) \rangle.
\end{alignat}
Note that the only difference between $| \Phi_\alpha (\mathbf{p}) \rangle$ and $| \widehat{\Phi}_{\widehat{\alpha}} (\mathbf{p}) \rangle$ is the mass of the one-particle state. The states $| \Phi_\alpha (\mathbf{p}) \rangle$ and $| \Phi_\alpha^* (\mathbf{p}) \rangle$ are the eigenstates of three-momentum operator $\mathbf{P}$ and free Hamiltonian $H_0$, and they satisfy
\begin{alignat}{2}
	&\langle 0 | \Phi_{I \beta} (x) | \Phi_\alpha (\mathbf{p}) \rangle = \delta_{\beta \alpha} e^{-i p \cdot x} \big|_{p^0 = E_{\alpha \mathbf{p}}}, \qquad
	&&\langle 0 | \Phi_{I \beta} (x) | \Phi_\alpha^* (\mathbf{p}) \rangle = 0, \\
	&\langle 0 | \Phi_{I \beta}^\dag (x) | \Phi_\alpha (\mathbf{p}) \rangle = 0, \qquad
	&&\langle 0 | \Phi_{I \beta}^\dag (x) | \Phi_\alpha^* (\mathbf{p}) \rangle = \delta_{\beta \alpha} e^{-i p \cdot x} \big|_{p^0 = E_{\alpha \mathbf{p}}}.
\end{alignat}
The states $| \widehat{\Phi}_{\widehat{\alpha}} (\mathbf{p}) \rangle$ and $| \widehat{\Phi}_{\widehat{\alpha}}^* (\mathbf{p}) \rangle$ are the eigenstates of $\mathbf{P}$ and full Hamiltonian $H$, and they satisfy $\mathbf{P} | \widehat{\Phi}_{\widehat{\alpha}} (0) \rangle = \mathbf{P} | \widehat{\Phi}_{\widehat{\alpha}}^* (0) \rangle = 0$ as well as
\begin{alignat}{2}
	&\langle \Omega | \widehat{\Phi}_{\widehat{\beta}} (x) | \widehat{\Phi}_{\widehat{\alpha}} (\mathbf{p}) \rangle \propto \delta_{\widehat{\beta} \widehat{\alpha}}, \qquad
	&&\langle \Omega | \widehat{\Phi}_{\widehat{\beta}} (x) | \widehat{\Phi}_{\widehat{\alpha}}^* (\mathbf{p}) \rangle = 0, \\
	&\langle \Omega | \widehat{\Phi}_{\widehat{\beta}}^\dag (x) | \widehat{\Phi}_{\widehat{\alpha}} (\mathbf{p}) \rangle = 0, \qquad
	&&\langle \Omega | \widehat{\Phi}_{\widehat{\beta}}^\dag (x) | \widehat{\Phi}_{\widehat{\alpha}}^* (\mathbf{p}) \rangle \propto \delta_{\widehat{\beta} \widehat{\alpha}}.
\end{alignat}
As the eigenstates of Hermitian operators, these one-particle states belong to the orthogonal basis that spans the Fock space, and they are normalized such that
\begin{align}
	\langle \Phi (m_{\Phi_\beta}; \mathbf{q}) | \Phi (m_{\Phi_\alpha}; \mathbf{p}) \rangle
	&= \langle \Phi^* (m_{\Phi_\beta}; \mathbf{q}) | \Phi^* (m_{\Phi_\alpha}; \mathbf{p}) \rangle
	= (2 \pi)^3 2 E_{\alpha \mathbf{p}} \delta_{\beta \alpha} \delta^3 (\mathbf{q} - \mathbf{p}).
\end{align}
The completeness relation in the Fock space can be written as
\begin{align}
	1 = | \Omega \rangle \langle \Omega |
		+ \sum_{\widehat{\alpha}} \int \frac{d^3 \mathbf{p}}{(2 \pi)^3} \frac{1}{2 E_{\widehat{\alpha} \mathbf{p}}} | \widehat{\Phi}_{\widehat{\alpha}} (\mathbf{p}) \rangle \langle \widehat{\Phi}_{\widehat{\alpha}} (\mathbf{p}) |
		+ \sum_{\widehat{\alpha}} \frac{d^3 \mathbf{p}}{(2 \pi)^3} \frac{1}{2 E_{\widehat{\alpha} \mathbf{p}}} | \widehat{\Phi}_{\widehat{\alpha}}^* (\mathbf{p}) \rangle \langle \widehat{\Phi}_{\widehat{\alpha}}^* (\mathbf{p}) |
		+ \cdots,
	\label{eq:ComRelPhys}
\end{align}
where $E_{\widehat{\alpha} \mathbf{p}} \coloneqq \sqrt{m_{\widehat{\Phi}_{\widehat{\alpha}}}^2 + |\mathbf{p}|^2}$ and the ellipsis denotes the contribution of all the other one-particle and multiparticle states. Inserting this completeness relation into the two-point correlation function for $x^0 > y^0$, we obtain
\begin{align}
	\langle \Omega | \widehat{\Phi}_{\widehat{\alpha}} (x) \widehat{\Phi}_{\widehat{\alpha}}^\dag (y) | \Omega \rangle
	= \int \frac{d^3 \mathbf{p}}{(2 \pi)^3} \frac{1}{2 E_{\widehat{\alpha} \mathbf{p}}}
		\langle \Omega | \widehat{\Phi}_{\widehat{\alpha}} (x) | \widehat{\Phi}_{\widehat{\alpha}} (\mathbf{p}) \rangle
		\langle \widehat{\Phi}_{\widehat{\alpha}} (\mathbf{p}) | \widehat{\Phi}_{\widehat{\alpha}}^\dag (y) | \Omega \rangle.
	\label{eq:CorrDiag1}
\end{align}
Following the discussion in section 7 of referece \cite{Peskin&Schroeder}, let us introduce a unitary Lorentz boost operator $U$ from $\mathbf{p}$ to $0$. Since $\widehat{\Phi}_{\widehat{\alpha}} (x)$ is a scalar field, we have $U \widehat{\Phi}_{\widehat{\alpha}} (0) U^{-1} = \widehat{\Phi}_{\widehat{\alpha}} (0)$. Then,
\begin{align}
	\langle \Omega | \widehat{\Phi}_{\widehat{\alpha}} (x) | \widehat{\Phi}_{\widehat{\alpha}} (\mathbf{p}) \rangle
	&= \langle \Omega | e^{i P \cdot x} \widehat{\Phi}_{\widehat{\alpha}} (0) e^{-i P \cdot x} | \widehat{\Phi}_{\widehat{\alpha}} (\mathbf{p}) \rangle
	= \langle \Omega | \widehat{\Phi}_{\widehat{\alpha}} (0) | \widehat{\Phi}_{\widehat{\alpha}} (\mathbf{p}) \rangle e^{-i p \cdot x} \big|_{p^0 = E_{\widehat{\alpha} \mathbf{p}}} \nonumber \\
	&= \langle \Omega | U^{-1} U \widehat{\Phi}_{\widehat{\alpha}} (0) U^{-1} U | \widehat{\Phi}_{\widehat{\alpha}} (\mathbf{p}) \rangle e^{-i p \cdot x} \big|_{p^0 = E_{\widehat{\alpha} \mathbf{p}}} \nonumber \\
	&= \langle \Omega | \widehat{\Phi}_{\widehat{\alpha}} (0) | \widehat{\Phi}_{\widehat{\alpha}} (0) \rangle e^{-i p \cdot x} \big|_{p^0 = E_{\widehat{\alpha} \mathbf{p}}}.
\end{align}
We can therefore write
\begin{align}
	\langle \Omega | \widehat{\Phi}_{\widehat{\alpha}} (x) \widehat{\Phi}_{\widehat{\alpha}}^\dag (y) | \Omega \rangle
	= \int \frac{d^4 p}{(2 \pi)^4} e^{-i p \cdot (x - y)} \frac{i}{p^2 - m_{\widehat{\Phi}_{\widehat{\alpha}}}^2} \bigg|_{p^0 = E_{\widehat{\alpha} \mathbf{p}}}
		\big| \langle \Omega | \widehat{\Phi}_{\widehat{\alpha}} (0) | \widehat{\Phi}_{\widehat{\alpha}} (0) \rangle \big|^2 \quad (x^0 > y^0).
	\label{eq:CorrDiag2}
\end{align}
Note that $\Phi$ has been renormalized such that $\big| \langle \Omega | \widehat{\Phi}_{\widehat{\alpha}} (0) | \widehat{\Phi}_{\widehat{\alpha}} (0) \rangle \big|^2 = 1$. Hence, by redefining the phase of $| \widehat{\Phi}_{\widehat{\alpha}} (\mathbf{p}) \rangle$ if necessary, we can write
\begin{align}
	\boxed{\langle \Omega | \widehat{\Phi}_{\widehat{\beta}} (x) | \widehat{\Phi}_{\widehat{\alpha}} (\mathbf{p}) \rangle = \delta_{\widehat{\beta} \widehat{\alpha}} e^{-i p \cdot x} \big|_{p^0 = E_{\widehat{\alpha} \mathbf{p}}},}
	\label{eq:1PSPhys}
\end{align}
Similarly, from $\langle \Omega | \widehat{\Phi}_{\widehat{\alpha}}^\dag (y) \widehat{\Phi}_{\widehat{\alpha}} (x) | \Omega \rangle$ for $y^0 > x^0$, we can also obtain
\begin{align}
	\langle \Omega | \widehat{\Phi}_{\widehat{\beta}}^\dag (x) | \widehat{\Phi}_{\widehat{\alpha}}^* (\mathbf{p}) \rangle = \delta_{\widehat{\beta} \widehat{\alpha}} e^{-i p \cdot x} \big|_{p^0 = E_{\widehat{\alpha} \mathbf{p}}}.
\end{align}
Note that the pole should real-valued to obtain equation \ref{eq:CorrDiag2} from equation \ref{eq:CorrDiag1}, and thus this derivation is valid only for stable particles. \\

The current approach, in which the canonically quantized fields are different from the fields of physical particles, loses consistency when the scalar particles are external states of a perturbation theory. The states $| \widehat{\Phi}_{\widehat{\alpha}} (\mathbf{p}) \rangle$ and $| \widehat{\Phi}_{\widehat{\alpha}}^* (\mathbf{p}) \rangle$ are the particles that will be observed to propagate like free particles as the eigenstates of the full Hamiltonian $H$, and thus they should be the in- and out-states of a physical process. However, a basic assumption of a perturbation theory is that such external states are asymptotic states free from any interactions, \textit{i.e.}, $\lim_{t \to \pm \infty} H_I (t) = 0$, which requires that the asymptotic states are the eigenstates of the free Hamiltonian $H_0$. Hence, the on-shell renormalization scheme should be used for external stable particles, and in that scheme $\widehat{\Phi}_{\widehat{\alpha}} (\mathbf{x})$ are the canonically quantized fields and $\widehat{f}_{i \widehat{\alpha}}$ are the Yukawa couplings. The canonical quantization of $\Phi_\alpha (x)$ and its conjugate momentum density $\pi_{\Phi_\alpha} (x)$ implies that the field $\widehat{\Phi}_{\widehat{\alpha}} (x) = \sum_\beta (C_{\widehat{\Phi}}^{-1})_{\widehat{\alpha} \beta} \Phi_\beta (x)$ and its conjugate momentum density $\pi_{\widehat{\Phi}_{\widehat{\alpha}}} (x)$ are canonically quantized as well. To show that, we first write
\begin{align}
	\pi_{\widehat{\Phi}_{\widehat{\alpha}}} (x) &\coloneqq \frac{\partial \mathcal{L} (x)}{\partial [\partial_0 \widehat{\Phi}_{\widehat{\alpha}} (x)]}
	= \sum_\beta \frac{\partial \mathcal{L} (x)}{\partial [\partial_0 \Phi_\beta (x)]} \frac{\partial [\partial_0 \Phi_\beta (x)]}{\partial [\partial_0 \widehat{\Phi}_{\widehat{\alpha}} (x)]}
	= \sum_\beta \frac{\partial \mathcal{L} (x)}{\partial [\partial_0 \Phi_\beta (x)]} \frac{\partial \Phi_\beta (x)}{\partial \widehat{\Phi}_{\widehat{\alpha}} (x)} \nonumber \\
	&= \sum_\beta \pi_{\Phi_\beta} (x) \, (C_{\widehat{\Phi}})_{\beta \widehat{\alpha}},
\end{align}
and thus
\begin{align}
	[\widehat{\Phi}_{\widehat{\beta}} (\mathbf{y}), \pi_{\widehat{\Phi}_{\widehat{\alpha}}} (\mathbf{x})]
	&= \sum_{\gamma, \delta} (C_{\widehat{\Phi}}^{-1})_{\widehat{\beta} \delta} \
		[\Phi_\delta (\mathbf{y}), \pi_{\Phi_\gamma} (\mathbf{x})] \
		(C_{\widehat{\Phi}})_{\gamma \widehat{\alpha}}
	= i \delta_{\widehat{\beta} \widehat{\alpha}} \delta^3 (\mathbf{y} - \mathbf{x}).
\end{align}
In the on-shell scheme, the Hamiltonian is decomposed into
\begin{align}
	H (t) = \widehat{H}_0 + \widehat{H}_I (t),
\end{align}
where $\widehat{H}_0$ and $\widehat{H}_I$ are free and interaction Hamiltonians associated with $\widehat{\Phi}_{\widehat{\alpha}} (x)$. The one-particle states $| \widehat{\Phi}_{\widehat{\alpha}} (\mathbf{p}) \rangle$ are eigenstates of both $H$ and $\widehat{H}_0$, and thus
\begin{align}
	\widehat{\Phi}_{\widehat{\alpha}} (x) = e^{i H t} \widehat{\Phi}_{\widehat{\alpha}} (\mathbf{x}) e^{-i H t}, \qquad
	\widehat{\Phi}_{I \widehat{\alpha}} (x) = e^{i \widehat{H}_0 t} \widehat{\Phi}_{\widehat{\alpha}} (\mathbf{x}) e^{-i \widehat{H}_0 t},
\end{align}
and, for example,
\begin{align}
	\langle \widehat{0} | \widehat{\Phi}_{I \widehat{\beta}} (x) | \widehat{\Phi}_{\widehat{\alpha}} (\mathbf{p}) \rangle
	= \langle \Omega | \widehat{\Phi}_{\widehat{\beta}} (x) | \widehat{\Phi}_{\widehat{\alpha}} (\mathbf{p}) \rangle
	= \delta_{\widehat{\beta} \widehat{\alpha}} e^{-i p \cdot x} \big|_{p^0 = E_{\widehat{\alpha} \mathbf{p}}},
	\label{eq:1PSOS}
\end{align}
where $| \widehat{0} \rangle$ is the vacuum of the free field $\widehat{\Phi}_I (x)$. The on-shell renormalization scheme will be further discussed in section \ref{sec:RenScheme}.

The relation between the $S$-matrix element of a physical process and the associated correlation function is determined by the LSZ reduction formula, as is well-known. In the case of scattering $\widehat{\Phi}_{\widehat{\alpha}} \widehat{\Phi}_{\widehat{\beta}}^* \to \widehat{\Phi}_{\widehat{\gamma}} \widehat{\Phi}_{\widehat{\delta}}^*$, for example, the LSZ reduction formula in the renormalized theory is given by
\begin{align}
	\langle \widehat{\Phi}_{\widehat{\delta}}^* &(\mathbf{p}_{\widehat{\delta}}) \, \widehat{\Phi}_{\widehat{\gamma}} (\mathbf{p}_{\widehat{\gamma}}) | \, S \, | \widehat{\Phi}_{\widehat{\beta}}^* (\mathbf{p}_{\widehat{\beta}}) \, \widehat{\Phi}_{\widehat{\alpha}} (\mathbf{p}_{\widehat{\alpha}}) \rangle \nonumber \\
	&= \bigg[ i \int d^4 x_{\widehat{\delta}} \, e^{i p_{\widehat{\delta}} \cdot x_{\widehat{\delta}}} (\partial_{\widehat{\delta}}^2 + m_{\widehat{\Phi}_{\widehat{\delta}}}^2) \bigg] \cdots
		\bigg[ i \int d^4 x_{\widehat{\alpha}} \, e^{-i p_{\widehat{\alpha}} \cdot x_{\widehat{\alpha}}} (\partial_{\widehat{\alpha}}^2 + m_{\widehat{\Phi}_{\widehat{\alpha}}}^2) \bigg] \nonumber \\
		&\quad \qquad \langle \Omega | T \big\{ \widehat{\Phi}_{\widehat{\delta}}^\dag (x_{\widehat{\delta}}) \widehat{\Phi}_{\widehat{\gamma}} (x_{\widehat{\gamma}}) \widehat{\Phi}_{\widehat{\beta}} (x_{\widehat{\beta}}) \widehat{\Phi}_{\widehat{\alpha}}^\dag (x_{\widehat{\alpha}}) \big\} | \Omega \rangle.
	\label{eq:LSZ}
\end{align}
In the correlation function on the right-hand side, there exist divergences at the physical poles $m_{\widehat{\Phi}_{\widehat{\alpha}}}^2$ of the external states, and they are canceled by the factors $\int d^4 x_{\widehat{\alpha}} \, e^{-i p_{\widehat{\alpha}} \cdot x_{\widehat{\alpha}}} (\partial_{\widehat{\alpha}}^2 + m_{\widehat{\Phi}_{\widehat{\alpha}}}^2)$. In consequence, we can obtain a non-zero finite $S$-matrix element from the diagram whose external fields in the on-shell renormalization scheme are amputated.

\subsection{Unstable particles}
\begin{figure}[t]
	\centering
	\includegraphics[width = 40 mm]{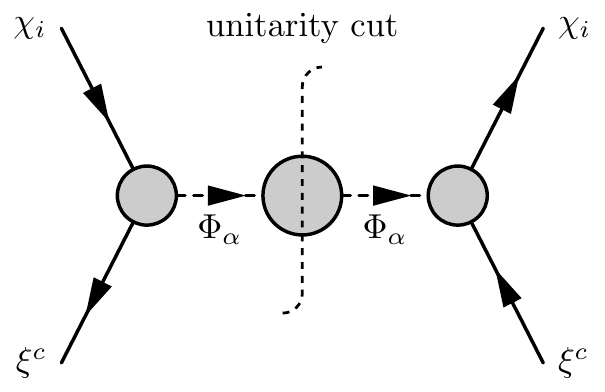}
	\caption{Applying the unitarity cut to the scattering $\chi_i \xi^c \to \chi_i \xi^c$.}
	\label{fig:UnitCut}
\end{figure}

The case of unstable particles is more subtle. Each of physical unstable particles cannot be related to a single renormalized field as discussed above, and thus we cannot obtain a relation such as equation \ref{eq:1PSPhys} simply by inserting the completeness relation such as equation \ref{eq:ComRelPhys} into the two-point correlation function. For example, the derivation of equation \ref{eq:1PSPhys} is valid only for stable particles whose propagator has real-valued poles, as mentioned above. In fact, the one-particle states such as $| \Phi_\alpha (\mathbf{p}) \rangle$ and $| \widehat{\Phi}_{\widehat{\alpha}} (\mathbf{p}) \rangle$ do not contribute to the unitarity cut of a scattering process mediated by $\Phi$, and it is determined by the multiparticle states such as $| \chi_j (\mathbf{p}_{\chi_j}) \, \xi^c (\mathbf{p}_{\xi^c}) \rangle$, where $\chi_j$ and $\xi$ are renormalized in the on-shell scheme. In other words, only cutting through stable particles $\chi_j$ and $\xi^c$ contribute to the unitarity cut. This is a general result in the case of an unstable particle, regardless of the number of flavors. To illustrate that, let us consider the diagram in figure \ref{fig:UnitCut}.
The cutting through $\Phi_\alpha$ generates the sum of diagrams as follows:
\begin{align*}
	\parbox{30 mm}{\includegraphics[width = 30 mm]{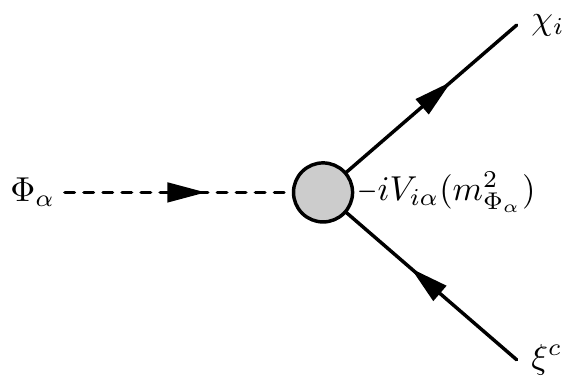}} \ + \
	\parbox{37.5 mm}{\includegraphics[width = 37.5 mm]{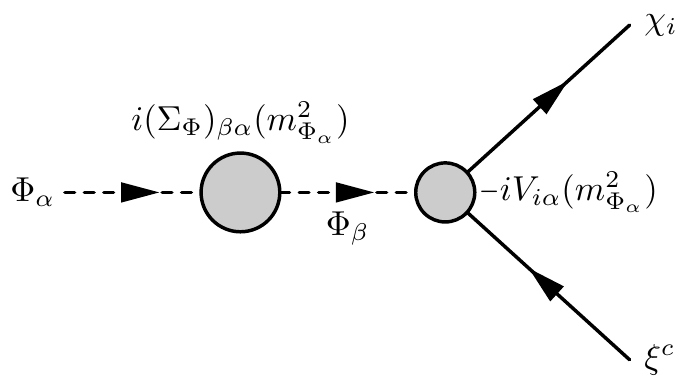}} \ + \
	\parbox{47.5 mm}{\includegraphics[width = 47.5 mm]{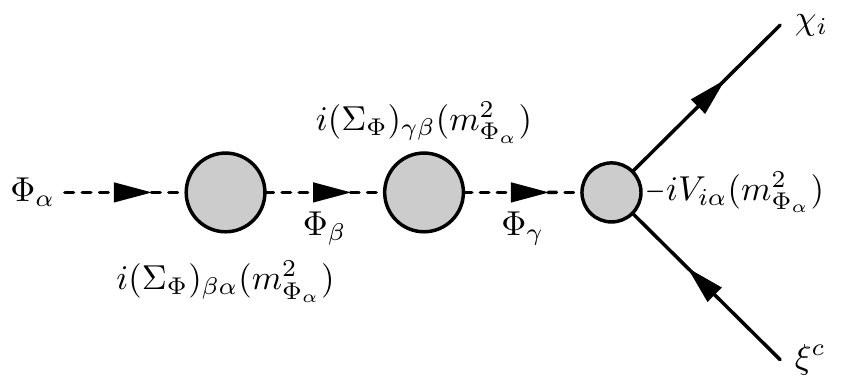}} \ + \ \cdots.
\end{align*}
The sum of these diagrams times $i$ is written as
\begin{align}
	V_{i \alpha} (p^2) + \sum_\beta V_{i \beta} (p^2) \frac{i}{p^2 - m_{\Phi_\beta}^2} i \Sigma_{\beta \alpha} (p^2) + \sum_{\beta, \gamma} V_{i \gamma} (p^2) \frac{i}{p^2 - m_{\Phi_\gamma}^2} i \Sigma_{\gamma \beta} (p^2) \frac{i}{p^2 - m_{\Phi_\beta}^2} i \Sigma_{\beta \alpha} (p^2) + \cdots.
	\label{eq:UnitCut1PS}
\end{align}
To calculate this sum at $p^2 = m_{\Phi_\alpha}^2$, we have to take the summation in the region around $p^2 = m_{\Phi_\alpha}^2$ where the series is convergent, and analytically continue the result to the region $p^2 = m_{\Phi_\alpha}^2$ after summation. Using equation \ref{eq:PropResum} and $p^2 = m_{\Phi_\alpha}^2$ for on-shell $\Phi_\alpha$, we can rewrite it as
\begin{align}
	\sum_\beta V_{i \beta} (p^2) &\bigg[ \delta_{\beta \alpha} + \sum_\gamma i \Delta_{\beta \gamma} (p^2) i \Sigma_{\gamma \alpha} (p^2) \bigg] \nonumber \\
	&= \sum_\beta V_{i \beta} (p^2) \bigg[ \delta_{\beta \alpha} - \sum_\gamma \Delta_{\beta \gamma} (p^2) \big\{ (p^2 - m_{\Phi_\alpha}^2) \delta_{\gamma \alpha} + \Sigma_{\gamma \alpha} (p^2) \big\} \bigg] \nonumber \\
	&= \sum_\beta V_{i \beta} (p^2) \bigg[ \delta_{\beta \alpha} - \sum_\gamma \Delta_{\beta \gamma} (p^2) (\Delta^{-1})_{\gamma \alpha} (p^2) \bigg]
	= 0.
\end{align}
In other words, each of physical particles corresponding to the complex pole is irrelevant to any one-particle state $| \Phi_\alpha (\mathbf{p}_{\Phi_\alpha}) \rangle$ as an external state, and its property can be rigorously studied only when the unstable particle appears as an intermediate state. This is a reasonable result consistent with the nature of unstable particles and external states. The external states in a perturbation theory are asymptotic states that existed or will exist well before or after the interaction occurs at around $t = 0$, and thus unstable particles as asymptotic states should have already decayed at $t = 0$. \\

In advance of discussing how to read the properties of unstable particles from scattering, it is worth mentioning the difference between stable and unstable particles with regard to the unitarity cut. If we take the cutting through a stable particle, then the result is also proportional to series \ref{eq:UnitCut1PS}. The sum also vanishes by the same reason, unless the fields of stable particles are renormalized in the on-shell scheme. This is a natural result since each component of the diagonalized propagator of stable particles can be related to a single one-particle state with the pole mass as discussed above. Furthermore, in the on-shell scheme, series \ref{eq:UnitCut1PS} is divergent at the pole mass, and it is nothing but the divergence in the correlation function in the LSZ reduction formula. Those divergences of external states are canceled by the factors $\int d^4 x_{\widehat{\alpha}} \, e^{-i p_{\widehat{\alpha}} \cdot x_{\widehat{\alpha}}} (\partial_{\widehat{\alpha}}^2 + m_{\widehat{\Phi}_{\widehat{\alpha}}}^2)$ to result in a non-zero finite $S$-matrix element associated with the diagram obtained by cutting through the stable particles. On the contrary, in the case of unstable particles whose dressed propagator has complex-valued poles, only multiparticle states such as $| \chi_j (\mathbf{p}_{\chi_j}) \, \xi^c (\mathbf{p}_{\xi^c}) \rangle$ contribute to the unitarity cut, as mentioned before.

The difference between single and multiple flavors should also be mentioned. The decaying particle is often regarded as an asymptotic state with a pole mass. In that approach, the external field of the unstable particle is amputated so that only the first term $V_{i \alpha} (p^2)$ in series \ref{eq:UnitCut1PS} is taken into account. In the case of a single flavor, the well-known formula of the decay width
\begin{align}
	\Gamma_{\Phi \to \chi \xi^c} = \frac{1}{2 m_\Phi} \int d\Pi_\chi \int d\Pi_{\xi^c} \, (2 \pi)^4 \delta^4 (p_\chi + p_{\xi^c} - p_\Phi) |\mathcal{M} (\Phi \to \chi \xi^c)|^2
	\label{eq:DWFormula}
\end{align}
can be derived in such a way from the $S$-matrix element $\langle \chi (\mathbf{p}_\chi) \, \xi^c (\mathbf{p}_{\xi^c}) | \, S \, | \Phi (\mathbf{p}_\Phi) \rangle$. However, amputating the external fields can never be justified for unstable particles, since the divergences in the correlation function at the complex-valued poles in equation \ref{eq:LSZ} cannot be canceled by the factors $\int d^4 x_{\widehat{\alpha}} \, e^{-i p_{\widehat{\alpha}} \cdot x_{\widehat{\alpha}}} (\partial_{\widehat{\alpha}}^2 + m_{\widehat{\Phi}_{\widehat{\alpha}}}^2)$ and thus we cannot obtain a non-vanishing $S$-matrix element for real-valued $p^2$. In fact, the LSZ reduction formula as given by equation \ref{eq:LSZ} is inapplicable to unstable particles, since the factors $\int d^4 x_{\widehat{\alpha}} \, e^{-i p_{\widehat{\alpha}} \cdot x_{\widehat{\alpha}}} (\partial_{\widehat{\alpha}}^2 + m_{\widehat{\Phi}_{\widehat{\alpha}}}^2)$ can be obtained only when the unstable particles can be treated as asymptotically free states which are the solutions of the Klein-Gordon equation. In contrast to stable particles, however, the unstable particles can never exist as such asymptotic states. The decay widths can be calculated in a well-defined way only from the scattering such as figure \ref{fig:UnitCut} that is mediated by the unstable particles, and in the case of a single flavor, the commonly accepted method by amputating the external field turns out to give the same result. Such a method, however, cannot be applied to the case of multiple flavors. As we have discussed, each of physical particles after mixing cannot be related to a single renormalized field, and thus amputating an external field, whichever renormalized field it is, cannot correctly describe the particle. To the decay of $\widehat{\Phi}_{\widehat{\alpha}}$ to $\chi_i \xi^c$, for example, two different types of effective vertices, \textit{i.e.}, $\widehat{V}_{i \widehat{\alpha}}$ and $\widehat{V}_{i \widehat{\alpha}}^c$ which will be defined below, must somehow contribute, since both are involved in the unitarity cut of the diagram mediated by $\widehat{\Phi}_{\widehat{\alpha}}$. However, only one of them should be chosen for each amputated external field $\widehat{\Phi}_{\widehat{\alpha}}$ to obtain the $S$-matrix element of the decay, and the partial decay width will then be given in terms of a single effective vertex $\widehat{V}_{i \widehat{\alpha}}$ by
\begin{align}
	\Gamma^\text{wrong}_{\widehat{\Phi}_{\widehat{\alpha}} \to \chi_i \xi^c}
	\coloneqq \frac{m_{\widehat{\Phi}_{\widehat{\alpha}}}}{16 \pi} |\widehat{V}_{i \widehat{\alpha}}|^2.
	\label{eq:TotDWWrong}
\end{align}
In section \ref{sec:Example} where numerical examples are given, we will confirm that this method is indeed wrong. \\

To study the properties of physical unstable particles, let us return to the scattering of figure \ref{fig:UnitCut} and calculate their partial decay widths to $| \chi_i \xi^c \rangle$ using the unitarity cut. The basic strategy is to take into account all possible initial configurations that contribute to $| \chi_i \xi^c \rangle$ through the physical particles. To that purpose, we will calculate a transition rate denoted by $\sum_{j, \widehat{\alpha}} \Gamma^\text{scat}_{\chi_j \xi^c \to \widehat{\Phi}_{\widehat{\alpha}} \to \chi_i \xi^c}$, and relate it to the decay widths. To be more specific, the cutting of figure \ref{fig:UnitCut} will generate all the scattering processes to $| \chi_i \xi^c \rangle$ mediated by $\Phi$ with the phase space of the initial states being integrated. Multiplying the factor of the initial spin average to each scattering and taking only the on-shell contributions of physical particles $\widehat{\Phi}$, we will obtain $\sum_{j, \widehat{\alpha}} \Gamma^\text{scat}_{\chi_j \xi^c \to \widehat{\Phi}_{\widehat{\alpha}} \to \chi_i \xi^c}$. The relation between this transition rate and the decay widths will be clarified. In this paper, we will only briefly discuss the steps, and the details are provided in reference \cite{MixingQFT}.

The cutting through intermediate stable particles up to the one-loop order in the self-energy of $\Phi$ gives only one type of a diagram with the final multiparticle state $| \chi_i \xi^c \rangle$:
\begin{align*}
	\includegraphics[width = 30 mm]{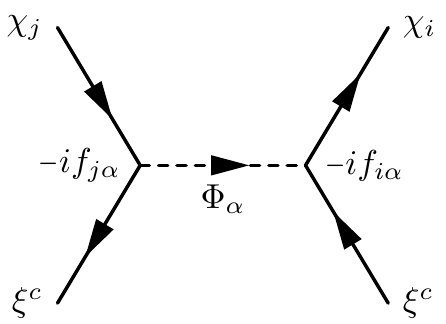},
\end{align*}
and we can calculate it using the $\mathcal{T}$ matrix which is the non-trivial part of the $S$ matrix, $S = 1 + i \mathcal{T}$:
\begin{align}
	\langle \chi_{i, r} \xi^c_s | \mathcal{T} \mathcal{T}^\dag | \chi_{i, r} \xi^c_s \rangle
	&= \sum_{X, a} \int d\Pi_X \, \langle \chi_{i, r} \xi^c_s | \mathcal{T} | X^a \rangle \langle X^a | \mathcal{T}^\dag | \chi_{i, r} \xi^c_s \rangle \nonumber \\
	&= (2 \pi)^4 \delta^4 (0)
		\sum_{\widehat{j}, k, l} \int d\Pi_{\chi_j}' \int d\Pi_{\xi^c}' \, (2 \pi)^4 \delta^4 (p_{\chi_j}' + p_{\xi^c}' - p_{\chi_i} - p_{\xi^c}) \nonumber \\
		&\qquad \qquad \qquad \qquad \qquad \big| \mathcal{M} [\chi_{j, k} (p_{\chi_j}') \xi^c_l (p_{\xi^c}') \to \chi_{i, r} \xi^c_s] \big|^2 + \cdots,
	\label{eq:UnitCutExample}
\end{align}
where $X$ denotes the one- and multi-particle states in the Fock space and the subscripts $r$, $s$, and $a$ are the indices of the internal degrees of freedom such as spin. We have also used $\langle f | \mathcal{T} | i \rangle = (2 \pi)^4 \delta^4 (p_i - p_f) \mathcal{M} (i \to f)$ to obtain the final expression. The divergent factor $(2 \pi)^4 \delta^4 (0)$ originates from total time $T$ and volume $V$, \textit{i.e.}, $(2 \pi)^4 \delta^4 (0) = TV$, and it is canceled out in any observable quantities \cite{Peskin&Schroeder}. To acquire a quantity associated with decay widths, we integrate equation \ref{eq:UnitCutExample} over the phase space of the final state as well:
\begin{align}
	\int d\Pi_{\chi_i} \int d\Pi_{\xi^ c} \, (2 \pi)^3 \delta^3 (\mathbf{p} - \mathbf{p}_{\chi_i} - \mathbf{p}_{\xi^c}) \sum_{r, s} \langle \chi_{i, r} \xi^c_s | \mathcal{T} \mathcal{T}^\dag | \chi_{i, r} \xi^c_s \rangle.
\end{align}
For convenience, let us define
\begin{align}
	\sigma' (\chi_j \xi^c \to \chi_i \xi^c) &\coloneqq \int d\Pi_{\chi_i} \int d\Pi_{\xi^c} \, (2 \pi)^4 \delta^3 (\mathbf{p} - \mathbf{p}_{\chi_i} - \mathbf{p}_{\xi^c}) \delta (E_{\chi_j}' + E_{\xi^c}' - E_{\chi_i} - E_{\xi^c}) \nonumber \\
			&\qquad \qquad \qquad \sum_{r, s, k, l} \left| \mathcal{M} [\chi_{j, k} (p_{\chi_j}') \xi^c_l (p_{\xi^c}') \to \chi_{i, r} \xi^c_s] \right|^2,
\end{align}
which is a dimensionless quantity related to the scattering cross section $\sigma$ as follows:
\begin{align}
	\sigma' (\chi_j \xi^c \to \chi_i \xi^c) = 2 E_{\chi_j} \, 2 E_{\xi^c} \big| \mathbf{v}_{\chi_j} - \mathbf{v}_{\xi^c} \big| \, 4 \, \sigma (\chi_j \xi^c \to \chi_i \xi^c).
\end{align}
Now we define a transition rate and calculate it in the center-of-momentum (CM) frame as follows:
\begin{empheq}[box=\fbox]{align}
	\sum_{j, \widehat{\alpha}} &\Gamma^\text{scat}_{\chi_j \xi^c \to \widehat{\Phi}_{\widehat{\alpha}} \to \chi_i \xi^c} \nonumber \\
	&\coloneqq \frac{1}{4} \sum_j \int d\Pi_{\chi_j}' \int d\Pi_{\xi^c}' \, (2 \pi)^3 \delta^3 (\mathbf{p}_{\chi_j}' + \mathbf{p}_{\xi^c}') \,
			\sigma'_\text{CM} [\chi_j (p_{\chi_j}') \xi^c (p_{\xi^c}') \to \chi_i \xi^c] \bigg|_\text{OS}
		\label{eq:TRScat} \\
	&= \frac{i}{2^8 \pi^2} \sum_{j, \widehat{\alpha}, \widehat{\beta}} 
		\widehat{V}_{i \widehat{\beta}}^* \widehat{V}_{j \widehat{\beta}}^{c *} \widehat{V}_{i \widehat{\alpha}} \widehat{V}_{j \widehat{\alpha}}^c
		\frac{m_{\widehat{\Phi}_{\widehat{\beta}}}^3 + m_{\widehat{\Phi}_{\widehat{\alpha}}}^3}{p_{\widehat{\Phi}_{\widehat{\beta}}}^{* 2} - p_{\widehat{\Phi}_{\widehat{\alpha}}}^2},
		\label{eq:TRScatExp}
\end{empheq}
where $1/4$ is the spin average over the initial state $| \chi_j \xi^c \rangle$, and the subscript ``OS" means that only the on-shell contributions of $\widehat{\Phi}_{\widehat{\alpha}}$ have been taken from the scattering. In addition, the effective vertices are defined by
\begin{align}
	\widehat{V}_{i \widehat{\alpha}} \coloneqq (V C_{\widehat{\Phi}^f})_{i \widehat{\alpha}} (p_{\widehat{\Phi}_{\widehat{\alpha}}}^2), \qquad
	\widehat{V}_{i \widehat{\alpha}}^c \coloneqq (V^* C_{\widehat{\Phi}^i})_{i \widehat{\alpha}} (p_{\widehat{\Phi}_{\widehat{\alpha}}}^2).
\end{align}
The derivation of equation \ref{eq:TRScatExp} is given in appendix \ref{sec:TransRate}. It turns out that $\sum_{j, \widehat{\alpha}} \Gamma^\text{scat}_{\chi_j \xi^c \to \widehat{\Phi}_{\widehat{\alpha}} \to \chi_i \xi^c}$ is related to the total decay widths, which are determined by the imaginary parts of complex poles, in the following way:
\begin{align}
	\boxed{\sum_{\widehat{\alpha}} \Gamma_{\widehat{\Phi}_{\widehat{\alpha}}}
	= \sum_{i, j, \widehat{\alpha}} \Gamma^\text{scat}_{\chi_j \xi^c \to \widehat{\Phi}_{\widehat{\alpha}} \to \chi_i \xi^c}.}
	\label{eq:DWTot}
\end{align}
This identity can be numerically verified. Note, however, that $\sum_j \Gamma^\text{scat}_{\chi_j \xi^c \to \widehat{\Phi}_{\widehat{\alpha}} \to \chi_i \xi^c}$ is a complex-valued quantity by itself in general, and thus the partial decay width cannot be defined by $\Gamma_{\widehat{\Phi}_{\widehat{\alpha}} \to \chi_i \xi^c} = \sum_j \Gamma^\text{scat}_{\chi_j \xi^c \to \widehat{\Phi}_{\widehat{\alpha}} \to \chi_i \xi^c}$. Moreover, we have $\Gamma_{\widehat{\Phi}_{\widehat{\alpha}}} \neq \sum_{i, j} \Gamma^\text{scat}_{\chi_j \xi^c \to \widehat{\Phi}_{\widehat{\alpha}} \to \chi_i \xi^c}$. At best, we can define a collective partial decay width as
\begin{align}
	\boxed{\sum_{\widehat{\alpha}} \Gamma_{\widehat{\Phi}_{\widehat{\alpha}} \to \chi_i \xi^c}
	\coloneqq \sum_{j, \widehat{\alpha}} \Gamma^\text{scat}_{\chi_j \xi^c \to \widehat{\Phi}_{\widehat{\alpha}} \to \chi_i \xi^c}.}
	\label{eq:DWPartial}
\end{align}
In addition, we have $\sum_{\widehat{\alpha}} \Gamma^\text{wrong}_{\widehat{\Phi}_{\widehat{\alpha}} \to \chi_i \xi^c} \neq \sum_{\widehat{\alpha}} \Gamma_{\widehat{\Phi}_{\widehat{\alpha}} \to \chi_i \xi^c}$, although $\Gamma_{\widehat{\Phi}_{\widehat{\alpha}}} = \sum_i \Gamma^\text{wrong}_{\widehat{\Phi}_{\widehat{\alpha}} \to \chi_i \xi^c}$, where $\Gamma^\text{wrong}_{\widehat{\Phi}_{\widehat{\alpha}} \to \chi_i \xi^c}$ is defined by equation \ref{eq:TotDWWrong}. Hence, equation \ref{eq:TotDWWrong} is not the correct partial decay width, since it is inconsistent with unitarity. These relations can also be numerically verified, as we will see in section \ref{sec:Example}.

In fact, the partial decay width of each of physical particles cannot be defined since it is a quasiparticle that is not separately observable, \textit{i.e.}, each of them is in a sense unphysical. In any experiment that tries to observe $\Phi$, we can only see the results generated by multiple physical particles. In scattering mediated by $\Phi$, for example, the interferences between physical particles also contribute to the resonance peak at each pole mass of $\widehat{\Phi}$. Even when the total decay widths of physical particles have a hierarchy such that it seems possible to observe one of them after the others decayed, we actually cannot observe a single quasiparticle as it is, since the interferences between the particle with the longest lifetime and the others should have decayed as well. The details were discussed in reference \cite{MixingQFT} with the example of neutral kaons. Some alternative transition rates were also calculated in reference \cite{MixingQFT}, and it was shown that the results are mutually consistent in the sense that all the transition rates satisfy equations \ref{eq:DWTot} and \ref{eq:DWPartial}, although their values for each of physical particles are different.

To obtain the transition rate given by equation \ref{eq:TRScatExp}, the self-energy has been calculated up to the one-loop order. For large mass differences, such precision guarantees that equation \ref{eq:TRScatExp} is exact up to the next-leading order in perturbation, since the one-loop corrections to the self-energy and vertices are both of $\mathcal{O} (\alpha)$. For small mass differences, the one-loop corrections to the self-energy get enhanced by the non-perturbative effect mentioned in section \ref{sec:Quasi}, and they become effects of $\mathcal{O} (1)$. Hence, equation \ref{eq:TRScatExp} is exact only up to the leading order in perturbation, and the loop corrections to the vertices can therefore be neglected. Note that $p_{\widehat{\Phi}_{\widehat{\beta}}}^{* 2} - p_{\widehat{\Phi}_{\widehat{\alpha}}}^2 \sim m_{\widehat{\Phi}_{\widehat{\alpha}}}^2 \mathcal{O} (\alpha)$ in equation \ref{eq:TRScatExp}, and its next-leading contributions of $m_{\widehat{\Phi}_{\widehat{\alpha}}}^2 \mathcal{O} (\alpha^2)$ come from the two-loop diagrams of the self-energy. Accordingly, the leading contributions to equation \ref{eq:TRScatExp} indeed corresponds to the one-loop diagrams of the self-energy.

\section{Physical renormalization schemes and their limitations}		\label{sec:RenScheme}
Until now, we have discussed the toy model renormalized in non-physical schemes, whether the particles are stable or unstable. In this section, we specifically examine the on-shell and complex-mass renormalization schemes, whose common purpose is to choose the fields of physical particles as the canonically quantized fields in terms of which the Lagrangian density is expressed. The limitations of those physical schemes will be discussed.

\subsection{On-shell renormalization scheme}
In the on-shell renormalization scheme \cite{EWOSRen}, the renormalization conditions are derived from the requirement
\begin{align}
	(\Delta_\Phi)_{\beta \alpha} (p^2) \big|_{p^2 \approx m_{\Phi_\gamma}^2} = \frac{\delta_{\beta \gamma} \delta_{\gamma \alpha}}{p^2 - m_{\Phi_\gamma}^2} + \cdots,
	\label{eq:OSRenReq}
\end{align}
where $m_{\Phi_\gamma}$ is the pole mass of $\Phi_\gamma$ and the ellipsis denotes the higher order terms of the Laurent expansion around $p^2 = m_{\Phi_\gamma}^2$. This condition is supposed to guarantee that $\Phi_\gamma$ is the field of the physical particle with mass $m_{\Phi_\gamma}$. In the case of unstable particles, however, the propagator has a complex pole $p_{\Phi_\gamma}^2 = m_{\Phi_\gamma}^2 - i m_{\Phi_\gamma} \Gamma_{\Phi_\gamma}$, and thus the Laurent expansion around $m_{\Phi_\gamma}^2$ is invalid at $p^2 = m_{\Phi_\gamma}^2$ simply because $\Delta_\Phi$ is not divergent there. In other words, the ellipsis must also be divergent at $p^2 = m_{\Phi_\gamma}^2$ to make $\Delta_\Phi$ regular. Hence, the on-shell renormalization scheme cannot give a correct result for unstable particles as is well known, although it is nevertheless often applied to unstable particles as well in many works in the literature. Here, we restrict the analysis to the case of stable particles.

The renormalized self-energy of $\Phi$ is given by equation \ref{eq:SERenwCT}:
\begin{align}
	\Sigma_\Phi (p^2) = p^2 \big[ \Sigma_{0 \Phi}' (p^2) + \delta_\Phi^H \big] + \delta \Sigma_\Phi - \delta M_\Phi^2,
	\label{eq:SERenwCT1}
\end{align}
where we have used equations \ref{eq:SERen1Loop} and \ref{eq:SENonRen}. Since $\Phi$ is stable, $\Sigma_\Phi (p^2)$ and $\Sigma_{0 \Phi}' (p^2)$ have no absorptive parts, \textit{i.e.}, $\Sigma_\Phi (p^2)$ and $\Sigma_{0 \Phi}' (p^2)$ are Hermitian. The resulting renormalization conditions are
\begin{align}
	(\Sigma_\Phi)_{\beta \alpha} (m_{\Phi_\alpha}^2) = (\Sigma_\Phi)_{\alpha \beta} (m_{\Phi_\alpha}^2) = 0, \qquad
	\frac{d(\Sigma_\Phi)_{\alpha \alpha}}{dp^2} (m_{\Phi_\alpha}^2) = 0,
\end{align}
which are satisfied if the counterterms are
\begin{align}
	(\delta_\Phi)_{\beta \alpha} &= \frac{(\Sigma_{0 \Phi})_{\beta \alpha} (m_{\Phi_\alpha}^2)}{m_{\Phi_\beta}^2 - m_{\Phi_\alpha}^2} \quad (\beta \neq \alpha), \qquad
	(\delta_\Phi)_{\alpha \alpha} = \minus \frac{d(\Sigma_{0 \Phi})_{\alpha \alpha}}{dp^2} (m_{\Phi_\alpha}^2),
		\label{eq:OSRenCT1} \\
	(\delta M_\Phi^2)_{\beta \alpha} &= (\delta \Sigma_\Phi)_{\beta \alpha} - \frac{m_{\Phi_\beta}^2 m_{\Phi_\alpha}^2 \big[ (\Sigma_{0 \Phi}')_{\beta \alpha} (m_{\Phi_\beta}^2) - (\Sigma_{0 \Phi}')_{\beta \alpha} (m_{\Phi_\alpha}^2) \big]}{m_{\Phi_\beta}^2 - m_{\Phi_\alpha}^2} \quad (\beta \neq \alpha),
		\label{eq:OSRenCT2} \\
	(\delta M_\Phi^2)_{\alpha \alpha} &= (\delta \Sigma_\Phi)_{\alpha \alpha} + (\Sigma_{0 \Phi})_{\alpha \alpha} (m_{\Phi_\alpha}^2) - m_{\Phi_\alpha}^2 \frac{d(\Sigma_{0 \Phi})_{\alpha \alpha}}{dp^2} (m_{\Phi_\alpha}^2).
		\label{eq:OSRenCT3}
\end{align}
The on-shell renormalization conditions and counterterms given here are explicitly derived in appendix \ref{sec:OSRen}. In the on-shell renormalization scheme, the Lagrangian density of the toy model is given by equation \ref{eq:LagPhys} with $\widehat{\Phi}_{\widehat{\alpha}} = \widehat{\Phi}_{\widehat{\alpha}}^f = \widehat{\Phi}_{\widehat{\alpha}}^i$, and $\widehat{\Phi}_{\widehat{\alpha}}$ are the canonically quantized fields that create or annihilate orthonormal one-particle states according to equation \ref{eq:1PSOS}. The inverse propagator of $\widehat{\Phi}$ can be written as $\widehat{\Delta}_{\widehat{\Phi}}^{-1} (p^2) = C_{\widehat{\Phi}}^\dag \Delta_\Phi^{-1} (p^2) C_{\widehat{\Phi}}$, and it is straightforward to show that $\widehat{\Delta}_{\widehat{\Phi}} (p^2)$ satisfies equation \ref{eq:OSRenReq}.

\subsection{Complex-mass renormalization scheme}
Similarly to the on-shell scheme, the complex-mass renormalization scheme \cite{CompMass} is also supposed to express the Lagrangian density in terms of the fields of physical particles, and it was developed as a renormalization scheme appropriate for unstable particles.

In this scheme, the complex pole $p_{\Phi_\alpha}^2 = m_{\Phi_\alpha}^2 - i m_{\Phi_\alpha} \Gamma_{\Phi_\alpha}$ is chosen as the complex-valued squared-mass by introducing interaction terms into the Lagrangian density which compensate for the imaginary part of $p_{\Phi_\alpha}^2$:
\begin{align}
	\mathcal{L}_\text{mass} = \minus \sum_\alpha m_{\Phi_\alpha}^2 \Phi_\alpha^\dag \Phi_\alpha
	= \minus \sum_\alpha p_{\Phi_\alpha}^2 \Phi_\alpha^\dag \Phi_\alpha - \sum_\alpha i m_{\Phi_\alpha} \Gamma_{\Phi_\alpha} \Phi_\alpha^\dag \Phi_\alpha.
\end{align}
The non-renormalized self-energy $\Sigma_{0 \Phi}^\text{CMS} (p^2)$ in this scheme can be written in terms of the non-renormalized self-energy $\Sigma_{0 \Phi} (p^2)$ for a real squared-mass as
\begin{align}
	(\Sigma_{0 \Phi}^\text{CMS})_{\beta \alpha} (p^2) = (\Sigma_{0 \Phi})_{\beta \alpha} (p^2) + i m_{\Phi_\alpha} \Gamma_{\Phi_\alpha} \delta_{\beta \alpha},
\end{align}
and thus
\begin{align}
	\Sigma_{0 \Phi}'^{\text{CMS}} (p^2) = \Sigma_{0 \Phi}' (p^2), \qquad
	(\delta \Sigma_\Phi^\text{CMS})_{\beta \alpha} = (\delta \Sigma_\Phi)_{\beta \alpha} + i m_{\Phi_\alpha} \Gamma_{\Phi_\alpha} \delta_{\beta \alpha}.
\end{align}
By analytic continuation, $p^2$ is allowed to have a complex value, and the renormalization conditions are chosen such that the dressed propagator satisfies
\begin{align}
	(\Delta_\Phi)_{\beta \alpha} (p^2) \big|_{p^2 \approx p_{\Phi_\gamma}^2} = \frac{\delta_{\beta \gamma} \delta_{\gamma \alpha}}{p^2 - p_{\Phi_\gamma}^2} + \cdots.
	\label{eq:CMSRenReq}
\end{align}
Using equation \ref{eq:SERenwCT1} and following the same steps as in the on-shell scheme, we can find the renormalization conditions in the complex-mass scheme:
\begin{align}
	(\Sigma_\Phi^\text{CMS})_{\beta \alpha} (p_{\Phi_\alpha}^2)
		= (\Sigma_\Phi^\text{CMS})_{\alpha \beta} (p_{\Phi_\alpha}^2) = 0, \qquad
	\frac{d(\Sigma_\Phi^\text{CMS})_{\alpha \alpha}}{dp^2} (p_{\Phi_\alpha}^2) = 0.
\end{align}
and the associated counterterms:
\begin{align}
	(\delta_\Phi)_{\beta \alpha} &= \frac{(\Sigma_{0 \Phi}^\text{CMS})_{\beta \alpha} (m_{\Phi_\alpha}^2)}{m_{\Phi_\beta}^2 - m_{\Phi_\alpha}^2} \quad (\beta \neq \alpha), \qquad
	(\delta_\Phi)_{\alpha \alpha} = \minus \frac{d(\Sigma_{0 \Phi}^\text{CMS})_{\alpha \alpha}}{dp^2} (m_{\Phi_\alpha}^2),
		\label{eq:CMSRenCT1} \\
	(\delta M_\Phi^2)_{\beta \alpha} &= (\delta \Sigma_\Phi^\text{CMS})_{\beta \alpha} - \frac{m_{\Phi_\beta}^2 m_{\Phi_\alpha}^2 [\Sigma_{0 \Phi}'^{\text{CMS}})_{\beta \alpha} (m_{\Phi_\beta}^2) - \Sigma_{0 \Phi}'^{\text{CMS}})_{\beta \alpha} (m_{\Phi_\alpha}^2)]}{m_{\Phi_\beta}^2 - m_{\Phi_\alpha}^2} \quad (\beta \neq \alpha),
		\label{eq:CMSRenCT2} \\
	(\delta M_\Phi^2)_{\alpha \alpha} &= (\Sigma_{0 \Phi}^\text{CMS})_{\alpha \alpha} (m_{\Phi_\alpha}^2) - m_{\Phi_\alpha}^2 \frac{d(\Sigma_{0 \Phi}^\text{CMS})_{\alpha \alpha}}{dp^2} (m_{\Phi_\alpha}^2),
		\label{eq:CMSRenCT3}
\end{align}
where we have replaced $p_{\Phi_\alpha}^2$ with $m_{\Phi_\alpha}^2$, since $p_{\Phi_\alpha}^2 = m_{\Phi_\alpha}^2 [1 + \mathcal{O} (\alpha)]$ and $\Sigma_{0 \Phi}^\text{CMS} (p_{\Phi_\alpha}^2) = \Sigma_{0 \Phi}^\text{CMS} (m_{\Phi_\alpha}^2)$ up to $\mathcal{O} (\alpha)$. \\

However, the complex-mass renormalization scheme cannot be applied to the theories with flavor mixing of unstable particles in general. As we have seen in section \ref{sec:Quasi}, it is impossible to relate a single renormalized field to each of physical particles. The propagators are given by equations \ref{eq:PropPMDExp} and \ref{eq:PropAMDExp}, and thus the basic requirement of the complex-mass scheme given by equation \ref{eq:CMSRenReq} cannot be satisfied for unstable particles. Whether the mass difference is large or small, any strategy to renormalize a theory to obtain such a single renormalized field cannot succeed. In addition, when the mass differences are small, the deviations of the mixing matrices $C_{\widehat{\Phi}^f}$ and $C_{\widehat{\Phi}^i}$ from unitarity can be so large that the counterterms will make corrections much larger than the precision of the theory $\mathcal{O} (\alpha)$, as discussed in section \ref{sec:RevRen}. If we try to change the basis using such mixing matrices, the order-by-order renormalization can become too complicated to have any practical advantage. In such cases, even the expression of the self-energy given by equation \ref{eq:SERenwCT1}, which is one of the conventional assumptions of the complex-mass scheme, is invalid.

\section{Examples}		\label{sec:Example}
In this section, we discuss multiple examples of two flavors. The case of a large mass difference can be analytically studied, while the case of a small mass difference can only be numerically analyzed in general. For unstable particles with a small mass difference, we will indeed see that the deviations of mixing matices from unitarity can go beyond the typical perturbative corrections. When there are more than two flavors, it is better to apply the numerical approach since an analytical calculation is very complicated. \\

First let us derive an analytical expression of $C^{-1} (p^2)$ for two flavors. Defining
\begin{align}
	A (p^2) &\coloneqq \left( \begin{array}{cc} 1 + (\Sigma_\Phi')_{11} & (\Sigma_\Phi')_{12} \\
			(\Sigma_\Phi')_{21} & 1 + (\Sigma_\Phi')_{22} \end{array} \right), \\
	B (p^2) &\coloneqq M_\Phi A^{-1} M_\Phi
		= \frac{1}{\text{Det} [A]} \left( \begin{array}{cc} m_{\Phi_1}^2 [(1 + (\Sigma_\Phi')_{22}] & \minus m_{\Phi_1} m_{\Phi_2} (\Sigma_\Phi')_{12} \\
				\minus m_{\Phi_1} m_{\Phi_2} (\Sigma_\Phi')_{21} & m_{\Phi_2}^2 [1 + (\Sigma_\Phi')_{11}] \end{array} \right),
\end{align}
and
\begin{alignat}{2}
	&a_{11} (p^2) \coloneqq m_{\Phi_1}^2 [1 + (\Sigma_\Phi')_{22}], \qquad
	&&a_{22} (p^2) \coloneqq m_{\Phi_2}^2 [1 + (\Sigma_\Phi')_{11}], \\
	&a_{12} (p^2) \coloneqq \minus m_{\Phi_1} m_{\Phi_2} (\Sigma_\Phi')_{12}, \qquad
	&&a_{21} (p^2) \coloneqq \minus m_{\Phi_1} m_{\Phi_2} (\Sigma_\Phi')_{21},
\end{alignat}
we can write
\begin{align}
	\text{Det} [A (p^2)] = \frac{a_{11} a_{22} - a_{12} a_{21}}{m_{\Phi_1}^2 m_{\Phi_2}^2}, \qquad
	B (p^2) = \frac{m_{\Phi_1}^2 m_{\Phi_2}^2}{a_{11} a_{22} - a_{12} a_{21}}
		\left( \begin{array}{cc} a_{11} & a_{12} \\ a_{21} & a_{22} \end{array} \right).
\end{align}
Hence,
\begin{align}
	P_1^2 (p^2) = \frac{m_{\Phi_1}^2 m_{\Phi_2}^2}{a_{11} a_{22} - a_{12} a_{21}} \lambda_1, \qquad
	P_2^2 (p^2) = \frac{m_{\Phi_1}^2 m_{\Phi_2}^2}{a_{11} a_{22} - a_{12} a_{21}} \lambda_2,
\end{align}
where
\begin{align}
	\lambda_1 (p^2) &\coloneqq \frac{1}{2} \bigg[ (a_{11} + a_{22}) - \sqrt{(a_{22} - a_{11})^2 + 4 a_{12} a_{21}} \bigg],
		\label{eq:lambda1} \\
	\lambda_2 (p^2) &\coloneqq \frac{1}{2} \bigg[ (a_{11} + a_{22}) + \sqrt{(a_{22} - a_{11})^2 + 4 a_{12} a_{21}} \bigg].
		\label{eq:lambda2}
\end{align}
Then,
\begin{align}
	\boxed{C^{-1} (p^2) = \left( \begin{array}{cc} \frac{|a_{12}|}{\sqrt{|a_{11} - \lambda_1|^2 + |a_{12}|^2}} & \minus \frac{(a_{22} - \lambda_2) |a_{21}|}{a_{21} \sqrt{|a_{22} - \lambda_2|^2 + |a_{21}|^2}} \\
		\minus \frac{(a_{11} - \lambda_1) |a_{12}|}{a_{12} \sqrt{|a_{11} - \lambda_1|^2 + |a_{12}|^2}} & \frac{|a_{21}|}{\sqrt{|a_{22} - \lambda_2|^2 + |a_{21}|^2}} \end{array} \right).}
	\label{eq:Cinv}
\end{align}
For a large mass difference, $C^{-1}$ is in the form of $C^{-1} = 1 + \mathcal{O} (\alpha)$ since $a_{\beta \alpha} \sim \mathcal{O} (\alpha)$ and $a_{\alpha \alpha} - \lambda_\alpha \sim \mathcal{O} (\alpha^2)$, as we will see below. For a small mass difference, on the other hand, every component of $C^{-1}$ is $\mathcal{O} (1)$ in general since $a_{\beta \alpha} \sim \mathcal{O} (\alpha)$ and $a_{\alpha \alpha} - \lambda_\alpha \sim \mathcal{O} (\alpha)$, which means that the $\mathcal{O} (\alpha)$ contributions to $C^{-1}$ can correctly be calculated only when $\mathcal{O} (\alpha^2)$ contributions to $a_{\beta \alpha}$ and $\lambda_\alpha$ are considered as well. Since this requires calculations up to two-loop contributions to the self-energy, we consider $C^{-1}$ only up to $\mathcal{O} (1)$ when the mass difference is small.

\subsection{Analytical approach to the case of a large mass difference}		\label{sec:ExLarge}
When the mass difference is large, \textit{i.e.},
\begin{align}
	\boxed{|m_{\Phi_\beta} - m_{\Phi_\alpha}| \gg m_{\Phi_\beta} \mathcal{O} (\alpha),}
\end{align}
we have
\begin{align}
	\bigg| \frac{4 a_{12} a_{21}}{(a_{22} - a_{11})^2} \bigg|
	= \frac{4 m_{\Phi_1}^2 m_{\Phi_2}^2 \mathcal{O} (\alpha^2)}{(m_{\Phi_2}^2 - m_{\Phi_1}^2)^2} \ll 1.
\end{align}
Hence, appropriately choosing the branch cut of the square root, we can write up to $\mathcal{O} (\alpha^2)$
\begin{align}
	\lambda_1 (p^2) &= \frac{1}{2} \bigg[ (a_{11} + a_{22}) - (a_{22} - a_{11}) - \frac{2 a_{12} a_{21}}{a_{22} - a_{11}} \bigg]
	= a_{11} - \frac{a_{12} a_{21}}{m_{\Phi_2}^2 - m_{\Phi_1}^2}, \\
	\lambda_2 (p^2) &= \frac{1}{2} \bigg[ (a_{11} + a_{22}) + (a_{22} - a_{11}) + \frac{2 a_{12} a_{21}}{a_{22} - a_{11}} \bigg]
	= a_{22} + \frac{a_{12} a_{21}}{m_{\Phi_2}^2 - m_{\Phi_1}^2},
\end{align}
and thus up to $\mathcal{O} (\alpha)$
\begin{align}
	P_1^2 (p^2) &= \frac{m_{\Phi_1}^2 m_{\Phi_2}^2}{a_{11} a_{22} - a_{12} a_{21}} \lambda_1
		= \frac{m_{\Phi_1}^2 m_{\Phi_2}^2}{a_{22}}
		= m_{\Phi_1}^2 [1 - (\Sigma_\Phi')_{11}], \\
	P_2^2 (p^2) &= \frac{m_{\Phi_1}^2 m_{\Phi_2}^2}{a_{11} a_{22} - a_{12} a_{21}} \lambda_2
		= \frac{m_{\Phi_1}^2 m_{\Phi_2}^2}{a_{11}}
		= m_{\Phi_2}^2 [1 - (\Sigma_\Phi')_{22}].
\end{align}
This implies
\begin{align}
	(\Sigma_{\widehat{\Phi}}')_{\widehat{\alpha}} (p^2) = (\Sigma_\Phi')_{\widehat{\alpha} \widehat{\alpha}} (p^2)
\end{align}
and
\begin{align}
	\text{Re} [(\Sigma_\Phi')_{\widehat{\alpha} \widehat{\alpha}} (p_{\widehat{\Phi}_{\widehat{\alpha}}}^2)] = \frac{m_{\widehat{\Phi}_{\widehat{\alpha}}}^2}{m_{\Phi_{\widehat{\alpha}}}^2} - 1, \qquad
	\text{Im} [(\Sigma_\Phi')_{\widehat{\alpha} \widehat{\alpha}} (p_{\widehat{\Phi}_{\widehat{\alpha}}}^2)] = \frac{\Gamma_{\widehat{\Phi}_{\widehat{\alpha}}}}{m_{\Phi_{\widehat{\alpha}}}}.
\end{align}
Up to $\mathcal{O} (\alpha^2)$, we can write
\begin{align}
	\sqrt{|a_{11} - \lambda_1|^2 + |a_{21}|^2} = |a_{21}|, \qquad
	\sqrt{|a_{22} - \lambda_2|^2 + |a_{12}|^2} = |a_{12}|,
\end{align}
and therefore up to $\mathcal{O} (\alpha)$
\begin{align}
	C^{-1} (p^2) &= \left( \begin{array}{cc} 1 & \minus \frac{a_{22} - \lambda_2}{a_{21}} \\
			\minus \frac{a_{11} - \lambda_1}{a_{12}} & 1 \end{array} \right)
		= \left( \begin{array}{cc} 1 & \minus \frac{m_{\Phi_1} m_{\Phi_2} (\Sigma_\Phi')_{12}}{m_{\Phi_2}^2 - m_{\Phi_1}^2} \\
			\frac{m_{\Phi_1} m_{\Phi_2} (\Sigma_\Phi')_{21}}{m_{\Phi_2}^2 - m_{\Phi_1}^2} & 1 \end{array} \right), \\
	C^\mathsf{T} (p^2) &= \left( \begin{array}{cc} 1 & \minus \frac{m_{\Phi_1} m_{\Phi_2} (\Sigma_\Phi')_{21}}{m_{\Phi_2}^2 - m_{\Phi_1}^2} \\
			\frac{m_{\Phi_1} m_{\Phi_2} (\Sigma_\Phi')_{12}}{m_{\Phi_2}^2 - m_{\Phi_1}^2} & 1 \end{array} \right).
\end{align}
Accordingly, up to $\mathcal{O} (\alpha)$
\begin{empheq}[box=\fbox]{align}
	C_f (p^2) &= M_\Phi^{-1} C^{-1} M_\Phi
		= \left( \begin{array}{cc} 1 & \minus \frac{m_{\Phi_2}^2 (\Sigma_\Phi')_{12}}{m_{\Phi_2}^2 - m_{\Phi_1}^2} \\
			\frac{m_{\Phi_1}^2 (\Sigma_\Phi')_{21}}{m_{\Phi_2}^2 - m_{\Phi_1}^2} & 1 \end{array} \right), \\
	C_i (p^2) &= M_\Phi^{-1} C^\mathsf{T} M_\Phi
		= \left( \begin{array}{cc} 1 & \minus \frac{m_{\Phi_2}^2 (\Sigma_\Phi')_{21}}{m_{\Phi_2}^2 - m_{\Phi_1}^2} \\
			\frac{m_{\Phi_1}^2 (\Sigma_\Phi')_{12}}{m_{\Phi_2}^2 - m_{\Phi_1}^2} & 1 \end{array} \right).
\end{empheq}
Hence, for a large mass difference, we can always write
\begin{align}
	C (p^2),~C_f (p^2),~C_i (p^2) = 1 + \mathcal{O} (\alpha).
\end{align}
Furthermore, we have
\begin{align}
	C^\dag (p^2) C (p^2)
		&= \left( \begin{array}{cc} 1 & \frac{m_{\Phi_1} m_{\Phi_2} [(\Sigma_\Phi')_{12} - (\Sigma_\Phi')_{21}^*]}{m_{\Phi_2}^2 - m_{\Phi_1}^2} \\
			\frac{m_{\Phi_1} m_{\Phi_2} [(\Sigma_\Phi')_{12}^* - (\Sigma_\Phi')_{21}]}{m_{\Phi_2}^2 - m_{\Phi_1}^2} & 1 \end{array} \right), \\
	C_f^\dag (p^2) C_f (p^2)
		&= \left( \begin{array}{cc} 1 & \frac{m_{\Phi_1}^2 (\Sigma_\Phi')_{21}^* - m_{\Phi_2}^2 (\Sigma_\Phi')_{12}}{m_{\Phi_2}^2 - m_{\Phi_1}^2} \\
			\frac{m_{\Phi_1}^2 (\Sigma_\Phi')_{21} - m_{\Phi_2}^2 (\Sigma_\Phi')_{12}^*}{m_{\Phi_2}^2 - m_{\Phi_1}^2} & 1 \end{array} \right), \\
	C_i^\dag (p^2) C_i (p^2)
		&= \left( \begin{array}{cc} 1 & \frac{m_{\Phi_1}^2 (\Sigma_\Phi')_{12}^* - m_{\Phi_2}^2 (\Sigma_\Phi')_{21}}{m_{\Phi_2}^2 - m_{\Phi_1}^2} \\
			\frac{m_{\Phi_1}^2 (\Sigma_\Phi')_{12} - m_{\Phi_2}^2 (\Sigma_\Phi')_{21}^*}{m_{\Phi_2}^2 - m_{\Phi_1}^2} & 1 \end{array} \right).
\end{align}
The mixing matrix $C (p^2)$ is unitary if and only if $\Sigma_\Phi'^\dag (p^2) = \Sigma_\Phi' (p^2)$, \textit{i.e.}, if and only if $\Phi_\alpha$ are stable particles, while the other mixing matrices $C_f (p^2)$ and $C_i (p^2)$ are non-unitary in general. In addition, the effective Yukawa couplings are written up to $\mathcal{O} (\alpha)$ as
\begin{align}
	\widehat{f}_{i \widehat{\alpha}} &= (f C_{\widehat{\Phi}^f})_{i \widehat{\alpha}}
		= |R_{\widehat{\Phi}_{\widehat{\alpha}}}|^\frac{1}{2} \bigg[ f_{i \widehat{\alpha}}
			+ f_{i \widehat{\beta}} \frac{m_{\Phi_{\widehat{\alpha}}}^2 (\Sigma_\Phi')_{\widehat{\beta} \widehat{\alpha}} (m_{\Phi_{\widehat{\alpha}}}^2)}{m_{\Phi_{\widehat{\beta}}}^2 - m_{\Phi_{\widehat{\alpha}}}^2} \bigg] \quad (\widehat{\beta} \neq \widehat{\alpha}), \\
	\widehat{f}_{i \widehat{\alpha}}^c &= (f^* C_{\widehat{\Phi}^i})_{i \widehat{\alpha}}
		= |R_{\widehat{\Phi}_{\widehat{\alpha}}}|^\frac{1}{2} \bigg[ f_{i \widehat{\alpha}}^*
			+ f_{i \widehat{\beta}}^* \frac{m_{\Phi_{\widehat{\alpha}}}^2 (\Sigma_\Phi')_{\widehat{\alpha} \widehat{\beta}} (m_{\Phi_{\widehat{\alpha}}}^2)}{m_{\Phi_{\widehat{\beta}}}^2 - m_{\Phi_{\widehat{\alpha}}}^2} \bigg] \quad (\widehat{\beta} \neq \widehat{\alpha}),
\end{align}
where we have used $(\Sigma_\Phi')_{\widehat{\beta} \widehat{\alpha}} (p_{\widehat{\Phi}_{\widehat{\alpha}}}^2) = (\Sigma_\Phi')_{\widehat{\beta} \widehat{\alpha}} (m_{\Phi_{\widehat{\alpha}}}^2)$ which is valid up to $\mathcal{O} (\alpha)$.

\subsection{Numerical approach to the case of a small mass difference}

\subsubsection{Unstable particles}		\label{sec:ExSmallUnstable}
Now we examine a numerical example of unstable particles with a small mass difference in which the deviations of mixing matrices from unitarity are large. Let us choose tree-level masses and Yukawa couplings
\begin{align}
	m_{\Phi_1} = 1~\text{TeV}, \qquad
	m_{\Phi_2} - m_{\Phi_1} = 10^{-7}~\text{TeV}, \qquad
	f = \left( \begin{array}{cc} 1 & 0.9 \\ 1 & e^{i 0.1 \pi} \end{array} \right) \cdot 10^{-3},
\end{align}
such that
\begin{align}
	\boxed{m_{\Phi_2} - m_{\Phi_1} \sim m_{\Phi_1} \mathcal{O} (\alpha).}
\end{align}
The physical poles $p_{\widehat{\Phi}_{\widehat{\alpha}}}^2$ can be calculated from equation \ref{eq:PDiag}, \textit{i.e.}, $p_{\widehat{\Phi}_{\widehat{\alpha}}}^2 = P^2_{\widehat{\alpha}} (m_{\Phi_{\widehat{\alpha}}}^2)$, where we have used $\Sigma_\Phi (p_{\widehat{\Phi}_{\widehat{\alpha}}}^2) = \Sigma_\Phi (m_{\Phi_{\widehat{\alpha}}}^2)$ which is correct up to $\mathcal{O} (\alpha)$. The results are
\begin{align}
	p_{\widehat{\Phi}_1} = 1 + (0.361396 - 1.99680 \, i) \cdot 10^{-8}~\text{TeV}, \qquad
	p_{\widehat{\Phi}_2} = 1 + (9.63860 - 1.79308 \, i) \cdot 10^{-8}~\text{TeV},
\end{align}
\textit{i.e.},
\begin{alignat}{2}
	&m_{\widehat{\Phi}_1} = 1 + 3.61396 \cdot 10^{-9}~\text{TeV}, \qquad
	&&\Gamma_{\widehat{\Phi}_1} = 3.99360 \cdot 10^{-8}~\text{TeV}, \\
	&m_{\widehat{\Phi}_2} = 1 + 9.63860 \cdot 10^{-8}~\text{TeV}, \qquad
	&&\Gamma_{\widehat{\Phi}_2} = 3.58616 \cdot 10^{-8}~\text{TeV}.
\end{alignat}
Using these poles and equation \ref{eq:SE1LoopUnstable}, we obtain
\begin{align}
	\Sigma'_\Phi (p_{\widehat{\Phi}_1}^2) = \Sigma'_\Phi (p_{\widehat{\Phi}_2}^2) = \left( \begin{array}{cc} 3.97887 \, i & \minus 0.614770 + 3.68256 \, i \\ 0.614770 + 3.68256 \, i & 3.60088 \, i \end{array} \right) \cdot 10^{-8}
\end{align}
The mixing matrices up to $\mathcal{O} (1)$ are found from equations \ref{eq:Cinv}, \ref{eq:CfCi}, and \ref{eq:MixingMat}:
\begin{align}
	C^{-1} (p_{\widehat{\Phi}_1}^2) &= C^{-1} (p_{\widehat{\Phi}_2}^2)
		= \left( \begin{array}{cc} 0.981764 & 0.0274773 - 0.188108 \, i \\ 0.0351156 + 0.186833 \, i & 0.981764 \end{array} \right), \\
	C_{\widehat{\Phi}^f} &= C^{-1} (p_{\widehat{\Phi}_1}^2), \\
	C_{\widehat{\Phi}^i} &= \left( \begin{array}{cc} 1.05822 - 0.00167886 \, i & \minus 0.0381697 - 0.201322 \, i \\ \minus 0.0292953 + 0.202803 \, i & 1.05822 - 0.00167885 \, i \end{array} \right),
\end{align}
which indeed satisfy equation \ref{eq:CiTCf}:
\begin{align}
	C_{\widehat{\Phi}^i}^\mathsf{T} C_{\widehat{\Phi}^f} &= \left( \begin{array}{cc} 2.78287 - 0.00329649 \, i & 1.84276 - 3.79900 \, i \\ 0.508810 - 4.19158 \, i & \minus 0.370157 + 0.00329648 \, i \end{array} \right) \cdot 10^{-8}, \\
	C_{\widehat{\Phi}^i} C_{\widehat{\Phi}^f}^\mathsf{T} &= \left( \begin{array}{cc} 1.26561 - 0.484450 \, i & 0.665784 - 3.89879 \, i \\ 1.89612 - 3.47111 \, i & 1.14710 + 0.484450 \, i \end{array} \right) \cdot 10^{-8},
\end{align}
Note that, for a small mass difference, the mixing matrices up to $\mathcal{O} (\alpha)$ require calculations up to $\mathcal{O} (\alpha^2)$, \textit{i.e.}, two-loop contributions to the self-energy, as mentioned above. These mixing matrices are non-unitary:
\begin{align}
	(C C^\dag)^{-1} (p_{\widehat{\Phi}_1}^2) &= (C C^\dag)^{-1} (p_{\widehat{\Phi}_2}^2)
		= \left( \begin{array}{cc} 1.00000 & 0.0614515 - 0.368103 \, i \\ 0.0614515 + 0.368103 \, i & 1.00000 \end{array} \right), \\
	C_{\widehat{\Phi}^f}^\dag C_{\widehat{\Phi}^f} &= (C C^\dag)^{-1} (p_{\widehat{\Phi}_1}^2), \\
	C_{\widehat{\Phi}^i}^\dag C_{\widehat{\Phi}^i} &= \left( \begin{array}{cc} 1.16181 & \minus 0.0713951 - 0.427667 \, i \\ \minus 0.0713951 + 0.427667 \, i & 1.16181 \end{array} \right),
\end{align}
and their deviations from unitarity are indeed large:
\begin{align}
	\big| (C C^\dag)^{-1} (p_{\widehat{\Phi}_{\widehat{\alpha}}}^2) \big|_{12}, \quad
	\big| (C^\dag_{\widehat{\Phi}^f} C_{\widehat{\Phi}^f})_{12} \big|, \quad
	\big| (C^\dag_{\widehat{\Phi}^i} C_{\widehat{\Phi}^i})_{12} \big| \
	\sim \ \mathcal{O} (10^{-1}) \
	\gg \ \mathcal{O} (\alpha).
\end{align}
We can also calculate the residues from equations \ref{eq:SEDiag}, \ref{eq:P}, and \ref{eq:Res} to obtain
\begin{align}
	R_{\widehat{\Phi}_1} = 1 + (2.00449 - 3.51576 \, i) \cdot 10^{-8}, \qquad
	R_{\widehat{\Phi}_2} = 1 + (4.08224 - 4.06399 \, i) \cdot 10^{-8}.
\end{align}
For a small mass difference, the vertex-loop corrections are negligible so that we can use the approximation $(\widehat{V}_{\widehat{\Phi}^f})_{i \widehat{\alpha}} = \widehat{f}_{i \widehat{\alpha}}$ and $(\widehat{V}_{\widehat{\Phi}^i})_{i \widehat{\alpha}} = \widehat{f}_{i \widehat{\alpha}}^c$. The decay widths and transition rates are given in table \ref{tab:DW&TR}, and they indeed satisfy equation \ref{eq:DWTot}. The transition rate for each quasiparticle is complex-valued and thus unphysical, but their sum is consistent with the total decay widths obtained from the self-energy or complex poles. The collective partial decay widths defined by equation \ref{eq:DWPartial} are also given in the table. Moreover, the popular method of calculating the decay widths using equation \ref{eq:TotDWWrong} indeed gives a wrong result as shown in the table, since $\sum_{\widehat{\alpha}} \Gamma_{\widehat{\Phi}_{\widehat{\alpha}} \to \chi_i \xi^c} \neq \sum_{\widehat{\alpha}} \Gamma^\text{wrong}_{\widehat{\Phi}_{\widehat{\alpha}} \to \chi_i \xi^c}$ although $\Gamma_{\widehat{\Phi}_{\widehat{\alpha}}} = \sum_i \Gamma^\text{wrong}_{\widehat{\Phi}_{\widehat{\alpha}} \to \chi_i \xi^c}$.
\begin{table}[h]
	\center
	\begin{tabular}{|c|c|c|c|}
		\hline
		\makecell{Decay widths or \\ transition rates} & Value ($10^{-8}$ TeV) &
			\makecell{Decay widths or \\ transition rates} & Value ($10^{-8}$ TeV) \\ \hline
		$\Gamma_{\widehat{\Phi}_1}$ & $3.99360$ &
			$\Gamma_{\widehat{\Phi}_2}$ & $3.58616$ \\ \hline
		$\sum_{i, j} \Gamma^\text{scat}_{\chi_j \xi^c \to \widehat{\Phi}_1 \to \chi_i \xi^c}$ & $4.02656 + 1.50117 \, i$ &
			$\sum_{i, j} \Gamma^\text{scat}_{\chi_j \xi^c \to \widehat{\Phi}_2 \to \chi_i \xi^c}$ & $3.55319 - 1.50117 \, i$ \\ \hline
		$\sum_j \Gamma^\text{scat}_{\chi_j \xi^c \to \widehat{\Phi}_1 \to \chi_1 \xi^c}$ & $2.01975 + 0.709765 \, i$ &
			$\sum_j \Gamma^\text{scat}_{\chi_j \xi^c \to \widehat{\Phi}_2 \to \chi_1 \xi^c}$ & $1.58113 - 0.709765 \, i$ \\ \hline
		$\sum_j \Gamma^\text{scat}_{\chi_j \xi^c \to \widehat{\Phi}_1 \to \chi_2 \xi^c}$ & $2.00682 + 0.791407 \, i$ &
			$\sum_j \Gamma^\text{scat}_{\chi_j \xi^c \to \widehat{\Phi}_2 \to \chi_2 \xi^c}$ & $1.97206 - 0.791407 \, i$ \\ \hline
		$\sum_{\widehat{\alpha}} \Gamma_{\widehat{\Phi}_{\widehat{\alpha}} \to \chi_1 \xi^c}$ & $3.60088$ &
			$\sum_{\widehat{\alpha}} \Gamma_{\widehat{\Phi}_{\widehat{\alpha}} \to \chi_2 \xi^c}$ & $3.97887$ \\ \hline
		$\Gamma^\text{wrong}_{\widehat{\Phi}_1 \to \chi_1 \xi^c}$ & $2.09923$ &
			$\Gamma^\text{wrong}_{\widehat{\Phi}_2 \to \chi_1 \xi^c}$ & $1.72171$ \\ \hline
		$\Gamma^\text{wrong}_{\widehat{\Phi}_1 \to \chi_2 \xi^c}$ & $1.89437$ &
			$\Gamma^\text{wrong}_{\widehat{\Phi}_2 \to \chi_2 \xi^c}$ & $1.86445$ \\ \hline
		$\sum_{\widehat{\alpha}} \Gamma^\text{wrong}_{\widehat{\Phi}_{\widehat{\alpha}} \to \chi_1 \xi^c}$ & $3.82094$ &
			$\sum_{\widehat{\alpha}} \Gamma^\text{wrong}_{\widehat{\Phi}_{\widehat{\alpha}} \to \chi_2 \xi^c}$ & $3.75882$ \\ \hline
	\end{tabular}
	\caption{Decay widths and transition rates.}
	\label{tab:DW&TR}
\end{table}

In calculating the mixing matrices, it is important that no discontinuity exists in any relevant complex-valued functions between $p_{\widehat{\Phi}_1}^2$ and $p_{\widehat{\Phi}_2}^2$, since otherwise some erroneous results usually come out. To avoid such a subtlety, for example, it is safe to put the branch cut of the square root in equations \ref{eq:lambda1} and \ref{eq:lambda2} well outside the region where the poles exist. If we restrict to the case of a small mass difference, a simpler way is calculating them only for $p_{\widehat{\Phi}_1}^2$ and using $C (p_{\widehat{\Phi}_1}^2) = C (p_{\widehat{\Phi}_2}^2)$ which is correct up to $\mathcal{O} (\alpha)$, instead of separately calculating mixing matrices for $p_{\widehat{\Phi}_2}^2$ as well.

\subsubsection{Stable particles}		\label{sec:ExSmallStable}
As we have discussed, the mixing matrices for stable particles are always close to unitary matrices. To show that, let us choose the same parameters except for $m_\xi$ such that $\widehat{\Phi}_{\widehat{\alpha}}$ are stable:
\begin{align}
	m_{\Phi_1} = 1~\text{TeV}, \quad
	m_{\Phi_2} - m_{\Phi_1} = 10^{-7}~\text{TeV}, \quad
	m_\xi = 2~\text{TeV}, \quad
	f = \left( \begin{array}{cc} 1 & 0.9 \\ 1 & e^{i 0.1 \pi} \end{array} \right) \cdot 10^{-3}.
\end{align}
The pole masses $m_{\widehat{\Phi}_{\widehat{\alpha}}}^2$ up to $\mathcal{O} (\alpha)$ are again found from equation \ref{eq:PDiag}, \textit{i.e.}, $m_{\widehat{\Phi}_{\widehat{\alpha}}}^2 = P^2_{\widehat{\alpha}} (m_{\Phi_{\widehat{\alpha}}}^2)$:
\begin{align}
	m_{\widehat{\Phi}_1} = 1 + 4.01167 \cdot 10^{-8}~\text{TeV}, \qquad
	m_{\widehat{\Phi}_2} = 1 + 2.36877 \cdot 10^{-7}~\text{TeV}.
\end{align}
For stable particles, the renormalized self-energy is given by equation \ref{eq:SE1LoopStable}, and we find
\begin{align}
	\Sigma_\Phi' (m_{\widehat{\Phi}_1}^2) = \Sigma_\Phi' (m_{\widehat{\Phi}_2}^2)
		= \minus \left( \begin{array}{cc} 1.85820 & 1.71982 + 0.287108 \, i \\ 1.71982 - 0.287108 \, i & 1.68167 \end{array} \right) \cdot 10^{-7}.
\end{align}
The mixing matrices up to $\mathcal{O} (1)$ are again calculated from equations \ref{eq:Cinv}, \ref{eq:CfCi}, and \ref{eq:MixingMat}:
\begin{align}
	C^{-1} (m_{\widehat{\Phi}_1}^2) &= C^{-1} (m_{\widehat{\Phi}_2}^2)
		= \left( \begin{array}{cc} 0.855387 & 0.510919 + 0.0852933 \, i \\ \minus 0.510919 + 0.0852933 \, i & 0.855387 \end{array} \right), \\
	C_{\widehat{\Phi}} &= C^{-1} (m_{\widehat{\Phi}_1}^2),
\end{align}
where $C^{-1} (m_{\widehat{\Phi}_{\widehat{\alpha}}}^2)$ are unitary while $C_{\widehat{\Phi}}$ are non-unitary:
\begin{align}
	(C C^\dag)^{-1} (m_{\widehat{\Phi}_1}^2) - 1 &= (C C^\dag)^{-1} (m_{\widehat{\Phi}_2}^2) - 1 = 0, \\
	C_{\widehat{\Phi}}^\dag C_{\widehat{\Phi}} - 1 &= \minus \left( \begin{array}{cc} 0.530927 & 1.74654 + 0.291569 \, i \\ 1.74654 - 0.291569 \, i & 6.54236 \end{array} \right) \cdot 10^{-7}.
\end{align}
Even though the input parameters are almost identical to those of unstable particles in the example given above, the mixing matrix $C_{\widehat{\Phi}}$ for stable particles in this example is indeed almost unitary. The residues are calculated also from equations \ref{eq:SEDiag}, \ref{eq:P}, and \ref{eq:Res}, and they are found to be
\begin{align}
	R_{\widehat{\Phi}_1} = 1 + 5.69963 \cdot 10^{-10}, \qquad
	R_{\widehat{\Phi}_2} = 1 - 7.07898 \cdot 10^{-7}.
\end{align}

\section{Conclusion}
We have discussed the flavor mixing of unstable particles, focusing on renormalization of a theory and diagonalization of a dressed propagator. In the presence of mixing of multiple flavors, each of physical unstable particles cannot be related to a single renormalized field, and should be interpreted as a quasiparticle. It cannot be regarded as an external state of a physical process, and is not separately observable. Accordingly, several popular beliefs on renormalization have been disproved. In particular, it has been shown that the on-shell or complex-mass renormalization schemes, \textit{i.e.}, the physical renormalization schemes, cannot be applied to a theory of unstable particles with flavor mixing. Especially when the mass differences between flavors are small, there exists a non-perturbative effect which enhances the collective loop effects to the field-strength. In consequence, the fields of physical unstable particles must be much different from the canonically quantized fields, in order to keep the counterterms in the perturbative regime where a simple order-by-order renormalization can be applied, or equivalently in order not to lose any precision of perturbative calculations because of unnecessarily large interaction terms. We have also discussed how to study the properties of physical unstable particles from scattering mediated by their fields. In particular, the decay widths of unstable particles have been derived from scattering, and we have shown that a popular way of calculating decay widths by amputating the external field of an unstable particle and using effective vertices is wrong.

\section*{Acknowledgement}
This work was supported by the National Center for Theoretical Sciences, Hsinchu.

\appendix

\section{Derivation of the transition rate}		\label{sec:TransRate}
Here, we derive equation \ref{eq:TRScatExp}. The scattering cross section of $\chi_j \xi^c \to \chi_i \xi^c$ in the CM frame is
\begin{align}
	\sigma_\text{CM} (\chi_j \xi^c \to \chi_i \xi^c)
	= \frac{1}{16 \pi} \sum_{\widehat{\alpha}, \widehat{\beta}} \widehat{V}_{i \widehat{\beta}}^* \widehat{V}_{j \widehat{\beta}}^{c *} \widehat{V}_{i \widehat{\alpha}} \widehat{V}_{j \widehat{\alpha}}^c
		\frac{E^2}{(E^2 - p_{\widehat{\Phi}_{\widehat{\beta}}}^{* 2}) (E^2 - p_{\widehat{\Phi}_{\widehat{\alpha}}}^2)},
\end{align}
where $E$ is the total energy, and thus
\begin{align}
	\sigma'_\text{CM} (\chi_j \xi^c \to \chi_i \xi^c) &= 2 E_{\chi_j} \, 2 E_{\xi^c} \big| \mathbf{v}_{\chi_j} - \mathbf{v}_{\xi^c} \big| \, 4 \, \sigma_\text{CM} (\chi_j \xi^c \to \chi_i \xi^c) \nonumber \\
	&= \frac{1}{2 \pi} \sum_{\widehat{\alpha}, \widehat{\beta}} \widehat{V}_{i \widehat{\beta}}^* \widehat{V}_{j \widehat{\beta}}^{c *} \widehat{V}_{i \widehat{\alpha}} \widehat{V}_{j \widehat{\alpha}}^c
		\frac{E^4}{(E^2 - p_{\widehat{\Phi}_{\widehat{\beta}}}^{* 2}) (E^2 - p_{\widehat{\Phi}_{\widehat{\alpha}}}^2)}.
\end{align}
Hence,
\begin{align}
	&\sum_{j, \widehat{\alpha}} \Gamma^\text{scat}_{\chi_j \xi^c \to \widehat{\Phi}_{\widehat{\alpha}} \to \chi_i \xi^c}
	= \frac{1}{4} \sum_j \int d\Pi_{\chi_j}' \int d\Pi_{\xi^c}' \, (2 \pi)^3 \delta^3 (\mathbf{p}_{\chi_j}' + \mathbf{p}_{\xi^c}') \,
			\sigma'_\text{CM} [\chi_j (p_{\chi_j}') \xi^c (p_{\xi^c}') \to \chi_i \xi^c] \bigg|_\text{OS} \\ 
	&\quad = \frac{1}{8 \pi} \sum_{j, \widehat{\alpha}, \widehat{\beta}} \widehat{V}_{i \widehat{\beta}}^* \widehat{V}_{j \widehat{\beta}}^{c *} \widehat{V}_{i \widehat{\alpha}} \widehat{V}_{j \widehat{\alpha}}^c
		\int d\Pi_{\chi_j}' \int d\Pi_{\xi^c}' \, (2 \pi)^3 \delta^3 (\mathbf{p}_{\chi_j}' + \mathbf{p}_{\xi^c}') \,
		\frac{E^4}{(E^2 - p_{\widehat{\Phi}_{\widehat{\beta}}}^{* 2}) (E^2 - p_{\widehat{\Phi}_{\widehat{\alpha}}}^2)} \Bigg|_\text{OS},
\end{align}
where
\begin{align}
	&\int d\Pi_{\chi_j} \int d\Pi_{\xi^c} \, (2 \pi)^3 \delta^3 (\mathbf{p}_{\chi_j} + \mathbf{p}_{\xi^c}) \,
		\frac{E^4}{(E^2 - p_{\widehat{\Phi}_{\widehat{\beta}}}^{* 2}) (E^2 - p_{\widehat{\Phi}_{\widehat{\alpha}}}^2)} \nonumber \\
	&= \minus \frac{1}{16 \pi^2 (p_{\widehat{\Phi}_{\widehat{\beta}}}^{* 2} - p_{\widehat{\Phi}_{\widehat{\alpha}}}^2)} \int_0^\infty dE \,
		E^4 \bigg[ \frac{E^2 - m_{\widehat{\Phi}_{\widehat{\alpha}}}^2 - i m_{\widehat{\Phi}_{\widehat{\alpha}}} \Gamma_{\widehat{\Phi}_{\widehat{\alpha}}}}{(E^2 - m_{\widehat{\Phi}_{\widehat{\alpha}}}^2)^2 + (m_{\widehat{\Phi}_{\widehat{\alpha}}} \Gamma_{\widehat{\Phi}_{\widehat{\alpha}}})^2} - \frac{E^2 - m_{\widehat{\Phi}_{\widehat{\beta}}}^2 + i m_{\widehat{\Phi}_{\widehat{\beta}}} \Gamma_{\widehat{\Phi}_{\widehat{\beta}}}}{(E^2 - m_{\widehat{\Phi}_{\widehat{\beta}}}^2)^2 + (m_{\widehat{\Phi}_{\widehat{\beta}}} \Gamma_{\widehat{\Phi}_{\widehat{\beta}}})^2} \bigg].
	\label{eq:IntPhs}
\end{align}
Using the narrow-width approximation
\begin{align}
	\lim_{\frac{\Gamma_{\widehat{\Phi}_{\widehat{\alpha}}}}{m_{\widehat{\Phi}_{\widehat{\alpha}}}} \to 0} \frac{1}{(E^2 - m_{\widehat{\Phi}_{\widehat{\alpha}}}^2)^2 + (m_{\widehat{\Phi}_{\widehat{\alpha}}} \Gamma_{\widehat{\Phi}_{\widehat{\alpha}}})^2} = \frac{\pi}{m_{\widehat{\Phi}_{\widehat{\alpha}}} \Gamma_{\widehat{\Phi}_{\widehat{\alpha}}}} \delta(E^2 - m_{\widehat{\Phi}_{\widehat{\alpha}}}^2),
\end{align}
we obtain
\begin{align}
	\int d\Pi_{\chi_j} &\int d\Pi_{\xi^c} \, (2 \pi)^3 \delta^3 (\mathbf{p}_{\chi_j} + \mathbf{p}_{\xi^c}) \,
		\frac{E^4}{(E^2 - p_{\widehat{\Phi}_{\widehat{\beta}}}^{* 2}) (E^2 - p_{\widehat{\Phi}_{\widehat{\alpha}}}^2)} \Bigg|_\text{OS} \nonumber \\
	&= \frac{i}{32 \pi (p_{\widehat{\Phi}_{\widehat{\beta}}}^{* 2} - p_{\widehat{\Phi}_{\widehat{\alpha}}}^2)} \int_0^\infty dE^2 \,
		E^3 \big[ \delta(E^2 - m_{\widehat{\Phi}_{\widehat{\alpha}}}^2) + \delta(E^2 - m_{\widehat{\Phi}_{\widehat{\beta}}}^2) \big] \nonumber \\
	&= \frac{i (m_{\widehat{\Phi}_{\widehat{\beta}}}^3 + m_{\widehat{\Phi}_{\widehat{\alpha}}}^3)}{32 \pi (p_{\widehat{\Phi}_{\widehat{\beta}}}^{* 2} - p_{\widehat{\Phi}_{\widehat{\alpha}}}^2)}.
\end{align}
Note that equation \ref{eq:IntPhs} is divergent as $E \to \infty$. To consider only the on-shell contribution and justify the narrow-width approximation, we should have replaced $E$ in the numerators with $m_{\widehat{\Phi}_{\widehat{\alpha}}}$ or $m_{\widehat{\Phi}_{\widehat{\beta}}}$ before applying the approximation, which was called the on-shell prescription in reference \cite{MixingQFT}. Note also that the narrow-width approximation does not mean that the every propagator is treated as no more than a delta function, in which case there cannot exist any interference between physical particles. Through the on-shell prescription, the off-shell contribution of one propagator changed into the coefficient of the other on-shell propagator which is approximated by a delta function. We finally obtain
\begin{align}
	\boxed{\sum_{j, \widehat{\alpha}} \Gamma^\text{scat}_{\chi_j \xi^c \to \widehat{\Phi}_{\widehat{\alpha}} \to \chi_i \xi^c}
	= \frac{i}{2^8 \pi^2} \sum_{j, \widehat{\alpha}, \widehat{\beta}} \widehat{V}_{i \widehat{\beta}}^* \widehat{V}_{j \widehat{\beta}}^{c *} \widehat{V}_{i \widehat{\alpha}} \widehat{V}_{j \widehat{\alpha}}^c
		\frac{m_{\widehat{\Phi}_{\widehat{\beta}}}^3 + m_{\widehat{\Phi}_{\widehat{\alpha}}}^3}{p_{\widehat{\Phi}_{\widehat{\beta}}}^{* 2} - p_{\widehat{\Phi}_{\widehat{\alpha}}}^2}.}
\end{align}

\section{Derivations of renormalization conditions and counterterms}

\subsection{On-shell renormalization scheme}		\label{sec:OSRen}
In this section, let us derive the on-shell renormalization conditions and counterterms for stable $\Phi_\alpha$. We will follow the steps discussed in reference \cite{OSFlavMixing}. First defining
\begin{align}
	\varepsilon_\alpha \coloneqq p^2 - m_{\Phi_\alpha}^2
\end{align}
where $m_{\Phi_\alpha}$ is the physical mass of $\Phi_\alpha$, we write the Laurent expansion of the dressed propagator around $\varepsilon_\alpha =0$ as
\begin{align}
	(\Delta_\Phi)_{\beta \gamma} (\varepsilon_\alpha) \big|_{\varepsilon_\alpha \approx 0}
	= \frac{\delta_{\beta \alpha} \delta_{\alpha \gamma}}{\varepsilon_\alpha} + (\Delta_\Phi^{(0)})_{\beta \gamma} (\varepsilon_\alpha) + \mathcal{O} (\varepsilon_\alpha).
\end{align}
Let us denote the inverse propagator by $\minus i A (\varepsilon_\alpha)$ which satisfies
\begin{align}
	\delta_{\beta \gamma} = \sum_\delta (\Delta_\Phi)_{\beta \delta} A_{\delta \gamma}
	 = \sum_\delta A_{\beta \delta} (\Delta_\Phi)_{\delta \gamma},
\end{align}
and expand it around $\varepsilon_\alpha = 0$ as
\begin{align}
	A_{\beta \gamma} (\varepsilon_\alpha) \big|_{\varepsilon_\alpha \approx 0} = A_{\beta \gamma}^{\alpha, (0)} + \varepsilon_\alpha A_{\beta \gamma}^{\alpha, (1)} + \mathcal{O} (\varepsilon_\alpha^2).
\end{align}
Then,
\begin{align}
	\delta_{\beta \gamma} &= \sum_\delta (\Delta_\Phi)_{\beta \delta} A_{\delta \gamma} \big|_{\varepsilon_\alpha \approx 0}
		= \frac{1}{\varepsilon_\alpha} \delta_{\beta \alpha} A_{\alpha \gamma}^{\alpha, (0)} + \delta_{\beta \alpha} A_{\alpha \gamma}^{\alpha, (1)} + \sum_\delta (\Delta_\Phi^{(0)})_{\beta \delta} A_{\delta \gamma}^{\alpha, (0)} + \mathcal{O} (\varepsilon_\alpha) \\
	 &= \sum_\delta A_{\beta \delta} (\Delta_\Phi)_{\delta \gamma} \big|_{\varepsilon_\alpha \approx 0}
		= \frac{1}{\varepsilon_\alpha} A_{\beta \alpha}^{\alpha, (0)} \delta_{\alpha \gamma} + A_{\beta \alpha}^{\alpha, (1)} \delta_{\alpha \gamma} + \sum_\delta A_{\beta \delta}^{\alpha, (0)} (\Delta_\Phi^{(0)})_{\delta \gamma} + \mathcal{O} (\varepsilon_\alpha),
\end{align}
To avoid the singularity in the limit $\varepsilon_\alpha \to 0$, we must have
\begin{align}
	A_{\beta \alpha}^{\alpha, (0)} = A_{\alpha \gamma}^{\alpha, (0)} = 0,
	\label{eq:OSRenCond1}
\end{align}
which implies
\begin{align}
	\delta_{\beta \gamma} = \delta_{\beta \alpha} A_{\alpha \gamma}^{\alpha, (1)} + \sum_{\delta \neq \alpha} (\Delta_\Phi^{(0)})_{\beta \delta} A_{\delta \gamma}^{\alpha, (0)}
	= A_{\beta \alpha}^{\alpha, (1)} \delta_{\alpha \gamma} + \sum_{\delta \neq \alpha} A_{\beta \delta}^{\alpha, (0)} (\Delta_\Phi^{(0)})_{\delta \gamma}.
\end{align}
For $\beta = \gamma = \alpha$, we find
\begin{align}
	A_{\alpha \alpha}^{\alpha, (1)} = 1.
	\label{eq:OSRenCond2}
\end{align}
Equations \ref{eq:OSRenCond1} and \ref{eq:OSRenCond2} can be rewritten as the on-shell renormalization conditions on the inverse renormalized propagator:
\begin{align}
	A_{\beta \alpha} (0) = A_{\alpha \beta} (0) = 0, \qquad
	\frac{dA_{\alpha \alpha}}{dp^2} (0) = 1.
\end{align}
These conditions can also be rewritten in terms of the self-energy as
\begin{align}
	\boxed{(\Sigma_\Phi)_{\beta \alpha} (m_{\Phi_\alpha}^2) = (\Sigma_\Phi)_{\alpha \beta} (m_{\Phi_\alpha}^2) = 0, \qquad
	\frac{d(\Sigma_\Phi)_{\alpha \alpha}}{dp^2} (m_{\Phi_\alpha}^2) = 0.}
	\label{eq:OSRenCond}
\end{align}
Now we derive the counterterms corresponding to these on-shell conditions. Since the renormalized self-energy is given by equation \ref{eq:SERenwCT}, the first condition in equation \ref{eq:OSRenCond} implies
\begin{align}
	(\Sigma_\Phi)_{\beta \alpha} (m_{\Phi_\alpha}^2) &= m_{\Phi_\alpha}^2 \big[ (\Sigma_{0 \Phi}')_{\beta \alpha} (m_{\Phi_\alpha}^2) + (\delta_\Phi^H)_{\beta \alpha} \big] + (\delta \Sigma_\Phi)_{\beta \alpha} - (\delta M_\Phi^2)_{\beta \alpha} = 0,
		\label{eq:OSRenCondCT} \\
	(\Sigma_\Phi)_{\alpha \beta} (m_{\Phi_\alpha}^2) &= m_{\Phi_\alpha}^2 \big[ (\Sigma_{0 \Phi}')_{\beta \alpha}^* (m_{\Phi_\alpha}^2) + (\delta_\Phi^H)_{\beta \alpha}^* \big] + (\delta \Sigma_\Phi)_{\beta \alpha}^* - (\delta M_\Phi^2)_{\beta \alpha}^* = 0,
\end{align}
where we have used the fact that $\delta_\Phi^H$, $\delta M_\Phi^2$, $\Sigma_{0 \Phi}'$, and $\delta \Sigma_\Phi$ are Hermitian. Solving these coupled equations for the off-diagonal components of $\delta_\Phi^H$ and $\delta M_\Phi^2$, we obtain
\begin{align}
	(\delta_\Phi^H)_{\beta \alpha} &= \minus \frac{m_{\Phi_\beta}^2 (\Sigma_{0 \Phi}')_{\beta \alpha} (m_{\Phi_\beta}^2) - m_{\Phi_\alpha}^2 (\Sigma_{0 \Phi}')_{\beta \alpha} (m_{\Phi_\alpha}^2)}{m_{\Phi_\beta}^2 - m_{\Phi_\alpha}^2} \quad (\beta \neq \alpha),
		\label{eq:OSRenCT1H} \\
	(\delta M_\Phi^2)_{\beta \alpha} &= (\delta \Sigma_\Phi)_{\beta \alpha} - \frac{m_{\Phi_\beta}^2 m_{\Phi_\alpha}^2 [(\Sigma_{0 \Phi}')_{\beta \alpha} (m_{\Phi_\beta}^2) - (\Sigma_{0 \Phi}')_{\beta \alpha} (m_{\Phi_\alpha}^2)]}{m_{\Phi_\beta}^2 - m_{\Phi_\alpha}^2} \quad (\beta \neq \alpha).
\end{align}
Moreover, taking the derivative of equation \ref{eq:SERenwCT} with repect to $p^2$ and using the first condition of equation \ref{eq:OSRenCond} for $\beta = \alpha$, we can write the second condition of equation \ref{eq:OSRenCond} as
\begin{align}
	(\delta_\Phi^H)_{\alpha \alpha} = \minus \frac{d(\Sigma_{0 \Phi})_{\alpha \alpha}}{dp^2} (m_{\Phi_\alpha}^2).
\end{align}
This and equation \ref{eq:OSRenCondCT} in turn imply
\begin{align}
	(\delta M_\Phi^2)_{\alpha \alpha} = (\Sigma_{0 \Phi})_{\alpha \alpha} (m_{\Phi_\alpha}^2) - m_{\Phi_\alpha}^2 \frac{d(\Sigma_{0 \Phi})_{\alpha \alpha}}{dp^2} (m_{\Phi_\alpha}^2).
\end{align}
Hence, we may choose the counterterms as
\begin{empheq}[box=\fbox]{align}
	(\delta_\Phi)_{\beta \alpha} &= \frac{(\Sigma_{0 \Phi})_{\beta \alpha} (m_{\Phi_\alpha}^2)}{m_{\Phi_\beta}^2 - m_{\Phi_\alpha}^2} \quad (\beta \neq \alpha), \qquad
	(\delta_\Phi)_{\alpha \alpha} = \minus \frac{d(\Sigma_{0 \Phi})_{\alpha \alpha}}{dp^2} (m_{\Phi_\alpha}^2), \\
	(\delta M_\Phi^2)_{\beta \alpha} &= (\delta \Sigma_\Phi)_{\beta \alpha} - \frac{m_{\Phi_\beta}^2 m_{\Phi_\alpha}^2 [(\Sigma_{0 \Phi}')_{\beta \alpha} (m_{\Phi_\beta}^2) - (\Sigma_{0 \Phi}')_{\beta \alpha} (m_{\Phi_\alpha}^2)]}{m_{\Phi_\beta}^2 - m_{\Phi_\alpha}^2} \quad (\beta \neq \alpha), \\
	(\delta M_\Phi^2)_{\alpha \alpha} &= (\Sigma_{0 \Phi})_{\alpha \alpha} (m_{\Phi_\alpha}^2) - m_{\Phi_\alpha}^2 \frac{d(\Sigma_{0 \Phi})_{\alpha \alpha}}{dp^2} (m_{\Phi_\alpha}^2).
\end{empheq}
The Hermitian part of $\delta_\Phi$ is indeed identical to equation \ref{eq:OSRenCT1H}. Note that $\delta_\Phi$ cannot be uniquely determined since its skew-Hermitian part has no role in renormalization, while $\delta M_\Phi^2$ is uniquely determined by the renormalization conditions. These are the counterterms consistent with the expressions derived for real scalar fields in reference \cite{OSFlavMixing}. Note that the sign of $\Sigma (p^2)$ is oppositely defined in this paper.

\subsection{Complex-mass renormalization scheme}		\label{sec:CMSen}
First let us introduce the complex-valued squared mass $p_{\Phi_\alpha}^2 = m_{\Phi_\alpha}^2 - i m_{\Phi_\alpha} \Gamma_{\Phi_\alpha}$ and compensating interaction terms as follows:
\begin{align}
	\mathcal{L}_\text{mass} = \minus \sum_\alpha m_{\Phi_\alpha}^2 \Phi_\alpha^\dag \Phi_\alpha
	= \minus \sum_\alpha p_{\Phi_\alpha}^2 \Phi_\alpha^\dag \Phi_\alpha - \sum_\alpha i m_{\Phi_\alpha} \Gamma_{\Phi_\alpha} \Phi_\alpha^\dag \Phi_\alpha.
\end{align}
The self-energy in the complex-mass scheme is written as
\begin{align}
	(\Sigma_{0 \Phi}^\text{CMS})_{\beta \alpha} (p^2) = (\Sigma_{0 \Phi})_{\beta \alpha} (p^2) + i m_{\Phi_\alpha} \Gamma_{\Phi_\alpha} \delta_{\beta \alpha},
\end{align}
where $\Sigma_{0 \Phi} (p^2)$ is the self-energy calculated with real-valued squared mass $m_{\Phi_\alpha}^2$. In this scheme, we use the expansion parameter
\begin{align}
	\varepsilon_\alpha \coloneqq p^2 - p_{\Phi_\alpha}^2
\end{align}
to find the conditions that satisfy
\begin{align}
	\Delta_{\beta \gamma} (\varepsilon_\alpha ) \big|_{\varepsilon_\alpha \approx 0}
	= \frac{\delta_{\beta \alpha} \delta_{\alpha \gamma}}{\varepsilon_\alpha} + \Delta_{\beta \gamma}^{(0)} (\varepsilon_\alpha ) + \mathcal{O} (\varepsilon_\alpha).
\end{align}
Expanding the inverse propagator $\minus i A (\varepsilon_\alpha )$ around $\varepsilon_\alpha = 0$:
\begin{align}
	A_{\beta \gamma} (\varepsilon_\alpha) = A_{\beta \gamma}^{\alpha, (0)} + \varepsilon_\alpha A_{\beta \gamma}^{\alpha, (1)} + \mathcal{O} (\varepsilon_\alpha^2),
\end{align}
and using
\begin{align}
	\delta_{\beta \gamma} = \sum_\delta \Delta_{\beta \delta} A_{\delta \gamma}
	 = \sum_\delta A_{\beta \delta} \Delta_{\delta \gamma},
\end{align}
we again obtain the renormalization conditions on the inverse propagator:
\begin{align}
	A_{\beta \alpha} (0) = A_{\alpha \beta} (0) = 0, \qquad
	\frac{dA_{\alpha \alpha}}{dp^2} (0) = 1,
\end{align}
which are equivalent to the conditions on the self-energy as follows:
\begin{align}
	\boxed{(\Sigma_\Phi^\text{CMS})_{\beta \alpha} (p_{\Phi_\alpha}^2) = (\Sigma_\Phi^\text{CMS})_{\alpha \beta} (p_{\Phi_\alpha}^2) = 0, \qquad
	\frac{d(\Sigma_\Phi^\text{CMS})_{\alpha \alpha}}{dp^2} (p_{\Phi_\alpha}^2) = 0.}
\end{align}
The renormalized self-energy up to $\mathcal{O} (\alpha)$ is written as
\begin{align}
	\Sigma_\Phi^\text{CMS} (p^2) = p^2 \big[ \Sigma_{0 \Phi}'^{\text{CMS}} (p^2) + \delta_\Phi^H \big] + \delta \Sigma_\Phi^\text{CMS} - \delta M_\Phi^2,
\end{align}
where $\delta \Sigma_\Phi^\text{CMS} = 0$ for $m_\xi = 0$. Since $p_{\Phi_\alpha}^2 = m_{\Phi_\alpha}^2 [1 + \mathcal{O} (\alpha)]$ and $\Sigma_\Phi^\text{CMS} (p_{\Phi_\alpha}^2) = \Sigma_\Phi^\text{CMS} (m_{\Phi_\alpha}^2)$ up to $\mathcal{O} (\alpha)$, we can write
\begin{align}
	(\Sigma_\Phi^\text{CMS})^\text{disp}_{\beta \alpha} (p_{\Phi_\alpha}^2) &= m_{\Phi_\alpha}^2 \big[ (\Sigma_{0 \Phi}'^{\text{CMS}})^\text{disp}_{\beta \alpha} (m_{\Phi_\alpha}^2) + (\delta_\Phi^H)_{\beta \alpha} \big] + (\delta \Sigma_\Phi^\text{CMS})_{\beta \alpha} - (\delta M_\Phi^2)_{\beta \alpha} = 0, \\
	(\Sigma_\Phi^\text{CMS})^\text{disp}_{\alpha \beta} (p_{\Phi_\alpha}^2) &= m_{\Phi_\alpha}^2\big[ \Sigma_{0 \Phi}'^{\text{CMS}})^{\text{disp} *}_{\beta \alpha} (m_{\Phi_\alpha}^2) + (\delta_\Phi^H)_{\beta \alpha}^* \big] + (\delta \Sigma_\Phi^\text{CMS})_{\beta \alpha}^* - (\delta M_\Phi^2)_{\beta \alpha}^* = 0, \\
	(\Sigma_\Phi^\text{CMS})^\text{abs}_{\beta \alpha} (m_{\Phi_\alpha}^2) &= 0
\end{align}
up to $\mathcal{O} (\alpha)$. Here, $(\Sigma_\Phi^\text{CMS})^\text{disp} (p^2)$ and $(\Sigma_\Phi^\text{CMS})^\text{abs} (p^2)$ are the dispersive and absorptive parts of $\Sigma_\Phi^\text{CMS} (p^2)$, respectively, and $\delta \Sigma_\Phi^\text{CMS}$ belongs to $(\Sigma_\Phi^\text{CMS})^\text{disp} (p^2)$. We have used the fact that, for real-valued $p^2$, the matrices $(\Sigma_\Phi^\text{CMS})^\text{disp} (p^2)$, $\delta_\Phi^H$, and $\delta M_\Phi^2$ are Hermitian while $(\Sigma_\Phi^\text{CMS})^\text{abs} (p^2)$ is skew-Hermitian. Note also that, for an arbitrary matrix $X$, we have $X = X^H + X^S = 0$ if and only if $X^H = X^S = 0$. Following the same steps as in the on-shell scheme, we again find
\begin{align}
	(\delta_\Phi^H)_{\beta \alpha} &= \minus \frac{m_{\Phi_\beta}^2 (\Sigma_{0 \Phi}'^{\text{CMS}})^\text{disp}_{\beta \alpha} (m_{\Phi_\beta}^2) - m_{\Phi_\alpha}^2 (\Sigma_{0 \Phi}'^{\text{CMS}})^\text{disp}_{\beta \alpha} (m_{\Phi_\alpha}^2)}{m_{\Phi_\beta}^2 - m_{\Phi_\alpha}^2} \quad (\beta \neq \alpha), \\
	(\delta_\Phi^H)_{\alpha \alpha} &= \minus \frac{d(\Sigma_{0 \Phi}^\text{CMS})^\text{disp}_{\alpha \alpha}}{dp^2} (m_{\Phi_\alpha}^2),
\end{align}
and
\begin{empheq}[box=\fbox]{align}
	(\delta_\Phi)_{\beta \alpha} &= \frac{(\Sigma_{0 \Phi}^{\text{CMS}})^\text{disp}_{\beta \alpha} (m_{\Phi_\alpha}^2)}{m_{\Phi_\beta}^2 - m_{\Phi_\alpha}^2} \quad (\beta \neq \alpha), \qquad
	(\delta_\Phi)_{\alpha \alpha} = \minus \frac{d(\Sigma_{0 \Phi}^{\text{CMS}})^\text{disp}_{\alpha \alpha}}{dp^2} (m_{\Phi_\alpha}^2), \\
	(\delta M_\Phi^2)_{\beta \alpha} &= (\delta \Sigma_\Phi^\text{CMS})_{\beta \alpha} - \frac{m_{\Phi_\beta}^2 m_{\Phi_\alpha}^2 [(\Sigma_{0 \Phi}'^{\text{CMS}})^\text{disp}_{\beta \alpha} (m_{\Phi_\beta}^2) - (\Sigma_{0 \Phi}'^{\text{CMS}})^\text{disp}_{\beta \alpha} (m_{\Phi_\alpha}^2)]}{m_{\Phi_\beta}^2 - m_{\Phi_\alpha}^2} \quad (\beta \neq \alpha), \\
	(\delta M_\Phi^2)_{\alpha \alpha} &= (\Sigma_{0 \Phi}^\text{CMS})^\text{disp}_{\alpha \alpha} (m_{\Phi_\alpha}^2) - m_{\Phi_\alpha}^2 \frac{d(\Sigma_{0 \Phi}^\text{CMS})^\text{disp}_{\alpha \alpha}}{dp^2} (m_{\Phi_\alpha}^2).
\end{empheq}
As in the on-shell scheme, $\delta_\Phi^S$ can be arbitrarily chosen, while $\delta M_\Phi^2$ is uniquely determined.

\section{Calculations of one-loop diagrams}

\subsection*{Self-energy}		\label{sec:SECal}
For the self-energy of $\Phi_0 (p)$ by the loop of $\chi_{0 i} (p - k) \xi_0^c (k)$, we can write
\begin{align}
	i (\Sigma_{\Phi_0})_{\beta \alpha} (p^2)
	&\coloneqq \minus \frac{1}{2} \mu^{4 - d} \sum_i \int \frac{d^d k}{(2 \pi)^d} [\minus i (f_0)_{i \beta}^*] \ \text{Tr} \bigg[ \frac{i}{\slashed{k} - m_{\xi_0}} \frac{i}{\slashed{k} - \slashed{p}} \bigg] [\minus i (f_0)_{i \alpha}] \nonumber \\
	&= \mu^{4 - d} \frac{1}{2} \sum_i (f_0)_{i \beta}^* (f_0)_{i \alpha} \int \frac{d^d k}{(2 \pi)^d} \frac{\text{Tr} [(\slashed{k} + m_{\xi_0}) (\slashed{p} - \slashed{k})]}{(k^2 - m_{\xi_0}^2) (p - k)^2} \nonumber \\
	&= \mu^{4 - d} \frac{d}{2} \sum_i (f_0)_{i \beta}^* (f_0)_{i \alpha} \int \frac{d^d k}{(2 \pi)^d} \int_0^1 dx \frac{p \cdot k - k^2}{[(1 - x) (k^2 - m_{\xi_0}^2) + x (p - k)^2 ]^2},
\end{align}
where $d = 4 - \epsilon$ is the spacetime dimension and we have used $\text{Tr} [\gamma^\mu \gamma^\nu] = d \eta^{\mu \nu}$. The factor 1/2 is simply the coefficient of $\big[ \overline{\xi} (x_1) \chi_i (x_1) \Phi_\beta^\dag (x_1) \big] \big[ \overline{\chi_i} (x_2) \xi (x_2) \Phi_\alpha (x_2) \big]$ in the Taylor expansion of $e^{i \int dx \, \mathcal{L}_\text{int}}$ that appears in $\langle \Omega | T \{\Phi_\beta (y) \Phi_\alpha^\dag (x)\} | \Omega \rangle$. There exists no symmetric combination of contractions which cancels this factor. Manipulating the Feynman parameter and the momentum integration in the standard way, we can rewrite
\begin{align}
	\int \frac{d^d k}{(2 \pi)^d} \frac{p \cdot k - k^2}{[(1 - x) (k^2 - m_{\xi_0}^2) + x (p - k)^2]^2}
	= \int \frac{d^d \ell}{(2 \pi)^d} \frac{x (1 - x) p^2 - \ell^2}{[\ell^2 - (1 - x) (m_{\xi_0}^2 - x p^2)]^2},
\end{align}
where $\ell = k - x p$. Hence,
\begin{align}
	\mu^{4 - d} \frac{1}{2} &\int \frac{d^d k}{(2 \pi)^d} \frac{\text{Tr} [(\slashed{k} + m_{\xi_0}) (\slashed{p} - \slashed{k})]}{(k^2 - m_{\xi_0}^2) (p - k)^2}
	= \mu^{4 - d} \frac{d}{2} \int_0^1 dx \int \frac{d^d \ell}{(2 \pi)^d} \frac{x (1 - x) p^2 - \ell^2}{[\ell^2 - (1 - x) (m_{\xi_0}^2 - x p^2)]^2} \nonumber \\
	&= i \mu^{4 - d} \frac{d}{2} \bigg\{ p^2 \int_0^1 dx \, x (1 - x) \frac{\Gamma (2 - \frac{d}{2})}{(4 \pi)^{d/2}} [(1 - x) (m_{\xi_0}^2 - x p^2)]^{-(2 - d/2)} \nonumber \\
		&\qquad + \int_0^1 dx \, \frac{d}{2} \frac{\Gamma (1 - \frac{d}{2})}{(4 \pi)^{d/2}} [(1 - x) (m_{\xi_0}^2 - x p^2)]^{-(1 - d/2)} \bigg\} \nonumber \\
	&= i \frac{p^2}{16 \pi^2} \bigg[ \bigg( 1 - \frac{2 m_{\xi_0}^2}{p^2} \bigg) \frac{2}{\epsilon} + F (p^2) \bigg] + \mathcal{O} (\epsilon),
\end{align}
where $\widetilde{\mu}^2 \coloneqq 4 \pi e^{-\gamma_E} \mu^2$ and
\begin{align}
	F (p^2) &\coloneqq \minus \frac{1}{6} - 2 \int_0^1 dx \, (1 - x) \bigg( 3 x - \frac{2 m_{\xi_0}^2}{p^2} \bigg) \bigg[ \log{(1 - x)} + \log{\frac{m_{\xi_0}^2 - x p^2}{\widetilde{\mu}^2}} \bigg] \nonumber \\
	&= \bigg( \frac{3}{2} - \frac{2 m_{\xi_0}^2}{p^2} \bigg)
			- \frac{m_{\xi_0}^2}{p^2} \bigg( 3 - \frac{6 m_{\xi_0}^2}{p^2} + \frac{2 m_{\xi_0}^4}{p^4} \bigg) \log{\frac{m_{\xi_0}^2}{\widetilde{\mu}^2}}
			- \bigg( 1 - \frac{5 m_{\xi_0}^2}{p^2} + \frac{6 m_{\xi_0}^4}{p^4} - \frac{2 m_{\xi_0}^6}{p^6} \bigg) \log{\frac{m_{\xi_0}^2 - p^2}{\widetilde{\mu}^2}} \nonumber \\
	&= \bigg[ \frac{3}{2} - \log{\frac{m_{\xi_0}^2}{\widetilde{\mu}^2}} - \bigg( 1 - \frac{5 m_{\xi_0}^2}{p^2} + \frac{6 m_{\xi_0}^4}{p^4} - \frac{2 m_{\xi_0}^6}{p^6} \bigg) \log{\frac{m_{\xi_0}^2 - p^2}{m_{\xi_0}^2}} \bigg]
		- \frac{2 m_{\xi_0}^2}{p^2} \bigg( 1 - \log{\frac{m_{\xi_0}^2}{\widetilde{\mu}^2}} \bigg).
\end{align}
To obtain the bare self-energy up to the one-loop order, we may write $m_{\xi_0} = m_\xi$ and $f_0 = U_\chi f U_\Phi^\dag$ which are correct up to the leading order. Furthermore, $U_\chi$ can be neglected since $(f U_\Phi^\dag)^\dag (f U_\Phi^\dag) = (U_\chi f U_\Phi^\dag)^\dag (U_\chi f U_\Phi^\dag)$. We therefore obtain
\begin{align}
	(\Sigma_{\Phi_0})_{\beta \alpha} (p^2) &= p^2 \sum_i \frac{(f U_\Phi^\dag)_{i \beta}^* (f U_\Phi^\dag)_{i \alpha}}{16 \pi^2} \bigg[ \frac{2}{\epsilon} + \frac{3}{2} - \log{\frac{m_\xi^2}{\widetilde{\mu}^2}} - \bigg( 1 - \frac{5 m_\xi^2}{p^2} + \frac{6 m_\xi^4}{p^4} - \frac{2 m_\xi^6}{p^6} \bigg) \log{\frac{m_\xi^2 - p^2}{m_\xi^2}} \bigg] \nonumber \\
		&\qquad - \sum_i \frac{(f U_\Phi^\dag)_{i \beta}^* (f U_\Phi^\dag)_{i \alpha}}{8 \pi^2} m_\xi^2 \bigg( \frac{2}{\epsilon} + 1 - \log{\frac{m_\xi^2}{\widetilde{\mu}^2}} \bigg) + \mathcal{O} (\epsilon).
\end{align}
For complex-valued $p^2$, it is straightforward to apply analytic continuation to this expression, which we do not discuss since the effect of the imaginary part of $p^2$ is negligible up to $\mathcal{O} (\alpha)$ for $p^2 \approx p_{\widehat{\Phi}_{\widehat{\alpha}}}^2$. \\

Now we impose renormalization conditions, considering equation \ref{eq:SERen1Loop}:
\begin{align}
	\Sigma_\Phi' (p^2) = (U_\Phi^\dag \Sigma_{\Phi_0}' U_\Phi) (p^2) + \delta_\Phi^H, \qquad
	\delta \Sigma_\Phi = U_\Phi^\dag \delta \Sigma_{\Phi_0} U_\Phi.
\end{align}
When $m_{\Phi_\alpha} \gg m_\xi = m_{\chi_i} = 0$, we have
\begin{align}
	F (p^2) = \frac{3}{2} - \log{\frac{p^2}{\widetilde{\mu}^2}} + i \pi,
\end{align}
where we have used $\log{(\minus 1)} = \minus i \pi$. Choosing the counterterms
\begin{align}
	(\delta_\Phi)_{\beta \alpha} = \sum_i \frac{f_{i \beta}^* f_{i \alpha}}{16 \pi^2} \bigg( \minus \frac{2}{\epsilon} - \frac{3}{2} + \log{\frac{m_{\Phi_\beta} m_{\Phi_\alpha}}{\widetilde{\mu}^2}} \bigg), \qquad
	\delta M_\Phi^2 = 0,
\end{align}
we can write the renormalized self-energy as
\begin{align}
	\boxed{(\Sigma_\Phi)_{\beta \alpha} (p^2) = p^2 (\Sigma_\Phi')_{\beta \alpha} (p^2)
		= p^2 \sum_i \frac{f_{i \beta}^* f_{i \alpha}}{16 \pi^2} \bigg[ \minus \log{\bigg( \frac{p^2}{m_{\Phi_\beta} m_{\Phi_\alpha}} \bigg)} + i \pi \bigg], \quad
	\Sigma_{\Phi^*} (p^2) = \Sigma_\Phi^{\mathsf{T}} (p^2).}
\end{align}

When $m_\xi > m_{\Phi_\alpha} \gg m_{\chi_i} = 0$, we choose the counterterms
\begin{align}
	(\delta_\Phi)_{\beta \alpha} = \sum_i \frac{f_{i \beta}^* f_{i \alpha}}{16 \pi^2} \bigg( \minus \frac{2}{\epsilon} - \frac{3}{2} + \log{\frac{m_\xi^2}{\widetilde{\mu}^2}} \bigg), \quad
	(\delta M_\Phi^2)_{\beta \alpha} = \minus \sum_i \frac{f_{i \beta}^* f_{i \alpha}}{8 \pi^2} m_\xi^2 \bigg( \frac{2}{\epsilon} + 1 - \log{\frac{m_\xi^2}{\widetilde{\mu}^2}} \bigg),
\end{align}
such that
\begin{empheq}[box=\fbox]{align}
	(\Sigma_\Phi)_{\beta \alpha} (p^2) &= p^2 (\Sigma_\Phi')_{\beta \alpha} (p^2)
		= \minus p^2 \sum_i \frac{f_{i \beta}^* f_{i \alpha}}{16 \pi^2} \bigg( 1 - \frac{5 m_\xi^2}{p^2} + \frac{6 m_\xi^4}{p^4} - \frac{2 m_\xi^6}{p^6} \bigg) \log{\frac{m_\xi^2 - p^2}{m_\xi^2}}, \\
	\Sigma_{\Phi^*} (p^2) &= \Sigma_\Phi^{\mathsf{T}} (p^2).
\end{empheq}

\subsection*{Vertex}		\label{sec:VertexCal}
\begin{figure}[t]
	\centering
	\subfloat[$-i (V_{0 \Phi})_{i \alpha} (p^2)$]{
		\includegraphics[width = 40 mm]{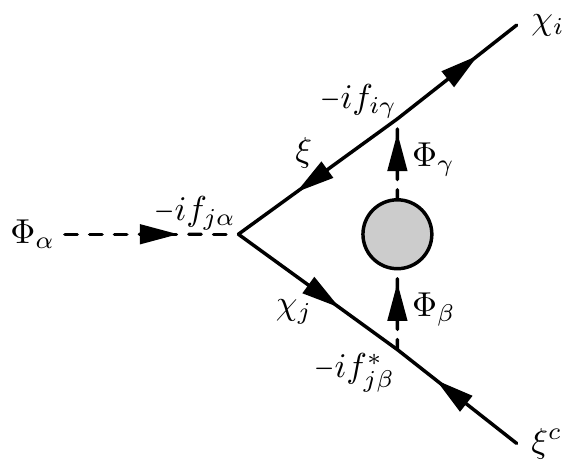}
		\label{fig:VP1Loop}
	} \qquad \quad
	\subfloat[$-i (V_{0 \Phi^*})_{i \alpha} (p^2)$]{
		\includegraphics[width = 40 mm]{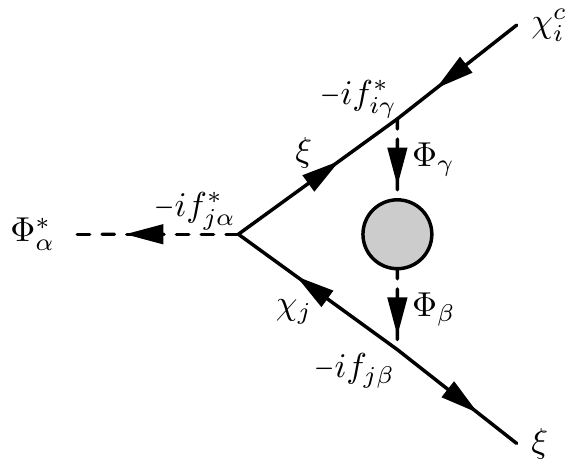}
		\label{fig:VPc1Loop}
	}
	\caption{One-loop contributions to the vertices. The dressed propagators are used for the internal $\Phi$ and $\Phi^*$ since they induce $\mathcal{O} (1)$ corrections to the associated Yukawa couplings in the case of small mass differences.}
	\label{fig:V1Loop}
\end{figure}

Now we calculate the vertex function up to the one-loop order. The associated one-loop diagrams are shown in figure \ref{fig:V1Loop}.
Note that dressed propagators rather than tree-level ones will be used for the internal $\Phi$ and $\Phi^*$, since they induce $\mathcal{O} (1)$ multiplicative corrections to the Yukawa coulings in the case of small mass differences. In fact, when the self-energy is calculated up to one loop as in this paper, the vertex-loop corrections are completely negligible for small mass differences since the effective Yukawa couplings and observable quantities are correct only up to $\mathcal{O} (1)$, as mentioned before, while the vertex-loop corrections induce $\mathcal{O} (\alpha)$ effects. In addition, when the mass differences are large, the tree-level propagators are good enough for the internal $\Phi$ and $\Phi^*$ up to $\mathcal{O} (\alpha)$. Hence, regardless of mass differences, the calculations using dressed propagators are actually needed only when the corrections to the self-energy beyond one loop are considered as well.

The non-renormalized vertex function for $\Phi_\alpha (p) \to \chi_i (k) \xi^c (p - k)$ up to one loop is given by
\begin{align}
	\minus i (V_{0 \Phi})_{i \alpha} (p^2) &= \minus i f_{i \alpha} + \mu^{4 - d} \sum_{j, \beta, \gamma} \int \frac{d^d \ell}{(2 \pi)^d} \frac{i}{\slashed{\ell} - \slashed{k}} (\minus i f_{i \gamma}) i (\Delta_{\Phi})_{\gamma \beta} (\minus i f_{j \beta}^*) \frac{i}{\slashed{\ell} + \slashed{p} - \slashed{k}} (\minus i f_{j \alpha}) \nonumber \\
	&= \minus i f_{i \alpha} + \mu^{4 - d} \sum_{j, \widehat{\beta}} \int \frac{d^d \ell}{(2 \pi)^d} \frac{i}{\slashed{\ell} - \slashed{k}} (\minus i \widehat{f}_{i \widehat{\beta}}) \frac{i}{\ell^2 - p_{\widehat{\Phi}_{\widehat{\beta}}}^2} (\minus i \widehat{f}_{j \widehat{\beta}}^c) \frac{i}{\slashed{\ell} + \slashed{p} - \slashed{k}} (\minus i f_{j \alpha}) \nonumber \\
	&= \minus i f_{i \alpha} + \mu^{4 - d} \sum_{j, \widehat{\beta}} \widehat{f}_{i \widehat{\beta}} \widehat{f}_{j \widehat{\beta}}^c f_{j \alpha} \int \frac{d^d \ell}{(2 \pi)^d} \frac{(\slashed{\ell} - \slashed{k}) (\slashed{\ell} + \slashed{p} - \slashed{k})}{(\ell^2 - p_{\widehat{\Phi}_{\widehat{\beta}}}^2) (\ell + p - k)^2 (\ell - k)^2} \nonumber \\
	&= \minus i f_{i \alpha} + 2 \mu^{4 - d} \sum_{j, \widehat{\beta}} \widehat{f}_{i \widehat{\beta}} \widehat{f}_{j \widehat{\beta}}^c f_{j \alpha} \int \frac{d^d \ell}{(2 \pi)^d} \int_0^1 dx \int_0^1 dy \int_0^1 dz \, \delta (1 - x - y - z) \nonumber \\
		&\qquad \qquad \qquad \qquad \qquad \frac{(\slashed{\ell} - \slashed{k}) (\slashed{\ell} + \slashed{p} - \slashed{k})}{[x (\ell^2 - p_{\widehat{\Phi}_{\widehat{\beta}}}^2) + y (\ell + p - k)^2 + z (\ell - k)^2]^3}.
\end{align}
Note that $\mu^{4 - d}$ should be used although three Yukawa couplings are involved because $V_{i \alpha} (p^2)$ is a dimensionless quantity corresponding to the loop corrections to $f_{i \alpha}$. The integration over $\ell$ can be rewritten as
\begin{align}
	\int \frac{d^d \ell}{(2 \pi)^d} &\frac{(\slashed{\ell} - \slashed{k}) (\slashed{\ell} + \slashed{p} - \slashed{k})}{[x (\ell^2 - p_{\widehat{\Phi}_{\widehat{\beta}}}^2) + y (\ell + p - k)^2 + z (\ell - k)^2]^3} \nonumber \\
	&= \int \frac{d^d q}{(2 \pi)^d} \frac{q^2 - y (1 - y) p^2 + (1 - y - z) \slashed{p} \slashed{k} - 2 (1 - y) (1 - y - z) (p \cdot k)}{[q^2 + 2 y z (p \cdot k) - x p_{\widehat{\Phi}_{\widehat{\beta}}}^2]^3} \nonumber \\
	&\to \int \frac{d^d q}{(2 \pi)^d} \frac{q^2 - (1 - y) (1 - z) p^2}{(q^2 + y z p^2 - x p_{\widehat{\Phi}_{\widehat{\beta}}}^2)^3},
\end{align}
where $q = \ell + y (p - k) - z k$ and we have used $2 p \cdot k = 2 (p - k) \cdot k = p^2$ and $\overline{u_{\chi_i}} \slashed{k} = 0$. Defining $r_{\widehat{\beta}} (p^2) \coloneqq p_{\widehat{\Phi}_{\widehat{\beta}}}^2 / p^2$,
we write
\begin{align}
	&2 \mu^{4 - d} \int_0^1 dx \int_0^1 dy \int_0^1 dz \, \delta (1 - x - y - z) \int \frac{d^d q}{(2 \pi)^d} \ \frac{q^2 - (1 - y) (1 - z) p^2}{(q^2 + y z p^2 - x m_{\Phi_{\widehat{\beta}}}^2)^3} \nonumber \\
	&= \frac{i}{(4 \pi)^2} \Bigg[ \frac{2}{\epsilon} - \log{\frac{p^2}{\widetilde{\mu}^2}}
			+1 - r_{\widehat{\beta}} + (1 + r_{\widehat{\beta}})^2 \log{(1 + r_{\widehat{\beta}}^{-1})} + s_{\widehat{\beta}} \nonumber \\
		&\qquad \qquad + i \pi \left\{ 1 + 2 r_{\widehat{\beta}} (2 + r_{\widehat{\beta}}) + (1 + r_{\widehat{\beta}}) \log{(1 + r_{\widehat{\beta}}^{-1})} \right\} \bigg] + \mathcal{O} (\epsilon),
\end{align}
where
\begin{align}
	s_{\widehat{\beta}} (p^2) \coloneqq \int_0^{r_{\widehat{\beta}}^{-1}} dx \, \bigg[ (1 - r_{\widehat{\beta}}) r_{\widehat{\beta}} - \frac{1 + r_{\widehat{\beta}}}{1 + x} + \frac{(1 + r_{\widehat{\beta}})^2}{(1 + x)^2} \bigg] \log{x}.
\end{align}
We therefore obtain
\begin{align}
	(V_{0 \Phi})_{i \alpha} (p^2) &= f_{i \alpha} - \frac{1}{16 \pi^2} \sum_{j, \widehat{\beta}} \widehat{f}_{i \widehat{\beta}} \widehat{f}_{j \widehat{\beta}}^c f_{j \alpha}
		\Bigg[ \frac{2}{\epsilon} - \log{\frac{p^2}{\widetilde{\mu}^2}}
			+1 - r_{\widehat{\beta}} + (1 + r_{\widehat{\beta}})^2 \log{(1 + r_{\widehat{\beta}}^{-1})} + s_{\widehat{\beta}} \nonumber \\
		&\qquad \qquad \qquad \qquad + i \pi \left\{ 1 + 2 r_{\widehat{\beta}} (2 + r_{\widehat{\beta}}) + (1 + r_{\widehat{\beta}}) \log{(1 + r_{\widehat{\beta}}^{-1})} \right\} \bigg] + \mathcal{O} (\epsilon) \nonumber \\
	&= f_{i \alpha} - \frac{1}{16 \pi^2} \sum_{j, \beta} f_{i \beta} f_{j \beta}^* f_{j \alpha} \bigg( \frac{2}{\epsilon} - \log{\frac{p^2}{\widetilde{\mu}^2}} \bigg) \nonumber \\
		&\qquad - \frac{1}{16 \pi^2} \sum_{j, \widehat{\beta}} \widehat{f}_{i \widehat{\beta}} \widehat{f}_{j \widehat{\beta}}^c f_{j \alpha}
		\Bigg[ 1 - r_{\widehat{\beta}} + (1 + r_{\widehat{\beta}})^2 \log{(1 + r_{\widehat{\beta}}^{-1})} + s_{\widehat{\beta}} \nonumber \\
		&\qquad \qquad \qquad \qquad + i \pi \left\{ 1 + 2 r_{\widehat{\beta}} (2 + r_{\widehat{\beta}}) + (1 + r_{\widehat{\beta}}) \log{(1 + r_{\widehat{\beta}}^{-1})} \right\} \bigg] + \mathcal{O} (\epsilon),
\end{align}
where, in the second identity, we have used equations \ref{eq:CfCi}, \ref{eq:CiTCf}, \ref{eq:MixingMat}, and
\begin{align}
	(\widehat{f} \widehat{f}^{c \mathsf{T}})_{ij}
	= \sum_{\beta, \gamma, \widehat{\alpha}} f_{i \gamma} (C_{\widehat{\Phi}^f})_{\gamma \widehat{\alpha}} (C_{\widehat{\Phi}^i})_{\beta \widehat{\alpha}} f_{j \beta}^*
	= (f f^\dag)_{ij} + \mathcal{O} (\alpha^2).
\end{align}
In addition,
\begin{align}
	(V_{0 \Phi^*})_{i \alpha} (p^2) &= f_{i \alpha}^* - \frac{1}{16 \pi^2} \sum_{j, \beta} f_{i \beta}^* f_{j \beta} f_{j \alpha}^* \bigg( \frac{2}{\epsilon} - \log{\frac{p^2}{\widetilde{\mu}^2}} \bigg) \nonumber \\
		&\qquad - \frac{1}{16 \pi^2} \sum_{j, \widehat{\beta}} \widehat{f}_{i \widehat{\beta}}^c \widehat{f}_{j \widehat{\beta}} f_{j \alpha}^*
		\Bigg[ 1 - r_{\widehat{\beta}} + (1 + r_{\widehat{\beta}})^2 \log{(1 + r_{\widehat{\beta}}^{-1})} + s_{\widehat{\beta}} \nonumber \\
		&\qquad \qquad \qquad \qquad + i \pi \left\{ 1 + 2 r_{\widehat{\beta}} (2 + r_{\widehat{\beta}}) + (1 + r_{\widehat{\beta}}) \log{(1 + r_{\widehat{\beta}}^{-1})} \right\} \bigg] + \mathcal{O} (\epsilon).
\end{align}
The renormalized vertex function is given by
\begin{align}
	V_\Phi (p^2) = V_{0 \Phi} (p^2) + \delta f, \qquad
	V_{\Phi^*} (p^2) = V_{0 \Phi^*} (p^2) + \delta f^*.
\end{align}
Choosing the vertex conterterm
\begin{align}
	\boxed{\delta f_{i \alpha} = \frac{1}{16 \pi^2} \sum_{j, \beta} f_{i \beta} f_{j \beta}^* f_{j \alpha} \bigg( \frac{2}{\epsilon} - \log{\frac{\widetilde{\mu}^2}{m_{\Phi_\alpha}^2}} \bigg),}
\end{align}
we can write the renormalized vertex function as
\begin{empheq}[box=\fbox]{align}
	(V_\Phi)_{i \alpha} (p^2) &= f_{i \alpha} - \frac{1}{16 \pi^2} \sum_{j, \widehat{\beta}} \widehat{f}_{i \widehat{\beta}} \widehat{f}_{j \widehat{\beta}}^c f_{j \alpha}
		\bigg[ \minus \log{\frac{p^2}{m_{\Phi_\alpha}^2}}
			+1 - r_{\widehat{\beta}} + (1 + r_{\widehat{\beta}})^2 \log{(1 + r_{\widehat{\beta}}^{-1})} + s_{\widehat{\beta}} \nonumber \\
		&\qquad \qquad \qquad \qquad + i \pi \left\{ 1 + 2 r_{\widehat{\beta}} (2 + r_{\widehat{\beta}}) + (1 + r_{\widehat{\beta}}) \log{(1 + r_{\widehat{\beta}}^{-1})} \right\} \bigg], \\
	(V_{\Phi^*})_{i \alpha} (p^2) &= f_{i \alpha}^* - \frac{1}{16 \pi^2} \sum_{j, \widehat{\beta}} \widehat{f}_{i \widehat{\beta}}^c \widehat{f}_{j \widehat{\beta}} f_{j \alpha}^*
		\bigg[ \minus \log{\frac{p^2}{m_{\Phi_\alpha}^2}}
			+1 - r_{\widehat{\beta}} + (1 + r_{\widehat{\beta}})^2 \log{(1 + r_{\widehat{\beta}}^{-1})} + s_{\widehat{\beta}} \nonumber \\
		&\qquad \qquad \qquad \qquad + i \pi \left\{ 1 + 2 r_{\widehat{\beta}} (2 + r_{\widehat{\beta}}) + (1 + r_{\widehat{\beta}}) \log{(1 + r_{\widehat{\beta}}^{-1})} \right\} \bigg],
\end{empheq}
where
\begin{align}
	\boxed{s_{\widehat{\beta}} (p^2) = \int_0^{r_{\widehat{\beta}}^{-1}} dx \, \bigg[ (1 - r_{\widehat{\beta}}) r_{\widehat{\beta}} - \frac{1 + r_{\widehat{\beta}}}{1 + x} + \frac{(1 + r_{\widehat{\beta}})^2}{(1 + x)^2} \bigg] \log{x}, \qquad
	r_{\widehat{\beta}} (p^2) = \frac{m_{\widehat{\Phi}_{\widehat{\beta}}}^2}{p^2}.}
\end{align}
Here, we have used $r_{\widehat{\beta}} = p_{\widehat{\Phi}_{\widehat{\beta}}}^2 / p^2 = m_{\widehat{\Phi}_{\widehat{\beta}}}^2 [1 + \mathcal{O} (\alpha)] / p^2$. Note that $s_{\widehat{\beta}}$ is infrared divergent because we have assumed that $\chi_i$ are massless in this calculation.


\end{document}